\documentclass[a4paper,11pt]{article}
\usepackage{jheppub}

\makeatletter
\def\@fpheader{}
\makeatother

\usepackage{units}
\usepackage{amsmath}
\usepackage{graphicx}
\usepackage{dcolumn}
\usepackage{pbox}
\usepackage{amssymb}
\usepackage{epsfig}
\usepackage{slashed}
\usepackage{gensymb}
\usepackage{amssymb}
\usepackage{mathrsfs}
\usepackage[usenames,dvipsnames]{xcolor}

\usepackage{multirow}
\usepackage{lineno}
\usepackage{autobreak}
\usepackage{lipsum}
\usepackage{cancel}
\usepackage{xparse}
\usepackage{physics}
\usepackage{booktabs}
\usepackage{bm}
\usepackage{comment}
\usepackage{subcaption}
\usepackage{graphicx}
\usepackage{bbold}
\usepackage{xcolor}
\usepackage{array}    
\usepackage{makecell}

\NewDocumentCommand\semiloop{O{black}mmmO{}O{above}}
{%
\draw[#1] let \p1 = ($(#3)-(#2)$) in (#3) arc (#4:({#4+180}):({0.5*veclen(\x1,\y1)})node[midway, #6] {#5};)
}

\definecolor{MyLightBlue}{rgb}{0.22,0.51,0.9}
\definecolor{BrickRed}{rgb}{0.8, 0.25, 0.33}
\RequirePackage{hyperref}
\hypersetup{colorlinks, citecolor=blue,linkcolor=BrickRed, urlcolor=MyLightBlue}

\usepackage{bbding}
\usepackage{pifont}
\usepackage{wasysym}
\usepackage{amssymb}
\usepackage{tabu}

\usepackage{autobreak}

\newcommand{\be}{\begin{equation}}
\newcommand{\ee}{\end{equation}}

\def\beq{\begin{equation}}
\def\eeq{\end{equation}}
\def\beqr{\begin{eqnarray}}
\def\eeqr{\end{eqnarray}}

\def\da{\dagger}

\title{Universal Seesaw Pati-Salam Model with\\
P for Strong CP}

\author[a]{\bf K.S. Babu,}
\emailAdd{babu@okstate.edu}
\affiliation[a]{Department of Physics, Oklahoma State University, Stillwater, OK 74078, USA}

\author[a]{\bf Sumit Biswas}
\emailAdd{sumit.biswas@okstate.edu}

\abstract{We develop a universal seesaw version of the Pati-Salam model wherein quarks and leptons of each family are unified into common multiplets transforming as 
\{$\psi_L(2,1,4)+ \psi_R(1,2,4)$\} under the $SU(2)_L \times SU(2)_R \times SU(4)_c$ gauge symmetry. Parity symmetry is spontaneously broken in the model, which helps in solving the strong CP problem without the axion. The Higgs sector of the model is very simple, consisting of a single pair of \{$H_L(2,1,4)+ H_R(1,2,4)$\} fields. Fermion masses arise through mixing of the chiral fermions with vector-like quarks and leptons contained in $(1,1,15)$ as well as \{$(1,1,10)_L+(1,1,10)_R$\} multiplets via a universal seesaw mechanism. Consistency of such a spectrum with the observed quark and lepton masses is established. The parity solution to the strong CP problem is shown to be effective in this framework, although there are new loop contributions to $\overline{\theta}$, compared to the analogous left-right symmetric model, arising from color sextet and octet fermions, as well as from diagrams mediated by leptoquark bosons.  We also find that, in this setup, although lepton number is broken, neutrino masses remain zero at the tree-level. Small and finite Majorana neutrino masses are induced via one-loop diagrams, which we analyze and show to be compatible with oscillation experiments. 
}

\begin{document}
\maketitle
\flushbottom

\section{Introduction}

One of the most elegant extensions of the Standard Model (SM) is the Pati-Salam (PS) model based on the non-Abelian gauge symmetry $SU(2)_L \times SU(2)_R \times SU(4)_c$~\cite{Pati:1974yy}. With the embedding of the color group  $SU(3)_c$ of the SM in $SU(4)_c$, quarks and leptons are unified into common multiplets. This enables the identification of lepton number as the fourth color. The non-Abelian nature of the symmetry group implies that electric charge is quantized, which explains one of the puzzles of the SM.  Parity is a good symmetry of the model, which is broken spontaneously, providing a symmetry-based rationale for the $({\rm V}-{\rm A})$ 
structure of weak interactions.  Furthermore, the gauge structure of the model dictates that the left-handed neutrino be accompanied by a right-handed partner, which paves the way for small neutrino masses via the seesaw mechanism.

The purpose of this paper is to develop a universal seesaw version of the Pati-Salam model, which also solves the strong CP problem by virtue of parity symmetry. The axion is not needed for solving the strong CP problem in this framework. 
The fermions of each family belong to $\{\psi_L(2,1,4)+\psi_R(1,2,4)\}$ of the PS gauge symmetry. In conventional PS models, symmetry breaking and fermion mass generation are accomplished by introducing three sets of Higgs fields: (i) $\{\Delta_L(3,1,10)+\Delta_R(1,3,10)\}$, used for the PS gauge symmetry breaking down to the SM and for the generation of right-handed neutrino Majorana masses; (ii) $\Phi(2,2,1)$ and (iii) $\Sigma(2,2,15)$, which are introduced for realistic fermion mass generation as well as for electroweak symmetry breaking. While such models are fully consistent, the proliferation of Higgs fields and the associated Higgs potential parameters may be viewed as a blemish of these realizations. For a detailed study of the Higgs sector of such models, see  Ref.~\cite{Saad:2017pqj}. With this Higgs sector, it is difficult to solve the strong CP problem via parity symmetry as the determinant of the quark mass matrix is not real and contributes to the CP-violating parameter $\overline{\theta}$ at tree-level.  

An alternative to the conventional Higgs sector mentioned above is to introduce a single set of $\{H_L(2,1,4)+H_R(1,2,4)\}$ fields, which is sufficient for the purpose of symmetry breaking.  $H_R$ would break $SU(2)_R \times SU(4)_c$ down to $U(1)_Y \times SU(3)_c$ at an energy scale well above the weak scale, while $H_L$ breaks the $SU(2)_L \times U(1)_Y$ down to $U(1)_{\rm em}$. Such a simple Higgs sector alone, however, cannot generate masses for quarks and leptons.  To remedy this issue, one can introduce vector-like fermions with which the chiral fermions mix, thus generating their masses via a universal seesaw mechanism~\cite{Berezhiani:1983hm,Berezhiani:1985in,Berezhiani:1985vnk,Davidson:1987mh}.   Such models have been studied in the context of left-right symmetry based on the gauge group $SU(2)_L \times SU(2)_R \times U(1)_{B-L} \times SU(3)_c$~\cite{Davidson:1987mh,Babu:1988mw,Babu:1989rb,Hall:2018let,Hall:2019qwx,Craig:2020bnv}, which is a subgroup of the PS symmetry. These models can solve the strong CP problem by parity symmetry without the need for the axion~\cite{Babu:1988mw}. Invariance of the Lagrangian under parity sets the QCD $\theta$-term to zero, and the parity-symmetric seesaw structure of the quark mass matrix makes its determinant  real at tree level, and also at the one-loop level, making the strong CP-violation parameter $\overline{\theta}$ sufficiently small to be consistent with neutron electric dipole moment (nEDM) limits~\cite{Babu:1989rb,Hall:2018let,
Hall:2019qwx,Craig:2020bnv,deVries:2021pzl,Hisano:2023izx}.\footnote{For alternative approaches to solve the strong CP problem with parity symmetry in left-right symmetric models, see Ref.~\cite{Beg:1979dp,Mohapatra:1978fy,
Barr:1991qx,Mohapatra:1995xd,Mohapatra:1997su,Babu:2001se,Kuchimanchi:2010xs,Kuchimanchi:2023imj}.} The same mechanism can work in the proposed PS model, although there are certain important differences compared to the left-right symmetric models, which we shall elaborate on in this paper. In particular, the $\overline{\theta}$ parameter receives contributions from loop corrections to the masses of color octet fermions contained in $(1,1,15)$ as well as from color sextets from $\{(1,1,10)_L+(1,1,10)_R\}$ vector-like fermions. In addition, there are new loop diagrams correcting the up-type and down-type quark mass matrices through the exchange of vector and scalar leptoquarks. In spite of these new diagrams, we find that $\overline{\theta}$ can be sufficiently small for a wide range of model parameters so as to be consistent with nEDM limits. Specifically, we find that the one-loop induced $\overline{\theta}$ parameter is parametrically suppressed if either of the vector-like fermions $(1,1,15)$ or $(1,1,10)$ has a bare mass much smaller than the parity restoration scale of order $5 \times 10^{13}$ GeV.

There is yet another motivation for the PS models with the simple Higgs sector belonging to $\{H_L(2,1,4)+H_R(1,2,4)\}$.  Once parity symmetry is imposed, the Higgs potential will have a single mass parameter in this setup. In the exact parity-symmetric limit, this setup does not introduce an additional independent Higgs-sector tuning beyond the usual tuning of the electroweak scale relative to a high cutoff, such as the Planck scale~\cite{Hall:2018let}. This point is further discussed in Sec.~\ref{sub:exact parity}. This is a special feature shared only by very few models, including this version of the PS model, its left-right symmetric subgroup, and a parity-symmetric mirror model~\cite{Foot:1995pa,Berezhiani:1995yi,Hall:2019qwx}.

We find the simplest set of vector-like fermions that can generate realistic quark and lepton masses through mixing in the PS model to be $(1,1,15)$ and $\{(1,1,10)_L+(1,1,10)_R\}$. The former real field contains a vector-like iso-singlet up-type quark, while the latter contains a vector-like down-type quark as well as a lepton. We show that with this spectrum, the closely related down-quark and charged lepton mass matrices can be consistently decoupled to realize correct masses for all fermions. A consequence of this setup is that, although lepton number is broken in the model, neutrino masses remain zero at the tree-level.  Small and finite Majorana masses for the neutrinos arise at one-loop, which we analyze and show consistency with neutrino oscillation data. 

Variations of the Pati-Salam model with vector-like fermions have been studied in Ref.~\cite{Dolan:2020doe,Clarke:2011aa,Babu:2025azv} to lower the PS scale down to multi-TeV. For other variants of the PS model, see Refs.~\cite{Kuznetsov:1994tt,Volkas:1995yn,Foot:1997pb,Foot:1999wv,Smirnov:2007hv,FileviezPerez:2013zmv,Calibbi:2017qbu,Greljo:2018tuh,Heeck:2018ntp,Iguro:2021kdw,FernandezNavarro:2022gst,Butterworth:2025ttw}. In contrast to these works, the PS scale in our proposal is high, of order $5 \times 10^{13}$ GeV, which is the scale where the two $SU(2)$ gauge couplings become equal, $g_{2L} = g_{2R}$.

\section{Pati-Salam model with universal seesaw}

The Pati-Salam model that we develop here is based on the gauge symmetry $SU(2)_L \times SU(2)_R \times SU(4)_c$. Each family of quarks and leptons transforms under the gauge group as
\begin{eqnarray}
\Psi_L(2,1,4) = \left(\begin{matrix}  u_1 & u_2 & u_3 & \nu  \cr d_1 & d_2 & d_3 & e \end{matrix} \right)_L,~~~\Psi_R(1,2,4) = \left(\begin{matrix}  u_1 & u_2 & u_3 & \nu  \cr d_1 & d_2 & d_3 & e \end{matrix} \right)_R~.
\label{eq:fermions}
\end{eqnarray}
Quarks are unified with leptons, with lepton number serving as the fourth color. The model developed here has a simple Higgs sector consisting of
\begin{eqnarray}
H_L(2,1,4) = \left(\begin{matrix}  \chi_1^u  & \chi_2^u & \chi_3^u & \chi^\nu  \cr \chi_1^d & \chi_2^d & \chi_3^d & \chi^e \end{matrix} \right)_L,~~~H_R(1,2,4) = \left(\begin{matrix}  \chi_1^u  & \chi_2^u & \chi_3^u & \chi^\nu  \cr \chi_1^d & \chi_2^d & \chi_3^d & \chi^e \end{matrix}\right)_R~.
\label{eq:bosons}
\end{eqnarray}
Spontaneous gauge symmetry breaking occurs when the neutral components of $H_L$ and $H_R$ acquire vacuum expectation values (VEVs), denoted as
\begin{eqnarray}
\left\langle \chi^\nu_{L}\right \rangle  = \kappa_L,~~~\left\langle \chi^\nu_{R} \right\rangle   = \kappa_R,
\label{eq:VEVs}
\end{eqnarray}
with the hierarchy $\kappa_R \gg \kappa_L$. 
The symmetry breaking occurs in two stages:
\begin{eqnarray}
SU(2)_L \times SU(2)_R \times SU(4)_c  \xrightarrow{\kappa_R} SU(2)_L \times U(1)_Y \times SU(3)_c \xrightarrow{\kappa_L} SU(3)_c \times U(1)_{\rm em}.
\end{eqnarray}
We note here that by $SU(2)_{L,R}$ rotations, the VEVs $(\kappa_L,\,\kappa_R)$ can be made real, a feature of this minimal Higgs sector that plays a role in the parity solution to the strong CP problem.

With this Higgs structure, while spontaneous gauge symmetry breaking can be achieved,  quarks and leptons will remain massless, as no Yukawa couplings are allowed between the fermions of Eq. (\ref{eq:fermions}) and the Higgs bosons of Eq. (\ref{eq:bosons}). Fermion mass generation can be achieved via a universal seesaw mechanism~\cite{Davidson:1987mh} where the chiral families mix with vector-like fermion families through Yukawa couplings involving the Higgs fields of Eq. (\ref{eq:bosons}). The simplest set of vector-like fermions that can lead to realistic fermion masses and mixings is found to be three families transforming as
\begin{eqnarray}
&&\Sigma_L(1,1,15) = \frac{1}{\sqrt{2}}\left(\begin{matrix} \frac{{\cal O}_3} {\sqrt{2}} + \frac{{\cal O}_8}{\sqrt{6}} + \frac{N}{\sqrt{12}}&  {\cal O}_{12} & {\cal O}_{13} & U_1 \cr {\cal O}_{12}^c & -\frac{{\cal O}_3}{\sqrt{2}}+\frac{{\cal O}_8}{\sqrt{6}} + \frac{N}{\sqrt{12}}&  {\cal O}_{23} & U_2 \cr {\cal O}_{13}^c & {\cal O}_{23}^c & -\sqrt{\frac{2}{3}} {\cal O}_8+ \frac{N}{\sqrt{12}}  & U_3 \cr U_1^c & U_2^c & U_3^c & -\frac{\sqrt{3}}{2}N    \end{matrix} \right)_{L} \label{eq:octet}\\
&&\Omega_{L,R}(1,1,10) = \frac{1}{\sqrt{2}}\left(\begin{matrix} \sqrt{2} {\cal S}_{11}&  {\cal S}_{12} & {\cal S}_{13} & D_1 \cr {\cal S}_{12} & \sqrt{2} {\cal S}_{22} &  {\cal S}_{23} & D_2 \cr {\cal S}_{13} & {\cal S}_{23} & \sqrt{2} {\cal S}_{33} & D_3 \cr D_1 & D_2 & D_3 & \sqrt{2} E^-    \end{matrix} \right)_{L,R}~.
\label{eq:sextet}
\end{eqnarray}

The $\Sigma_L(1,1,15)$ decomposes under $SU(3)_c \times SU(2)_L \times U(1)_Y$ as
\begin{equation}
\Sigma_L(1,1,15) = U(3,1,\frac{2}{3}) + U^c(\overline{3},1,-\frac{2}{3}) + {\cal O}(8,1,0) + N(1,1,0)
\end{equation}
which contains a vector-like up-type quark $U$, a Majorana singlet $N_L$ as well as a Majorana octet ${\cal O}$.  The up-type quarks will mix with the vector-like $U$ quarks and acquire their masses. Similarly, the $\nu_L$ and $(\nu_R)^c$ will mix with the $N_L$ from $\Sigma_L$ and generate light neutrino masses. The color octet, while unmixed with other fermions, can contribute to the strong CP parameter $\overline{\theta}$, which we shall address later.

The $\{\Omega_L(1,1,10) + \Omega_R(1,1,10)\}$ fields contain a vector-like $D$ quark, a vector-like charged lepton $E^-$ and a vector-like color sextet ${\cal S}$.  Its decomposition under $SU(3)_c \times SU(2)_L \times U(1)_Y$ is given by
\begin{equation}
\Omega_{L,R}(1,1,10) = D_{L,R}(3,1,-\frac{1}{3}) + E_{L,R}^-(1,1,-1) + {\cal S}_{L,R}(6,1,\frac{1}{3})~.
\end{equation}
The down-type quarks will mix with the $D$ quarks, and the charged leptons will mix with the $E^-$ leptons to acquire masses. Although the vector-like $D$
-quarks and $E^-$ leptons are unified in common multiplets, we shall show that consistent masses for the light down-type quarks and charged leptons can be realized within this setup.

Under parity symmetry, the fermion and scalar fields transform as
\begin{equation}
 \Psi_L \leftrightarrow \Psi_R,~~\Omega_L \leftrightarrow \Omega_R,~~\Sigma_L \leftrightarrow \Sigma^c_R,~~H_L \leftrightarrow H_R  
 \label{eq:parity}
\end{equation}
along with the $SU(2)_{L,R}$ gauge bosons transforming as $W_L \leftrightarrow W_R$. The action of parity on $\Sigma_L$ can be written explicitly as $N_L \leftrightarrow N^c_R$,  ${\cal O}_L \leftrightarrow {\cal O}^c_R$ and $U_L \leftrightarrow U_R$.  The parity-symmetric Yukawa/mass Lagrangian of the model can be written as
\begin{eqnarray}
{\cal L}_{\rm Yuk} &=& \sqrt{2} Y_{15}^\dagger \left\{ {\rm Tr}\left(\Sigma_L \Psi_L^T H_L^*\right) + {\rm Tr}\left(\overline{\Sigma}_L \Psi_R^T H_R^*\right)\right\}\nonumber\\
&-& \sqrt{2} Y_{10}\left\{ {\rm Tr}\left(\overline{\Psi}_L \tilde{H}_L \Omega_R\right) + {\rm Tr} \left(\overline{\Psi}_R \tilde{H}_R \Omega_L \right) \right\} \nonumber \\
&+& M_{10}\, \overline{\Omega}_L \Omega_R + M_{15} \,(\overline{\Sigma^c})_R  \Sigma_L + h.c.
\label{eq:Yuk}
\end{eqnarray}
Here we have defined $\tilde{H}_{L,R} = i \tau_2 H^*_{L,R}$. The Dirac spinor contraction in the first term of $Y_{15}$ is via the charge conjugation matrix $C$, and we have used the identity $\Sigma^c_R \,C =\overline{\Sigma}_L$. Note that the  $\overline{\Sigma}_L$ is defined with a transposition of the matrix of Eq. (\ref{eq:octet}). 
The gauge invariance of these Yukawa terms can be verified by applying special unitary rotations on various fields given as
\begin{eqnarray}
&&\Psi_L \rightarrow V_L \Psi_L V_c^T,~~\Psi_R \rightarrow V_R \Psi_R V_c^T,~~\Sigma_L \rightarrow V_c \Sigma_L V_c^\dagger,~~\Omega_{L,R}\rightarrow V_c \Omega_{L,R} V_c^T, \nonumber \\
&&\hspace*{1.1in} H_L \rightarrow V_L H_L V_c^T,~~H_R \rightarrow V_R H_R V_c^T~.
\label{eq:transform}
\end{eqnarray}
Here $V_L$, $V_R$, and $V_c$ are, respectively, the special unitary rotation matrices corresponding to $SU(2)_L$, $SU(2)_R,$ and $SU(4)_c$ gauge transformations.
Owing to parity symmetry, the mass matrices $M_{10}$ and $M_{15}$ of Eq. (\ref{eq:Yuk}) are hermitian:
\begin{equation}
M_{10}^\dagger = M_{10},~~M_{15}^\dagger = M_{15}~.
\end{equation}
Furthermore, $M_{15}$ is also symmetric, which follows from the Lorentz structure, implying that $M_{15}$ is real and symmetric.

The Yukawa/mass terms of Eq. (\ref{eq:Yuk}) can be expanded explicitly in component form as
\begin{eqnarray}
&&{\cal L}_{\rm Yuk}= Y_{15}^\dagger  \left(U_L^c u_L \chi^{\nu *}_{L} + U^c_L d_L \chi^{e *}_{L}  +  U_L e_L \chi_L^{d*} +  U_L \nu_L \chi_L^{u*} 
 -\frac{\sqrt{3}}{2} N_L \nu_L  \chi^{\nu *}_{L} - \frac{\sqrt{3}}{2}  N_Le_L  \chi^{e *}_L \right. \nonumber \\
&&\left. + \frac{1}{2 \sqrt{3}} N_L d_L \chi_L^{d*} + \frac{1}{2\sqrt{3}}  N_L u_L \chi_L^{u*} + \sqrt{2} {\cal O}^\alpha_{L \beta} d_{L \alpha} \chi_L^{d* \beta} + \sqrt{2}  {\cal O}^\alpha_{L \beta} u_{L \alpha} \chi_L^{u* \beta} \right) \nonumber \\
&& + Y_{15}^\dagger \left(\overline{U}_L u_R \chi_R^{\nu *} + \overline{U}_L d_R \chi^{e *}_R + \overline{U^c_L} e_R \chi_R^{d *} + \overline{U^c_L} \nu_R \chi_R^{u *} - \frac{\sqrt{3}}{2}\overline{N_L} \nu_R \chi_R^{\nu *} -\frac{\sqrt{3}}{2} \overline{N_L} e_R \chi_R^{e *}\right. \nonumber \\
&&\left. + \frac{1}{2\sqrt{3}} \overline{N_L} d_R \chi_R^{d *} + \frac{1}{2 \sqrt{3}} \overline{N_L} u_R \chi_R^{u *} 
+ \sqrt{2} \overline{{\cal O}}^\alpha_{L \beta} d_{R \alpha} \chi_R^{d* \beta} + \sqrt{2} \overline{{\cal O}}_{L \beta}^\alpha u_{R \alpha} \chi_R^{u * \beta}
\right) \nonumber \\
&& + Y_{10} \left( \overline{d}_L D_R \chi_L^{\nu *} - \overline{u}_L D_R \chi_L^{e *} + \sqrt{2} \overline{e}_L E_R \chi_L^{\nu *} - \sqrt{2} \overline{\nu}_L E_R \chi_L^{ e*} + \overline{e}_L D_R \chi_L^{u *} - \overline{\nu}_L D_R \chi_L^{ d*} \right. \nonumber \\
&&\left. -\overline{u}_L ^\alpha ({\cal S}_{R})_{ \alpha \beta} \chi_L^{d* \beta } +\overline{d}^\alpha_L ({\cal S}_{R})_ {\alpha \beta} \chi_L^{u* \beta}\right) +  Y_{10} \left( \overline{d}_R D_L \chi_R^{\nu *} - \overline{u}_R D_L \chi_R^{e *} + \sqrt{2} \overline{e}_R E_L \chi_R^{\nu *} \right. \nonumber \\
&&\left. - \sqrt{2} \overline{\nu}_R E_L \chi_R^{ e*} + \overline{e}_R D_L \chi_R^{u *} - \overline{\nu}_R D_L \chi_R^{ d*} - \overline{u}_R ^\alpha ({\cal S}_{L})_{ \alpha \beta} \chi_R^{d * \beta} +\overline{d}^\alpha_R ({\cal S}_{L})_ {\alpha \beta} \chi_R^{u* \beta}\right) \nonumber \\
&&\hspace*{-0.07in} + M_{10} \left(\overline{D}_L D_R + \overline{E}_L E_R + \overline{\cal S}_L {\cal S}_R \right) + M_{15} \left(U^c_L U_L+ \frac{1}{2}N_L N_L + \frac{1}{2}({\cal O}_L)_\alpha^\beta ({\cal O}_L)_\beta^\alpha  \right) +\rm {h.c}.
\label{eq:Yukexpand}
\end{eqnarray}
Here the color octet $({\cal O}_L)_{\beta}^\alpha$ and the color sextet ${\cal S}_{\alpha\beta}$ are defined in Eqs. (\ref{eq:octet})-(\ref{eq:sextet}) as the upper-left $3 \times 3$ block matrix with elements denoted as ${\cal O}$ and ${\cal S}$ respectively. Once the VEVs of the Higgs fields, as given in Eq. (\ref{eq:VEVs}), are inserted into Eq. (\ref{eq:Yukexpand}), fermion mass matrices with a universal seesaw structure will be generated. 

For the charged fermions, the mass matrices ${\cal M}_f$ written in a basis 
\begin{equation}
{\cal L}_{\rm mass} = \left( \overline{f}_L\, \overline{F}_L \right)  {\cal M}_f \left(\begin{matrix}  f_R \cr F_R \end{matrix}   \right),
\label{eqn:basis_def}
\end{equation}
where $f = (u,d,e)$ and $F=(U,D,E)$, take the form:
\begin{eqnarray}
\hspace*{-0.07in} {\cal M}_u = \left(\begin{matrix} 0 & Y_{15} \kappa_L \cr Y_{15}^\dagger \kappa_R & M_{15}^\dagger   \end{matrix} \right), ~{\cal M}_d = \left(\begin{matrix}  0 & Y_{10} \kappa_L \cr Y_{10}^\dagger \kappa_R & M_{10}  \end{matrix} \right), ~{\cal M}_e = \left(\begin{matrix}  0 & \sqrt{2} Y_{10} \kappa_L \cr \sqrt{2} Y_{10}^\dagger \kappa_R & M_{10}  \end{matrix} \right).
\label{eq:matrices}
\end{eqnarray}
In addition, the color-octet and color sextet fermion masses are given by the Hermitian mass matrices
\begin{equation}
{\cal M}_{\cal O} = M_{15}^\dagger,~~~{\cal M}_{\cal S} = M_{10}~.
\label{eqn:sextet-octet-mass}
\end{equation}
The light charged fermion masses arise via a universal seesaw, with the $3 \times 3$ mass matrices for $M_{u,d,e}$ given in the approximation 
$(Y_{15}^\dagger \kappa_R M_{15}^{-1}) \ll 1$ and $(Y_{10}^\dagger \kappa_R M_{10}^{-1}) \ll 1$ by
\begin{eqnarray}
M_u &\simeq& - \left(Y_{15} M_{15}^{-1} Y_{15}^\dagger\right) \kappa_L \kappa_R, 
\label{eq:uni1}\\
M_d &\simeq& - \left(Y_{10} M_{10}^{-1} Y_{10}^\dagger\right) \kappa_L \kappa_R,  
\label{eq:uni2} \\
M_e &\simeq& - 2 \left(Y_{10} M_{10}^{-1} Y_{10}^\dagger\right)\kappa_L \kappa_R~.
\label{eq:uni3}
\end{eqnarray}
This approximation  $(Y_{15}^\dagger \kappa_R M_{15}^{-1}) \ll 1$ and $(Y_{10}^\dagger \kappa_R M_{10}^{-1}) \ll 1$ may not be valid for the third family quarks and leptons, especially the top quark.  We shall allow for the possibility that for the third family, $(Y_{15}\,\kappa_R)_{33} \lesssim (M_{15})_{33}$ as well as $(Y_{10}\,\kappa_R)_{33} \lesssim (M_{10})_{33}$.

Note that the charged lepton mass matrix $M_e$ of Eq. (\ref{eq:uni3}) and the down-type quark mass matrix $M_d$ of Eq. (\ref{eq:uni2})  are proportional, with the constant of proportionality being 2, within the validity of the approximation used.  If the opposite limit is used instead, viz., $(Y_{10}\,\kappa_R)_{33} \ll (M_{10})_{33}$, the proportionality constant will become $\sqrt{2}$, rather than 2.  This latter possibility is consistent with the mass ratio $m_\tau/m_b$ when the low energy values of the respective Yukawa couplings are extrapolated to a high energy scale of about $2 \times 10^{12}$ GeV, which is close to the scale at which Pati-Salam symmetry breaks down to the SM gauge symmetry, $\mu_{\rm PS}\sim 5.2 \times10^{13}\,\rm GeV$. While the ratio $y_\tau/y_b \simeq \sqrt{2}$ is consistent with observed mass ratios, the ratios $ y_\mu/y_s$ and $ y_e/y_d$ are inconsistent with either $\sqrt{2}$ or 2, or any value in between.  Radiative corrections proportional to the top-quark or bottom-quark Yukawa couplings can correct these wrong relations and make them compatible with observed values. This issue will be addressed in the next section, after the symmetry-breaking sector of the model is analyzed.

Note that the determinants of ${\cal M}_u$ and ${\cal M}_d$ are real, a feature that helps solve the strong CP problem via parity symmetry. We shall delegate the discussion of the neutral lepton sector and the generation of small neutrino masses in the model to Sec.~\ref{sec:sec4}, where we show that the light neutrinos obtain their masses only via one-loop radiative corrections.

\section{Higgs potential analysis}
\label{sec:sec3}
The most general Higgs potential of the model involving the $\{H_L(2,1,4) + H_R(1,2,4)\}$ fields is given by\footnote{For construction of a similar potential see Ref.~\cite{Worah:1995sb}.}
\begin{eqnarray}
V&=&-\mu_L^2\mathrm{Tr}(H^{\dagger}_LH_L)-\mu_R^2\mathrm{Tr}(H^{\dagger}_RH_R)+ \lambda_1\left\{(\mathrm{Tr}(H^{\dagger}_LH_L))^2+(\mathrm{Tr}(H^{\dagger}_RH_R))^2\right\} \nonumber \\
&+&\lambda_2\left\{\mathrm{Tr}(H^{\dagger}_LH_LH^{\dagger}_LH_L)+\mathrm{Tr}(H^{\dagger}_RH_R H^{\dagger}_RH_R)\right\}+\lambda_3\mathrm{Tr}(H^{\dagger}_LH_L)\,\mathrm{Tr}(H^{\dagger}_R H_R) \nonumber \\
&+&\lambda_4\mathrm{Tr}(H^{\dagger}_LH_LH^{\dagger}_RH_R)+
\lambda_5 \left\{\mathrm{Tr}(H_L^T\tilde{H}_L^{*}H_R^{\dagger}\tilde{H}_R)+\mathrm{Tr}(H_R^T\tilde{H}_R^{*}H_L^{\dagger}\tilde{H}_L)\right\}  \nonumber \\
&+&
\left\{\lambda_6 \,H_L^{\alpha a}H_L^{\beta b}H_R^{\gamma c}H_R^{\delta d} \epsilon_{\alpha\beta\gamma\delta}~\epsilon_{ab}\epsilon_{cd} + \mathrm{h.c.} \right\} .
\label{eq:Higgspot}
\end{eqnarray}
Here we have imposed parity invariance, as defined in Eq. (\ref{eq:parity}), which makes the quartic coupling $\lambda_5$ real. We have allowed for the possibility that parity may be broken softly by the dimension-two terms of Eq. (\ref{eq:Higgspot}).  Exact parity may be realized by setting $\mu_L^2 = \mu_R^2$ in Eq. (\ref{eq:Higgspot}), which we shall also consider. The complex coupling $\lambda_6$ connects four different fields and therefore does not appear in the minimization conditions nor in the computation of the scalar spectrum. This term violates baryon number $B$, and will lead to dimension-nine $B$-violating operators that will lead to double nucleon decay, such as $n \rightarrow e^+ e^- \nu$, which will be discussed in Sec.~\ref{Baryon number violation}. 

The vacuum expectation values $\left\langle \chi^\nu_{L}\right \rangle  = \kappa_L$ and $\left\langle \chi^\nu_{R} \right\rangle   = \kappa_R$ can be determined by minimizing the potential of Eq. (\ref{eq:Higgspot}):
\begin{align}
  \mu_L^2 &= 2(\lambda_1+\lambda_2)\,\kappa_L^2
    + (\lambda_3+\lambda_4)\,\kappa_R^2\,,\\
  \mu_R^2 &= 2(\lambda_1+\lambda_2)\,\kappa_R^2
    + (\lambda_3+\lambda_4)\,\kappa_L^2 \, .
\end{align}
The masses of the scalar bosons can be readily computed in this vacuum.  Of the 32 real degrees of freedom contained in  $\{H_L(2,1,4) + H_R(1,2,4)\}$ fields, 12 are Goldstone degrees eaten up by the $(W_{L \mu}^\pm, W_{R \mu}^\pm, Z_\mu, Z'_\mu, X_\mu)$ gauge bosons that become massive. Here $X_\mu(3,1,\frac{2}{3})$ is the leptoquark gauge boson of Pati-Salam theory. The $\chi_L^e$ and $\chi_R^e$ fields become longitudinal modes of $W_\mu^\pm$ and $W_R^\pm$.  Writing $\chi_{L,R}^\nu$ as
\begin{equation}
\chi_L^\nu = \frac{\sigma_L + i a_L}{\sqrt{2}} + \kappa_L, ~~\chi_R^\nu = \frac{\sigma_R + i a_R}{\sqrt{2}} + \kappa_R
\end{equation}
leads to $a_L$ and $a_R$ becoming the longitudinal modes of $Z_\mu$ and $Z'_\mu$. The Goldstone boson associated with the leptoquark $X_\mu$ is a linear combination of $\chi_L^u$ and $\chi_R^u$, which mix with a mass matrix given by
\begin{eqnarray}
M^2_{\chi^u_{L,R}} = \left( \begin{matrix}    -\lambda_4 \kappa_R^2 & \lambda_4 \kappa_L \kappa_R \cr \lambda_4 \kappa_L \kappa_R & -\lambda_4 \kappa_L^2  \end{matrix}\right)~.
\end{eqnarray}
The linear combinations are defined as
\begin{equation}
\chi^u_1 = \frac{(\kappa_R \chi^u_L - \kappa_L \chi^u_R)}{\sqrt{\kappa_L^2+\kappa_R^2}}, ~~\chi^u_2 = \frac{(\kappa_L \chi^u_L + \kappa_R \chi^u_R)}{\sqrt{\kappa_L^2+\kappa_R^2}} 
\end{equation}
have masses given by
\begin{equation}
M^2_{\chi^u_1} = -\lambda_4 (\kappa_L^2+\kappa_R^2),~~~M_{\chi^u_2}^2 = 0
\label{eqn:chiu_goldstone}
\end{equation}
with $\chi^u_2$ identified as the longitudinal mode of $X_\mu$.

An analogous mixing matrix arises in the $(\chi_L^d,\,\chi_R^d)$ sector which is given by
\begin{eqnarray}
M^2_{\chi^d_{L,R}} = \left( \begin{matrix}  
-2 \lambda_2 \kappa_L^2 - \lambda_4 \kappa_R^2 & -2 \lambda_5 \kappa_L \kappa_R \cr -2 \lambda_5 \kappa_L \kappa_R & -2 \lambda_2 \kappa_R^2 - \lambda_4 \kappa_L^2
\end{matrix} \right)~.
\label{eq:chiLd}
\end{eqnarray}
The two mass eigenstates obtained by diagonalizing this matrix are physical, which we denote as $\chi^d_{1,2}$.

The remaining spectrum consists of the mixed states arising from the neutral scalars $(\sigma_L, \sigma_R)$ with a mass matrix given by
\begin{eqnarray}
 M^2_{\sigma_{L,R}} = \left(\begin{matrix}   4(\lambda_1 + \lambda_2) \kappa_L^2 & 2(\lambda_3+\lambda_4) \kappa_L \kappa_R \cr 2(\lambda_3+\lambda_4) \kappa_L \kappa_R & 4(\lambda_1+\lambda_2)\kappa_R^2 \end{matrix}   \right)~.  
\end{eqnarray}
The resulting two mass eigenstates are part of the spectrum, with the lighter one identified as the SM Higgs boson with a mass of 125 GeV.

We should note that consistent symmetry breaking can be achieved with all squared masses of the scalars being positive for suitable values of the quartic coupling parameters.
This Higgs spectrum will play a role in showing the consistency of the Yukawa sector by including one-loop radiative corrections, inducing small neutrino masses radiatively, and in the calculation of one-loop corrections to the strong CP parameter $\overline{\theta}$. We shall turn to addressing these issues in the subsequent sections. 

The gauge boson mass spectrum can be computed from the covariant derivatives of the scalar fields given as (Cf: Eq. (\ref{eq:transform}))
\begin{equation}
{\cal L}_{\rm kinetic} = {\rm Tr}\{(D_\mu H_L)^\dagger (D^\mu H_L)\} + {\rm Tr}\{(D_\mu H_R)^\dagger (D^\mu H_R)\},
\end{equation}
where 
\begin{eqnarray}
D_\mu H_{L} &=& \partial_\mu H_{L} -ig_{2L} (\vec{T}_L.\vec{W}_{L\mu})H_{L} - ig_4 H_{L}(\vec{\lambda}.\vec{G}_\mu)^T, \nonumber \\
D_\mu H_{R} &=& \partial_\mu H_{R} -ig_{2R} (\vec{T}_R.\vec{W}_{R\mu})H_{R} - ig_4 H_{R}(\vec{\lambda}.\vec{G}_\mu)^T~.
\label{eqn:Covariant-HLHR}
\end{eqnarray}
Here $\vec{T}_{L,R}$ stand for the normalized $SU(2)_{L,R}$ generators, while $\vec{\lambda}$ stands for the $SU(4)_c$ generators. (The matrix $\vec{\lambda}.\vec{G}_\mu$ has the same form as $\Sigma_L$ of Eq. (\ref{eq:octet}), but with ${\cal O}$ replaced by the gluon fields, $N$ by $G_\mu^{15}$, and $U$ replaced by $X_\mu$.)  The  masses of the charged gauge bosons are computed to be
\begin{eqnarray}
M_{W_L^{\mu \pm}}^2 &=& \frac{1}{2} g_{2L}^2 \kappa_L^2, \nonumber \\
M_{W_R^{\mu \pm}}^2 &=& \frac{1}{2} g_{2R}^2 \kappa_R^2, \nonumber \\
M_{X^\mu}^2 &=& \frac{1}{2} g_{4}^2 (\kappa_L^2 + \kappa_R^2)~.
\label{eqn:charged_gauge_boson_masses}
\end{eqnarray}
Note that there is no tree-level mixing between $W_L^{\mu \pm}$ and $W_R^{\mu \pm}$ gauge bosons.

In the neutral gauge boson sector there is mixing between $Z_{L \mu}$ and $Z_{R\mu}$ with the mass matrix given by
\begin{equation}
\renewcommand{\arraystretch}{1.5}
\setlength\arraycolsep{5pt}
M^2_{Z-Z'} =
\frac{1}{2} \, \left( \begin{matrix} (g_Y^2+g_{2L}^2)\, \kappa_L^2 & g_Y^2 \sqrt{\frac{g_Y^2 + g_{2L}^2}{g_{2R}^2-g_Y^2}} \,\kappa_L^2 \\
 g_Y^2 \sqrt{\frac{g_Y^2 + g_{2L}^2}{g_{2R}^2-g_Y^2}}\, \kappa_L^2 & \frac{g_Y^4}{g_{2R}^2-g_Y^2}\, \kappa_L^2 +\frac{g_{2R}^4}{g_{2R}^2-g_Y^2}\, \kappa_R^2 \end{matrix} \right).
 \label{eq:zprime}
\end{equation}
Here, one should use the matching condition among the gauge couplings at the Pati-Salam scale as
\begin{equation}
        \frac{1}{g_Y^2}=\frac{1}{g_{2R}^2}+\frac{2}{3g_4^2}.
        \label{eq:gaugerelation}
    \end{equation}
We have eliminated $g_4$ in favor of $g_Y$ and $g_{2R}$ in Eq. (\ref{eq:zprime}) using Eq. (\ref{eq:gaugerelation}). With parity symmetry, we have $g_{2R} = g_{2L}$ at the PS scale.
The $Z_{L\mu}$ and $Z_{R\mu}$ fields, as well as the massless photon field $A_\mu$, are defined in terms of the original gauge fields as
\begin{align}
    A_{\mu}&=\frac{\sqrt{3}g_4 g_{2R} W_{\mu L}^0+\sqrt{3} g_4 g_{2L} W_{\mu R}^0+\sqrt{2} g_{2L} g_{2R} G_\mu^{15}}{\sqrt{2 g_{2L}^2 g_{2R}^2 + 3 g_4^2 (g_{2L}^2+g_{2R}^2)}},\\
    Z_{L \mu}&=\frac{g_{2L}(3 g_4^2+2 g_{2R}^2) W_{\mu L}^0 - 3 g_4^2 g_{2R} W_{\mu R}^0 - \sqrt{6} g_{2R}^2 g_4 G_\mu^{15}}{\sqrt{(3 g_4^2+2 g_{2R}^2)(2 g_{2L}^2 g_{2R}^2 + 3 g_4^2 (g_{2L}^2 + g_{2R}^2))}},
    \label{eq:ZL basis}\\
    Z_{R \mu}&= \frac{-\sqrt{2} g_{2R} W_{\mu R}^0 +\sqrt{3} g_4  G_\mu^{15}}{\sqrt{3 g_4^2+2 g_{2R}^2}}
    \label{eq:ZR basis}.
\end{align}
Note that the $Z_{L\mu}-Z_{R\mu}$ mixing angle is of order $\kappa_L^2/\kappa_R^2 \ll 1$. The mass eigenstates arising from Eq. (\ref{eq:zprime}) will be denoted as $Z_\mu$ and $Z'_\mu$.

The gauge bosons couple to the fermions charged under the corresponding
gauge groups through their kinetic terms,
\begin{equation}
    \mathcal{L}_\mathrm{kin}
    = i\,\overline{\psi}_L\slashed{D}\psi_L
    + i\,\overline{\psi}_R\slashed{D}\psi_R
    + i\,\overline{\Sigma}_L\slashed{D}\Sigma_L
    + i\,\overline{\Omega}_{L}\slashed{D}\Omega_{L}
    + i\,\overline{\Omega}_{R}\slashed{D}\Omega_{R},
    \label{eq:kinetic-terms}
\end{equation}
where the covariant derivatives for the \((2,1,4)\) and \((1,2,4)\)
multiplets were given in Eq.~\eqref{eqn:Covariant-HLHR}.  For the
remaining \((1,1,10)_{L,R}\) and \((1,1,15)_L\) fields, using the
transformations in Eq.~\eqref{eq:transform} and with the same notation for
the SU(4)\(_c\) generators as in Eq.~\eqref{eqn:Covariant-HLHR}, we have
\begin{align}
  D_\mu \Sigma_L&= \partial_\mu \Sigma_L- i g_4 \big[\vec{\lambda}\!\cdot\!\vec{G}_\mu ,\, \Sigma_L\big],\\
  D_\mu \Omega_{L,R}&= \partial_\mu \Omega_{L,R}- i g_4 \vec{\lambda}\!\cdot\!\vec{G}_\mu\,\Omega_{L,R} -ig_4\Omega_{L,R}\big(\vec{\lambda}\!\cdot\!\vec{G}_\mu\big)^{T}.
  \label{eq:cov-OmegaR}
\end{align}
We can expand Eq.~\eqref{eq:kinetic-terms} to get the explicit couplings of $X_{\mu}$ to fermions in the gauge eigenbasis:
\begin{align}
   &\mathcal{L}_{X_{\mu}}=\frac{g_4}{\sqrt{2}} \Bigl[
(\overline{u}_L\gamma^\mu \nu_{L}+ \overline{d}_L\gamma^\mu e_L) X_{\mu} 
+\sqrt{2}\,\overline{D}_L\gamma^\mu E_{L}X_{\mu}+ \overline{\mathcal{S}}^{\alpha\beta}_L\gamma^\mu D_{L\beta} X_{\mu\alpha}+\mathrm{L \leftrightarrow R}\Bigr.\notag\\
\Bigl.& +\frac{1}{\sqrt{3}}\overline{N_L}\gamma^{\mu}U^c_LX_{\mu}+\frac{1}{\sqrt{3}}\overline{N_R^c}\gamma^{\mu}U_RX_{\mu}+\frac{1}{\sqrt{2}}\overline{(\mathcal{O}_L)^{\alpha }_{\beta}}
\gamma^{\mu}U^c_{L\alpha}X_{\mu}^{\beta}++\frac{1}{\sqrt{2}}\overline{(\mathcal{O}_R^c)^{\alpha }_{\beta}}
\gamma^{\mu}U_{R\alpha}X_{\mu}^{\beta}\Bigr]+\mathrm{h.c}. \label{eq:X}
\end{align}
We will require these couplings in the subsequent sections to study neutrino mass generation and the solution to the strong CP problem. Interactions with other gauge bosons can be extracted similarly when needed.

\section{Realistic down-quark and charged-lepton masses}

An important consistency check of the model is the generation of realistic quark and lepton masses with the particle content present in the model. The mass matrices ${\cal M}_d$ and ${\cal M}_e$ of Eq. (\ref{eq:matrices}) for the $6 \times 6$ down-type quarks and charged leptons involve the same two $3 \times 3$ sub-matrices, namely $M_{10}$ and $Y_{10}$.  This can lead to relations among the light fermion masses $(m_e)_i$ and $(m_d)_i$ that are potentially inconsistent with observations.  This is apparent in the light $3 \times 3$ mass matrices  $M_d$ and $M_e$ obtained by the seesaw formulas shown in Eqs. ({\ref{eq:uni2})-(\ref{eq:uni3}), which leads to the mass relations at the Pati-Salam scale $m_\tau = 2\, m_b$, $m_\mu = 2\, m_s$, $m_e = 2 \,m_d$.  These relations are incompatible with the values of the masses extrapolated to the parity restoration scale $\mu_{P} = 5.2 \times 10^{13}$ GeV.
In this section, we show how these relations can be corrected within the model without any modifications so as to be consistent with the observed fermion masses.

In Fig.~\ref{Fig:fig1} we present the running masses as a function of energy scale $\mu$ extrapolated from low to high energies. Here the running mass is defined as $m_i(\mu) = Y_i(\mu)\, v$ where $Y_i$ is the Yukawa couplings of fermion flavor $i$ and $v= 174$ GeV.  In obtaining this figure, we have used two-loop renormalization group (RG) equations for the running of the gauge couplings and the Yukawa couplings of the SM. As input at $\mu = m_t(m_t) = 163.6~\text{GeV}$, we have used the following running masses, based on the PDG fermion-mass averages~\cite{ParticleDataGroup:2024cfk} and the running-mass analysis of Ref.~\cite{Xing:2007fb}, with slight rounding for our numerical implementation:
\begin{eqnarray}
&&m_t(m_t) = 163.6~{\rm GeV},~~~ m_c(m_t) = 0.620~{\rm GeV}, ~~~m_u(m_t) = 1.30~{\rm MeV}, \nonumber \\
&&m_b(m_t) = 2.78~{\rm GeV}, ~~~~~m_s(m_t) = 55.0~{\rm MeV}, ~~~~~~~m_d(m_t) = 2.70~{\rm MeV}, \nonumber\\
&&m_\tau(m_t) = 1.78~{\rm GeV}, ~~~m_\mu(m_t) = 105.7~{\rm MeV},~~ m_e(m_t) = 0.511~{\rm MeV}. 
\end{eqnarray}
\begin{figure}[h!]
\begin{center}
\includegraphics[width=0.7\textwidth]{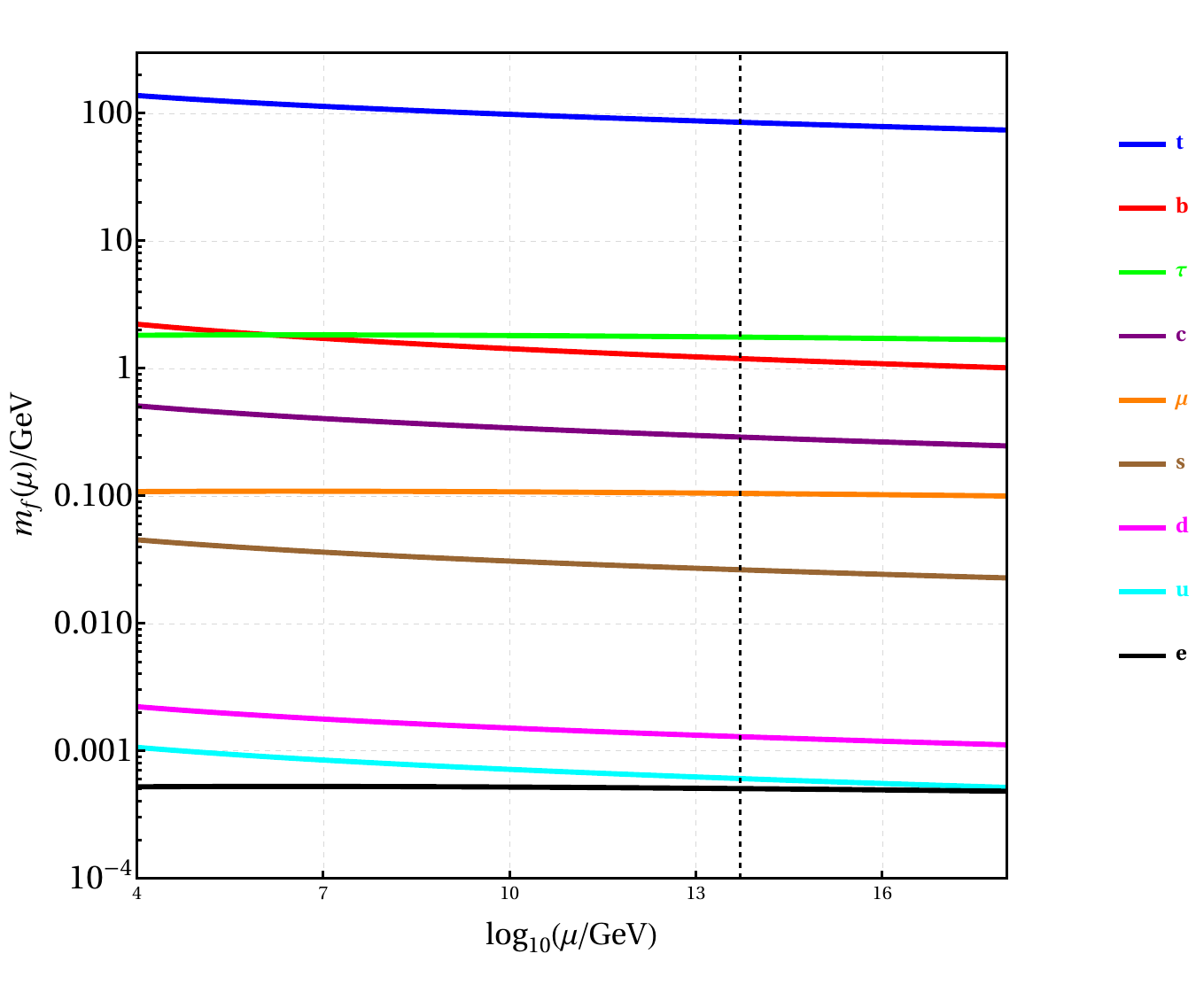}
\end{center}
\caption{Running masses of the quarks and leptons  \(m_f(\mu)\) as a function of the renormalization scale $\mu$.  The dashed vertical line represents the parity–restoration scale \(\mu_P=5.2\times 10^{13}\,\mathrm{GeV}\). The numerical values of the running masses at $\mu_P$ are listed in Table~\ref{tab:mf-muP}.}
\label{Fig:fig1}
\end{figure}
In addition, we have used $\alpha_Y(M_Z) = 0.017$, $ \alpha_2(M_Z) = 0.034$, and $\alpha_3(M_Z) = 0.118$ as input for the three gauge couplings of the SM in their two-loop RG evolution equations. We determine the scale of parity restoration, $\mu_P$, from the condition that $g_{2L}(\mu_P) = g_{2R}(\mu_P)$, aided by Eq. (\ref{eq:gaugerelation}).  We find $\mu_P = 5.2 \times 10^{13}$ GeV as the parity restoration scale. 

We also list the numerical values of the running fermion masses at the parity restoration scale in Table~\ref{tab:mf-muP}.  The lepton to quark mass ratios of the same generation are found to have values at $\mu_P$ given by
\begin{equation}
\frac{m_\tau}{m_b} = 1.47,~~~ \frac{m_\mu}{m_s} = 3.97,~~~\frac{m_e}{m_d} = 0.39~~~~~~(\mu = 5.2 \times 10^{13}~{\rm GeV}).
\label{eq:ratios}
\end{equation}
Thus, a constant value of $2$ for these ratios is implied by Eqs. (\ref{eq:uni2})-(\ref{eq:uni3}) needs correction of order one.

\begin{table}[h!]
  \centering
  \setlength{\tabcolsep}{10pt}     
  \renewcommand{\arraystretch}{1.2}
  \begin{tabular}{|c|c|}
    \hline
    $m_f(\mu_P)$ & [GeV] \\
    \hline
    $m_t$     & $8.50\times10^{1}$ \\
    $m_b$     & $1.19$ \\
    $m_\tau$  & $1.76$ \\
    $m_c$     & $2.89\times 10^{-1}$ \\
    $m_\mu$   & $1.04\times 10^{-1}$ \\
    $m_s$     & $2.63\times 10^{-2}$ \\
    $m_d$     & $1.29\times 10^{-3}$ \\
    $m_u$     & $6.06\times 10^{-4}$ \\
    $m_e$     & $5.05\times 10^{-4}$ \\
    \hline
  \end{tabular}
  \caption{Values of the running quark and lepton masses evaluated at the parity restoration scale $\mu_P=5.2 \times 10^{13}$ GeV.}
  \label{tab:mf-muP}
\end{table}
We now show that the mass ratios $(m_e)_i/(m_d)_i$ given in Eq. (\ref{eq:ratios}) can be consistently realized within the model once one-loop radiative corrections are included.  We present two scenarios where realistic fermion masses are generated: the first where the corrections arise through the Yukawa couplings $Y_{15}$, and another where they arise through the coupling $Y_{10}$. 

\subsection{Benchmark point with  $Y_{15}$-induced radiative corrections}
\label{sec:fermion_mass_benchmark_Y15}
Radiative corrections to $M_d$ and $M_e$ proportional to the Yukawa coupling matrix $Y_{15}$ from the up-type quark sector can be significant, as these would introduce a new flavor structure.  We present one-loop diagrams correcting the down-type quark and charged lepton masses through $Y_{15}$ in Fig.~\ref{Fig:fig2} through the exchange of $\chi^d_{L,R}$ leptoquarks.  There are similar diagrams that correct the up-type quark masses through $Y_{10}$ Yukawa couplings, but we shall ignore them in this benchmark point by choosing all entries of $|(Y_{10})_{ij}| \ll 1$. 
\begin{figure}[h!]
  \centering
  \includegraphics[width=0.45\textwidth]{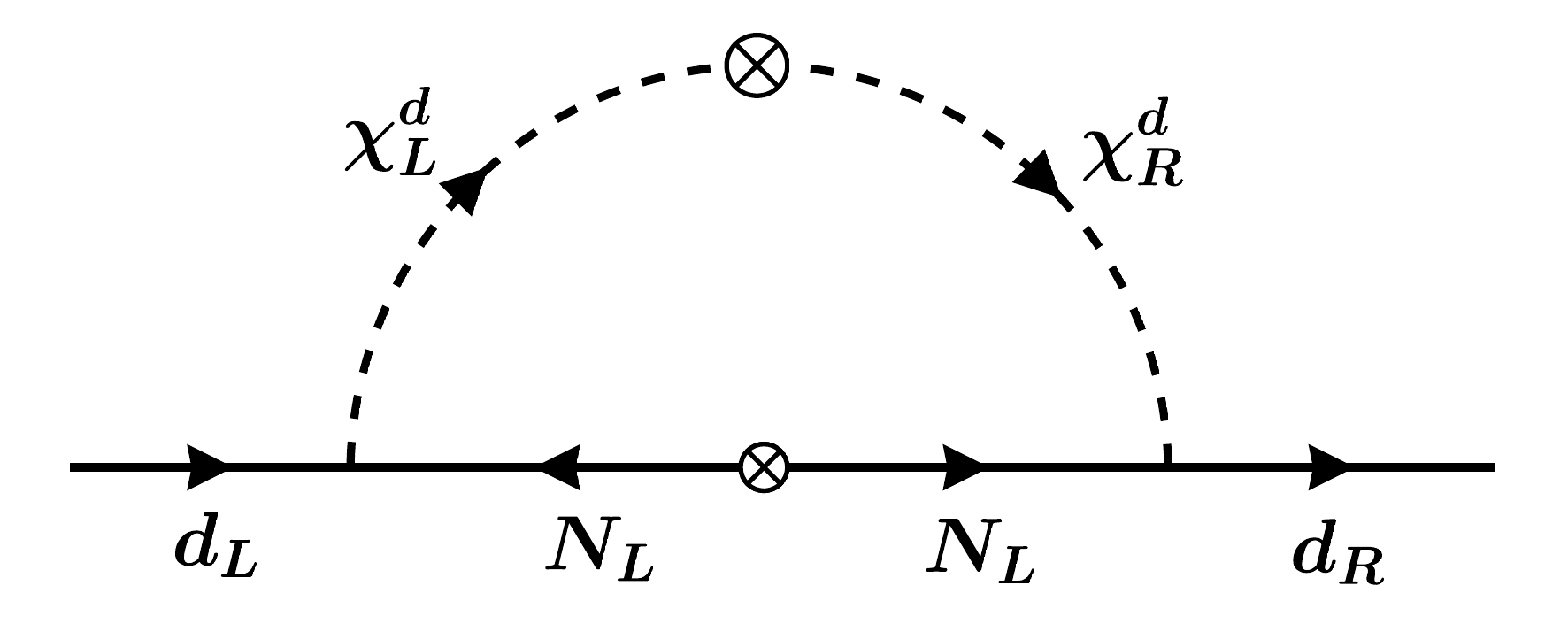}
  \includegraphics[width=0.45\textwidth]{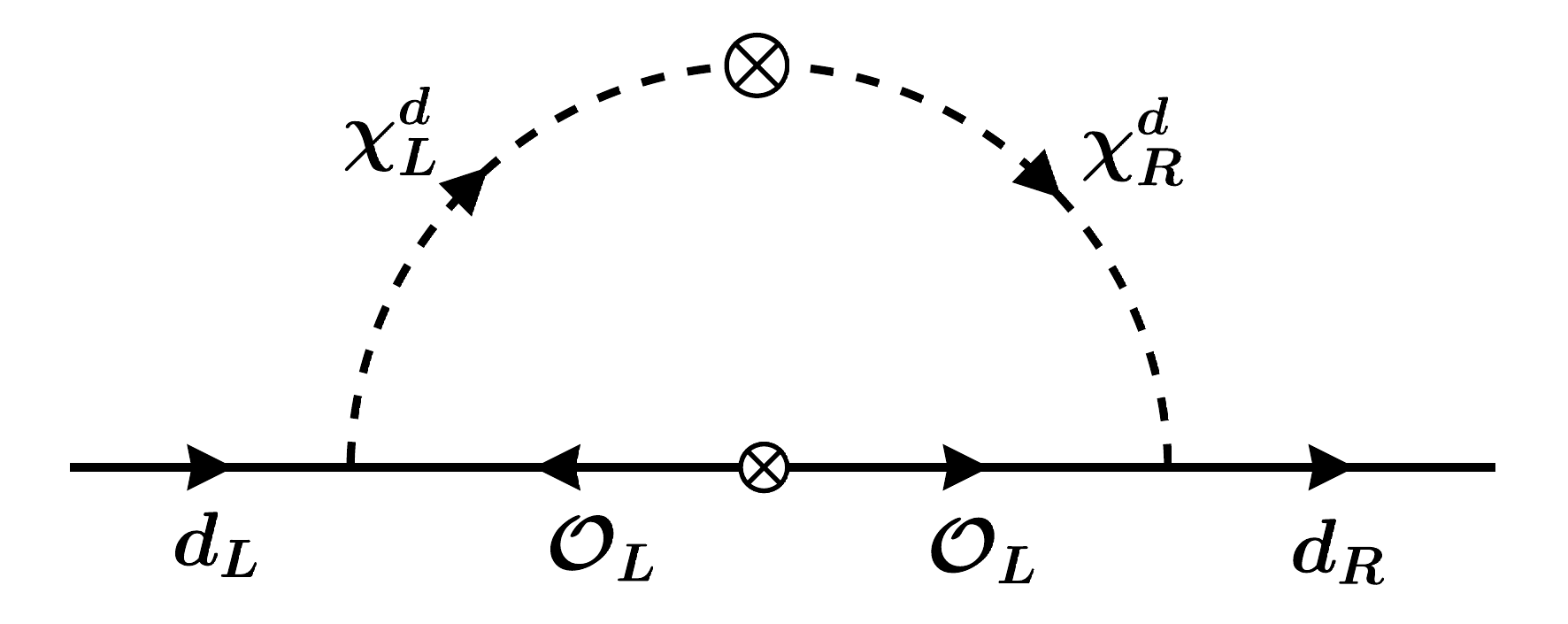}\\[4pt]
  \includegraphics[width=0.45\textwidth]{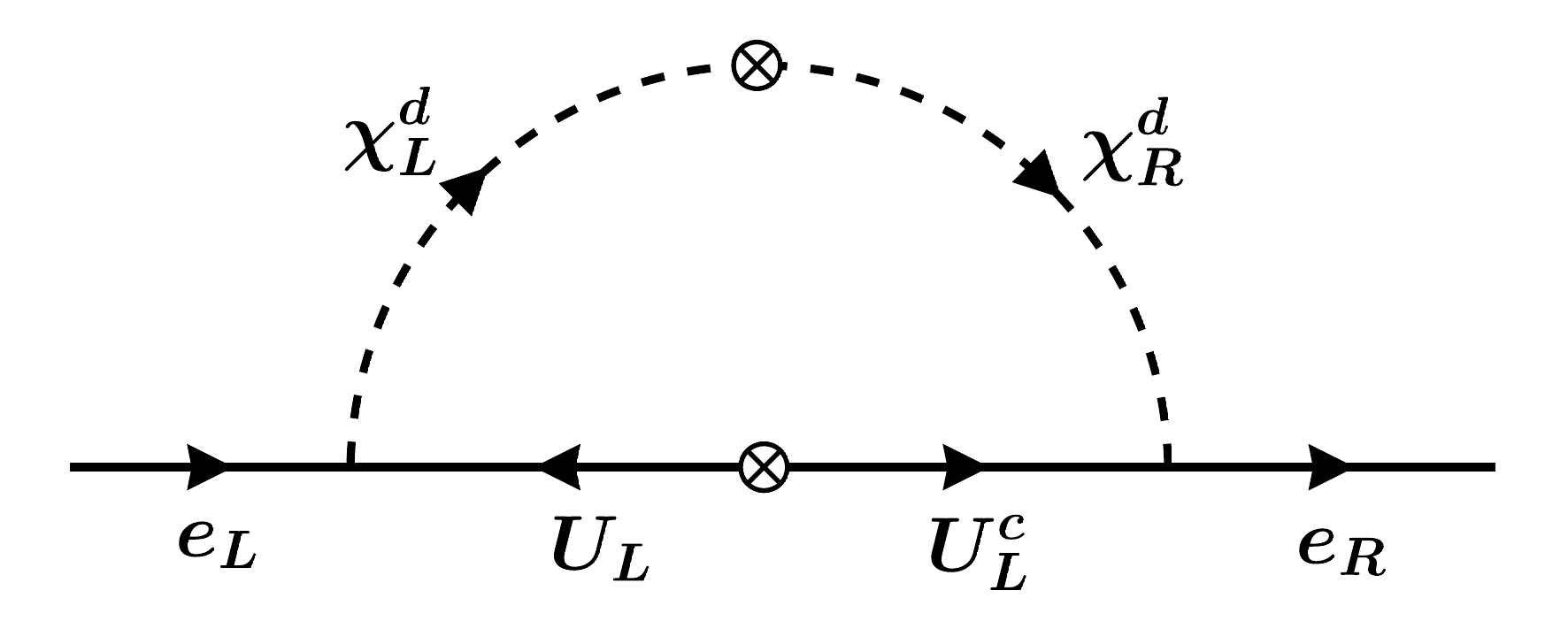}
  \caption{One–loop radiative corrections to the down–quark and charged–lepton masses induced by the $Y_{15}$ coupling.  The diagrams involve the exchange of $\chi^d_{L,R}$ leptoquark scalars together with the neutral fermion $N_L$, the color octet $\mathcal{O}_L$, and the up–type quarks and generate new flavor structures in $M_d$ and $M_e$ beyond the tree–level contribution.}
  \label{Fig:fig2}
\end{figure}
To evaluate the loop diagrams of Fig.~\ref{Fig:fig2}} that contribute to $M_d$ and $M_e$ we work in a basis where $M_{15}$ is diagonal and real, a choice that is most general, with the diagonal entries of $M_{15}$  denoted as $(M_{15})_i$. In addition, we assume that only those contributions to $M_d$ and $M_e$ that are proportional to $(M_{15})_3$ are significant. This is a reasonable assumption since realistic top-quark mass generation requires $(M_{15})_3$ to be not much larger than $\kappa_R$. In contrast, $u$ and $c$-quark masses can be induced even with $(M_{15})_{1,2} \gg \kappa_R$, implying that corrections to $M_d$ and $M_e$ proportional to $(M_{15})_{1,2}$ will be suppressed, which shall be elaborated below. We also adopt the seesaw approximation for the tree-level mass matrices as shown in Eqs. (\ref{eq:uni1})-(\ref{eq:uni3}).  Since the $\chi^d_L-\chi^d_R$ mixing is small, of order $\kappa_L/\kappa_R$ (see Eq. (\ref{eq:chiLd})), we keep only the lowest order term in this mixing angle. The loop corrections to $M_d$ and $M_e$  arising from Fig.~\ref{Fig:fig2} are then found to be
\begin{eqnarray}
\Delta M_d &\simeq& \left(\frac{11}{4}\right) \frac{Y_{15} M_{15} Y_{15}^\dagger}{16 \pi^2}
(2 \lambda_5 \kappa_L \kappa_R) \,F(M_{15_3}^2, M_{\chi_1^d}^2, M^2_{\chi_2^d})
\label{eq:correct-0}\nonumber \\
\Delta M_e &\simeq& (3) \frac{Y_{15} M_{15} Y_{15}^\dagger}{16 \pi^2}
(2 \lambda_5 \kappa_L \kappa_R)\, 
\,F(M_{15_3}^2, M_{\chi_1^d}^2, M^2_{\chi_2^d})~.
\label{eq:correct}
\end{eqnarray}
These corrections should be added to the tree-level mass matrices $M_d$ and $M_e$ given in Eqs. (\ref{eq:uni2})-(\ref{eq:uni3}). Here the symmetric function $F(x,y,z)$ is  defined as
\begin{equation}
F(x,y,z) = \frac{x \,{\rm log} x}{(x-y)(x-z)} +  \frac{y\, {\rm log} y}{(y-x)(y-z)} +  \frac{z\, {\rm log} z}{(z-x)(z-y)} ~. 
\label{eqn:F(x,y,z)}
\end{equation}
In Eq. (\ref{eq:correct}) the factors (11/4) and (3) are color factors, with the factor (11/4) for $\Delta M_d$ arising from the two diagrams with the exchange of $N$ (which gives a factor of 1/12) and the color-octet ${\cal O}$ (which yields a factor of 8/3).

To illustrate consistency of the down-type quark and charged lepton masses, including these radiative corrections, we observe that in the up-type quark sector, generation of the top quark requires $(M_{15})_3$ to be of the same order as $(Y_{15})_{33}\, \kappa_R$. This is because of the matching condition for the top quark Yukawa coupling, which is given by
\begin{equation}
Y_t(\mu_P) = \frac{|(Y_{15})_{33}|^2 \,\kappa_R}{\sqrt{|(M_{15})_{3}|^2+ |(Y_{15})_{33}|^2 \,\kappa_R^2}},
\label{eq:Yukbound}
\end{equation}
where we have ignored the small mixing of the third family with the first two. Numerical evolution of the top quark Yukawa coupling of the SM yields $Y_t(\mu_P) \simeq 0.5$, which implies that the ratio $|(M_{15})_{3}|/|(Y_{15})_{33}|$ cannot exceed a factor of 4 or so, so that the correct value of $Y_t(\mu_P)$ is reproduced while the coupling $(Y_{15})_{33}$ remains perturbative. In contrast, the lighter up-quark and charm-quark masses allow the masses $(M_{15})_2$ and $(M_{15})_1$ to be much larger than $\kappa_R$.  Consequently, in the correction terms for $M_d$ and $M_e$ given in Eq. (\ref{eq:correct}) we may ignore the contributions that are proportional to $(M_{15})_1$ and $(M_{15})_2$.  This is because for $x \gg y,z$, the function $F(x,y,z)$ takes the form 
\begin{equation}
F(x,y,z) \simeq \frac{y \,{\rm log}\left(\frac{x}{y} \right) - z \,{\rm log}\left(\frac{x}{z} \right)}{x(y-z)},~~~x \gg y,z~,
\end{equation}
leading to a suppression of $\kappa_R/(M_{15})_1$ for the contribution arising from the diagram proportional to $(M_{15})_1$ of Fig.~\ref{Fig:fig2}, with a similar suppression for the $(M_{15})_2$ contribution. We also note the approximation 
\begin{equation}
F(x,y,z) \simeq \frac{1}{x}\left({\rm log}\left(\frac{x}{y}\right) -1 \right), ~~~x \gg y=z~.
\end{equation}

Keeping only the contributions proportional to $(M_{15})_3$ in Eq. (\ref{eq:correct}), we see that $\Delta M_d$ and $\Delta M_e$ are of rank-one.  We can parametrize the radiatively corrected $M_d$ and $M_e$ as
\begin{eqnarray}
M_d = \left(\begin{matrix} m_{11} & m_{12} & m_{13} \cr m_{12}^* & m_{22} & m_{23} \cr m_{13}^* & m_{23}^* & m_{33}   \end{matrix}  \right) + \frac{11}{5} m_0 \left(\begin{matrix} |\delta_1|^2 & \delta_1 \delta_2^* & \delta_1 \delta_3^* \cr \delta_1^* \delta_2 & |\delta_2|^2 & \delta_2 \delta_3^* \cr \delta_1^* \delta_3 & \delta_2^* \delta_3 & |\delta_3|^2    \end{matrix}   \right) ,\nonumber\\
M_e = 2 \left(\begin{matrix} m_{11} & m_{12} & m_{13} \cr m_{12}^* & m_{22} & m_{23} \cr m_{13}^* & m_{23}^* & m_{33}   \end{matrix}  \right) + \frac{12}{5} m_0 \left(\begin{matrix} |\delta_1|^2 & \delta_1 \delta_2^* & \delta_1 \delta_3^* \cr \delta_1^* \delta_2 & |\delta_2|^2 & \delta_2 \delta_3^* \cr \delta_1^* \delta_3  & \delta_2^* \delta_3 & |\delta_3|^2    \end{matrix}   \right) .
\label{eq:paramet}
\end{eqnarray}
Here, the first matrices on the right-hand sides are the hermitian tree-level contributions, while the second ones are the loop corrections.  We have defined $\delta_i$ and $m_0$ as
\begin{eqnarray}
\delta_i &=& (Y_{15})_{i3}, \nonumber\\
m_0 &=& \left(\frac{5}{4}\right) \frac{(M_{15})_3} {{16 \pi^2}}
(2 \lambda_5 \kappa_L \kappa_R) \,F(M_{15_3}^2, M_{\chi_1^d}^2, M^2_{\chi_2^d}).
\label{eq:deltai}
\end{eqnarray}

A benchmark scenario where simple analytic formulas can be obtained for fitting the masses is obtained by setting the off-diagonal elements of $M_e$ to zero by appropriate choice of the off-diagonal $m_{ij}$. The mass matrices of Eq. (\ref{eq:paramet}) take the form, for this benchmark scenario, given as
\begin{eqnarray}
M_e = \left(\begin{matrix}  m_e & ~ & ~ \cr ~ & m_\mu & ~ \cr ~ & ~ & m_\tau \end{matrix}    \right),~~~M_d = \left(\begin{matrix} 
\frac{m_e}{2} + m_0 |\delta_1|^2 & m_0 \delta_1 \delta_2^* & m_0 \delta_1 \delta_3^* \cr
m_0 \delta_1^* \delta_2 & \frac{m_\mu}{2} + m_0 |\delta_2|^2 & m_0 \delta_2 \delta_3^* \cr
m_0 \delta_1^* \delta_3 & m_0 \delta_2^* \delta_3 & \frac{m_\tau}{2} + m_0 |\delta_3|^2
\end{matrix}  \right).
\label{eq:simp}
\end{eqnarray}
In this form, the charged lepton mass matrix is diagonal, and we can diagonalize $M_d$ using the approximation $m_\tau, |\delta_3|^2 m_0 \gg m_\mu, |\delta_2|^2 m_0 \gg m_e,|\delta_1|^2 m_0$. We obtain for the down-type quark masses the relations
\begin{eqnarray}
m_b &\simeq& \frac{m_\tau}{2} + |\delta_3|^2 m_0 \nonumber \\
m_s m_b &\simeq& \frac{m_\mu m_\tau}{4} + \frac{1}{2} |\delta_2|^2 m_0 m_\tau + \frac{1}{2} |\delta_3|^2 m_0 m_\mu \nonumber \\
m_d m_s m_b &\simeq& \frac{m_e m_\mu m_\tau}{8} + \frac{1}{4}|\delta_3|^2 m_0 m_e m_\mu + \frac{1}{4} |\delta_2|^2 m_0 m_e m_\tau + \frac{1}{4} |\delta_1|^2 m_0 m_\mu m_\tau.
\end{eqnarray}
These relations can be solved for $|\delta_i|^2 m_0$ as:
\begin{eqnarray}
|\delta_3|^2 m_0 &\simeq & \left(m_b - \frac{m_\tau}{2}\right), \nonumber\\
|\delta_2|^2 m_0 &\simeq& \frac{2 m_b}{m_\tau}\left(m_s - \frac{m_\mu}{2}\right), \nonumber\\
|\delta_1|^2 m_0 &\simeq& \frac{4 m_s m_b}{m_\mu m_\tau}\left(m_d - \frac{m_e}{2}\right)~.
\label{eq:radiative}
\end{eqnarray}

Now we use the values of the charged lepton masses and down-type quark masses from Table~\ref{tab:mf-muP} to determine the required values of the radiative corrections  $|\delta_i|^2 m_0$ from Eqs. (\ref{eq:radiative}). One should note that the mass parameters can take either positive or negative signs when applied to Eqs. (\ref{eq:radiative}).  Since the left-hand sides of these equations should have the same signs, certain restrictions occur for the choice of the signs of $(m_d)_i$ and $(m_e)_i$. A consistent choice results if all masses have positive signs, except for $m_\mu$ and $m_s$, which have negative signs.  With these sign choices and the values of the masses given in Table~\ref{tab:mf-muP}, we obtain
\begin{eqnarray}
|\delta_3|^2 m_0 = 0.313~{\rm GeV},~~~|\delta_2|^2 m_0 = 0.035~ {\rm GeV},~~~|\delta_1|^2 m_0 = 7.09 \times 10^{-4}~{\rm GeV}~.  
\label{eq:deltavalues}
\end{eqnarray}
Note that these corrections are of the same order as $m_b, m_s, m_d$ respectively.  We have verified that when these values are inserted into Eq. (\ref{eq:simp}) and the resulting $M_d$ is diagonalized numerically, the eigenvalues are correctly reproduced within a percent or so accuracy.

Now we turn to the realization of these corrections in terms of the model parameters given in Eq. (\ref{eq:deltai}). To explore this, we set the masses of $\chi_1^d$ and $\chi_2^d$ equal to $\kappa_R$, and  $(M_{15})_3 = 4 \kappa_R$.  Furthermore, if we set $\lambda_5 = 0.125$, we find that the  required radiative corrections are reproduced when the Yukawa couplings take values
\begin{equation}
(Y_{15})_{33} = 1.434, ~~(Y_{15})_{23} = 0.481,~~ (Y_{15})_{13} = 0.068~.
\label{eq:Yuk_suggest}
\end{equation}
This is a consistent choice of parameters. We also note that the seesaw approximation used for the top quark mass is reasonable, since for this choice we have $|(Y_{15})_{33}/(M_{15})_3| = 0.358$ along with $Y_t = 0.514$.  Corrections to the top-quark seesaw is of order $\frac{1}{2}|(Y_{15})_{33}/(M_{15})_3|^2 \simeq 0.064$, which is not excessive. Since the Yukawa couplings suggested in Eq. (\ref{eq:Yuk_suggest}) will lead to a contribution to charm-quark mass of order $m_c \sim \{(Y_{15})_{23}\}^4\,\kappa_L \kappa_R/(\{(Y_{15})_{33}\}^2 (M_{15})_2)$, we see that $(M_{15})_2/\kappa_R \geq 15$ is needed.  The contribution to the charm quark mass from the term 
$|(Y_{15})_{22}|^2 \kappa_L \kappa_R/(M_{15})_2$ allows for $(M_{15})_{2}/\kappa_R \sim 600$. Such a value is consistent with our assumption that contributions proportional to $(M_{15})_2$ and $(M_{15})_1$ to $\Delta M_d$ and $\Delta M_e$ are negligible.

The choice of the $|\delta_i|^2 m_0$ and the Yukawa couplings as given in Eqs. (\ref{eq:deltavalues})-(\ref{eq:Yuk_suggest}) leads to a quark mixing matrix arising from $M_d$ which is found to be (with all $\delta_i$ taken to be real)
\begin{eqnarray}
 V_d = \left(\begin{matrix}    
 0.991 & 0.132 & -0.024 \cr 0.134 & -0.987 & 0.083 \cr 0.013 & 0.086 & 0.996
 \end{matrix}   \right) ~.  
\end{eqnarray}
The CKM mixing matrix is given by $V_{\rm CKM} = V_u^\dagger V_d$ with the elements of $V_u$, the unitary matrix that diagonalizes $M_u$, completely free. We see that the elements of $V_d$ have a hierarchy similar to the CKM matrix, and that realistic CKM angles can be reproduced with the suggested benchmark.

\subsection{A second benchmark point with $Y_{10}$-induced radiative corrections}
\label{sec:Y10 induced corrections}
We now present a second benchmark scenario where the radiative corrections to $M_d$ and $M_e$ are induced through the $Y_{10}$ Yukawa coupling matrix.  Here we ignore the contributions from the $Y_{15}$ Yukawa coupling matrix, which is justified in the approximation $(M_{15})_i \ll \kappa_R$.  Such a scheme is desirable since this choice makes the one-loop contributions to the strong CP parameter $\overline{\theta}$ parametrically suppressed by the factor $(M_{15})_i/\kappa_R$.  

\begin{figure}[t!]
    \centering
    \begin{subfigure}[b]{0.325\textwidth}
        \centering
        \includegraphics[width=\textwidth]{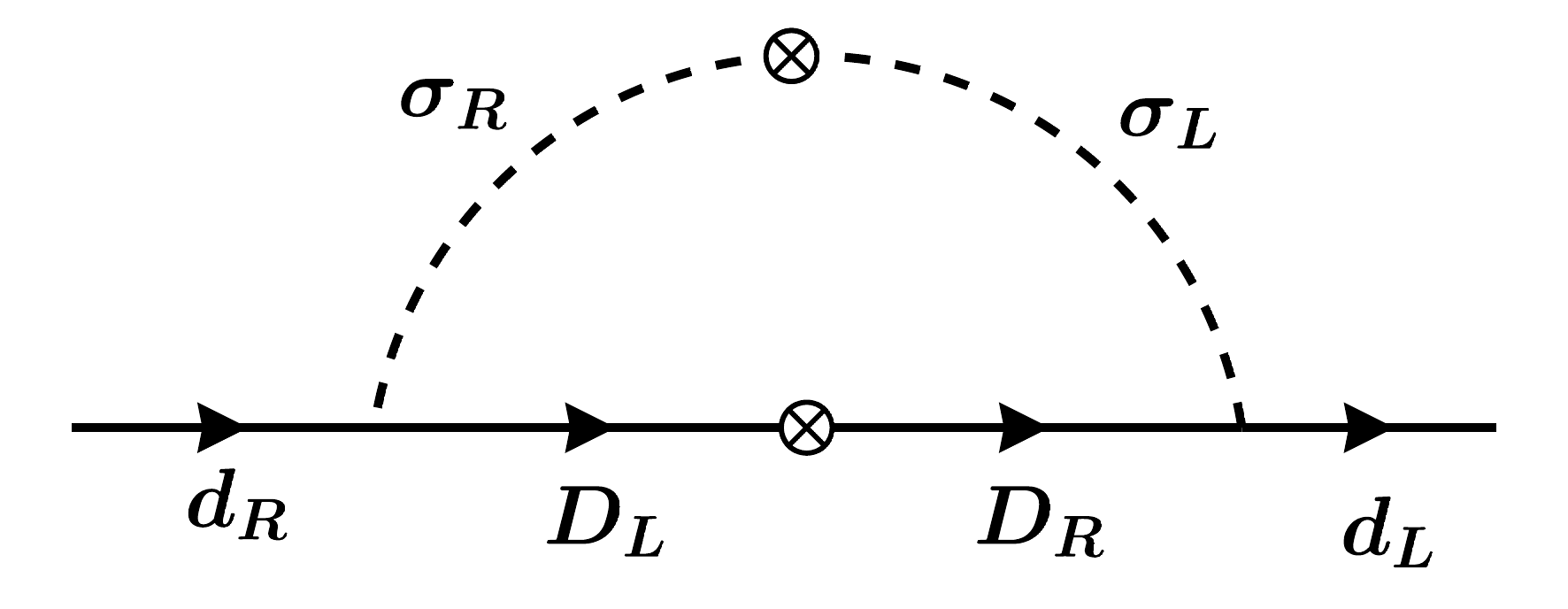}
        \caption{}
    \end{subfigure}
    \begin{subfigure}[b]{0.325\textwidth}
        \centering
        \includegraphics[width=\textwidth]{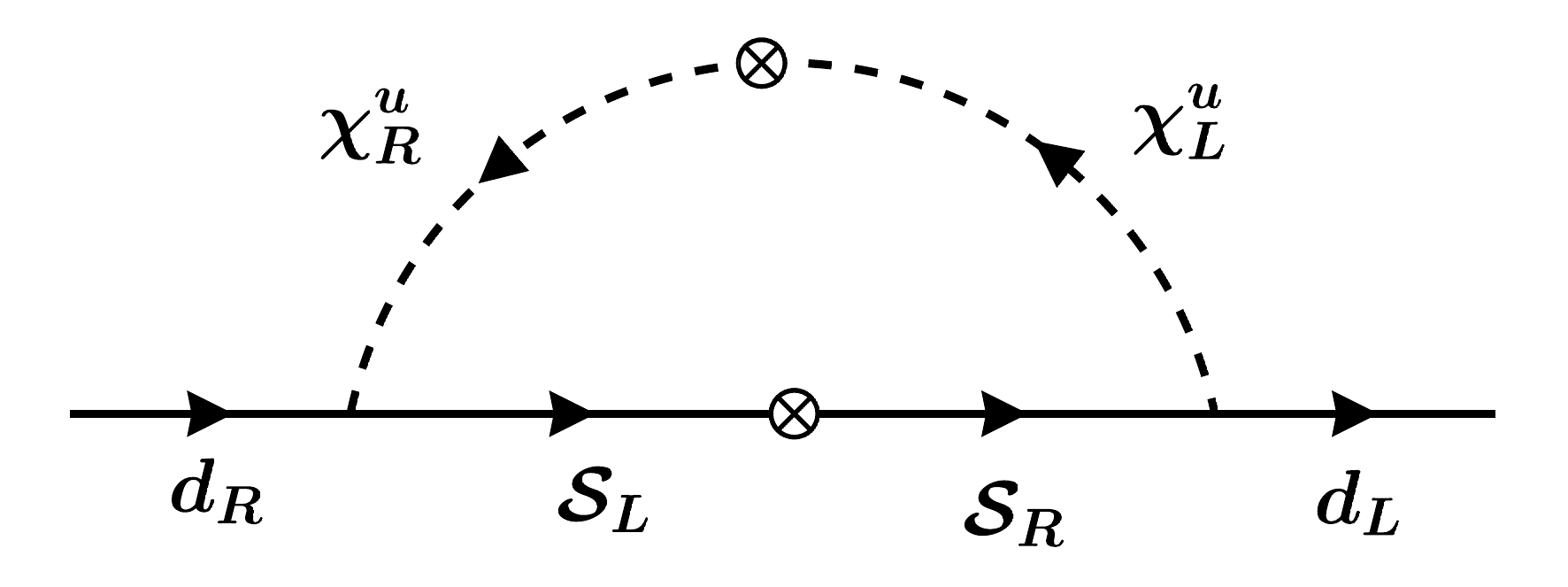}
        \caption{}
    \end{subfigure}
    \begin{subfigure}[b]{0.325\textwidth}
        \centering
        \includegraphics[width=\textwidth]{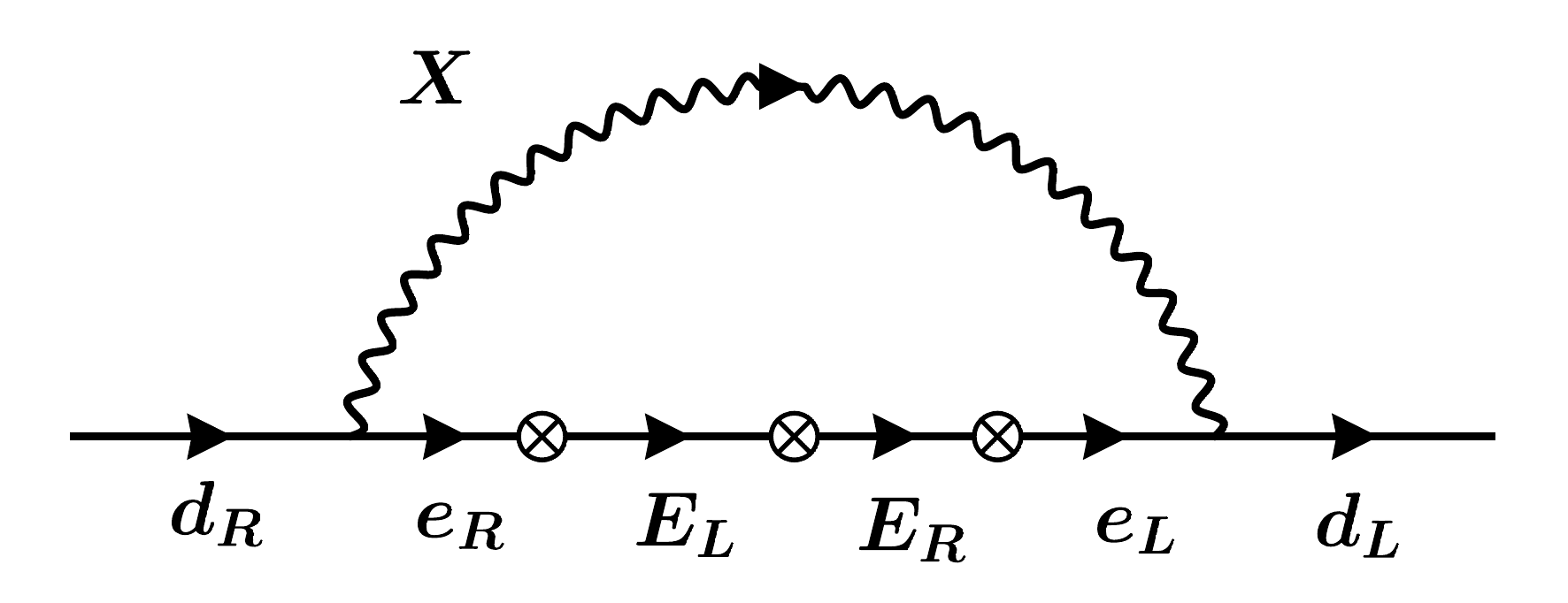}
        \caption{}
    \end{subfigure}

    \vspace{0.2cm} 

    \begin{subfigure}[b]{0.325\textwidth}
        \centering
        \includegraphics[width=\textwidth]{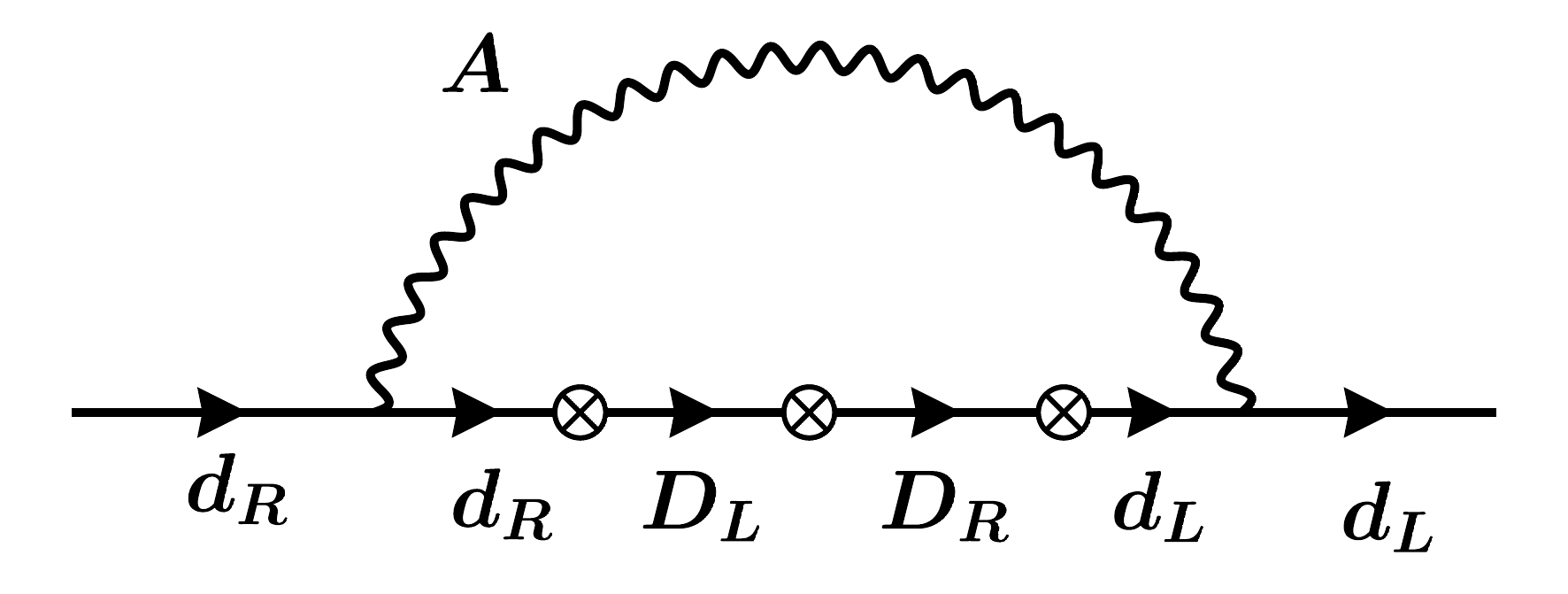}
        \caption{}
    \end{subfigure}
     \begin{subfigure}[b]{0.325\textwidth}
        \centering
        \includegraphics[width=\textwidth]{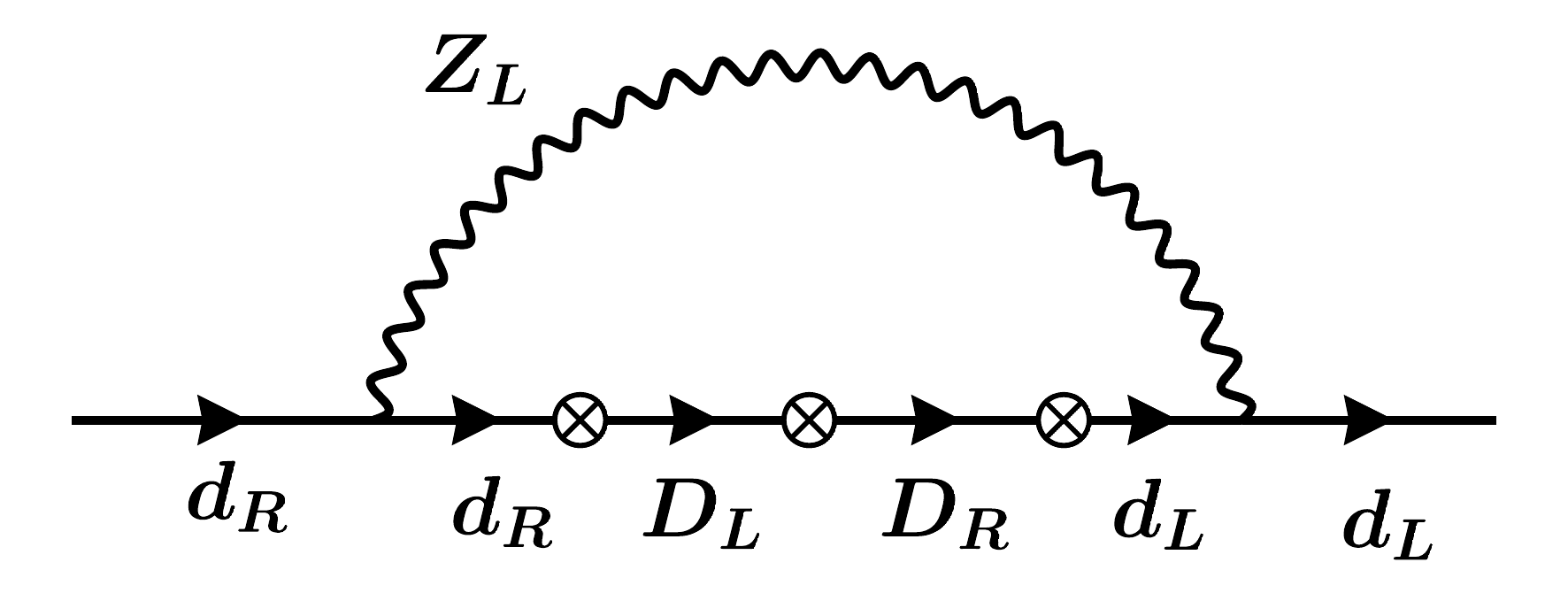}
        \caption{}
    \end{subfigure}
    \begin{subfigure}[b]{0.325\textwidth}
        \centering
        \includegraphics[width=\textwidth]{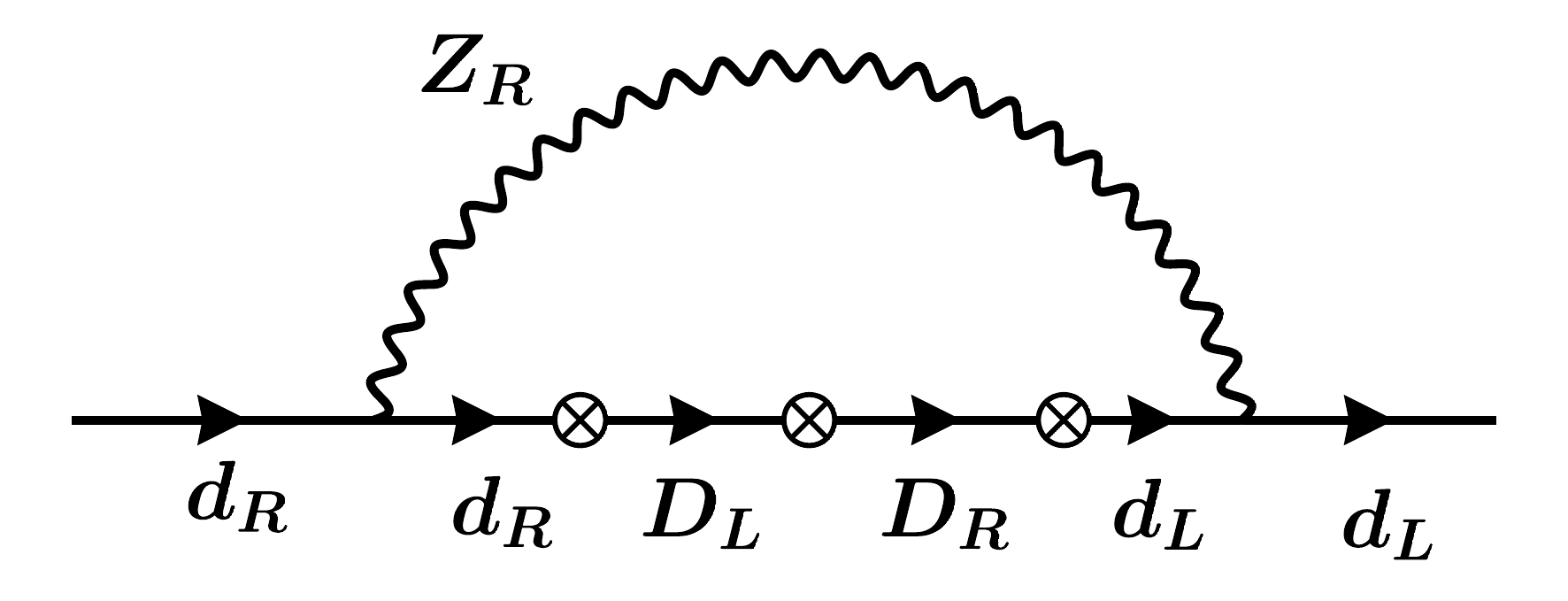}
        \caption{}
    \end{subfigure}

        \vspace{0.2cm} 

    \begin{subfigure}[b]{0.325\textwidth}
        \centering
        \includegraphics[width=\textwidth]{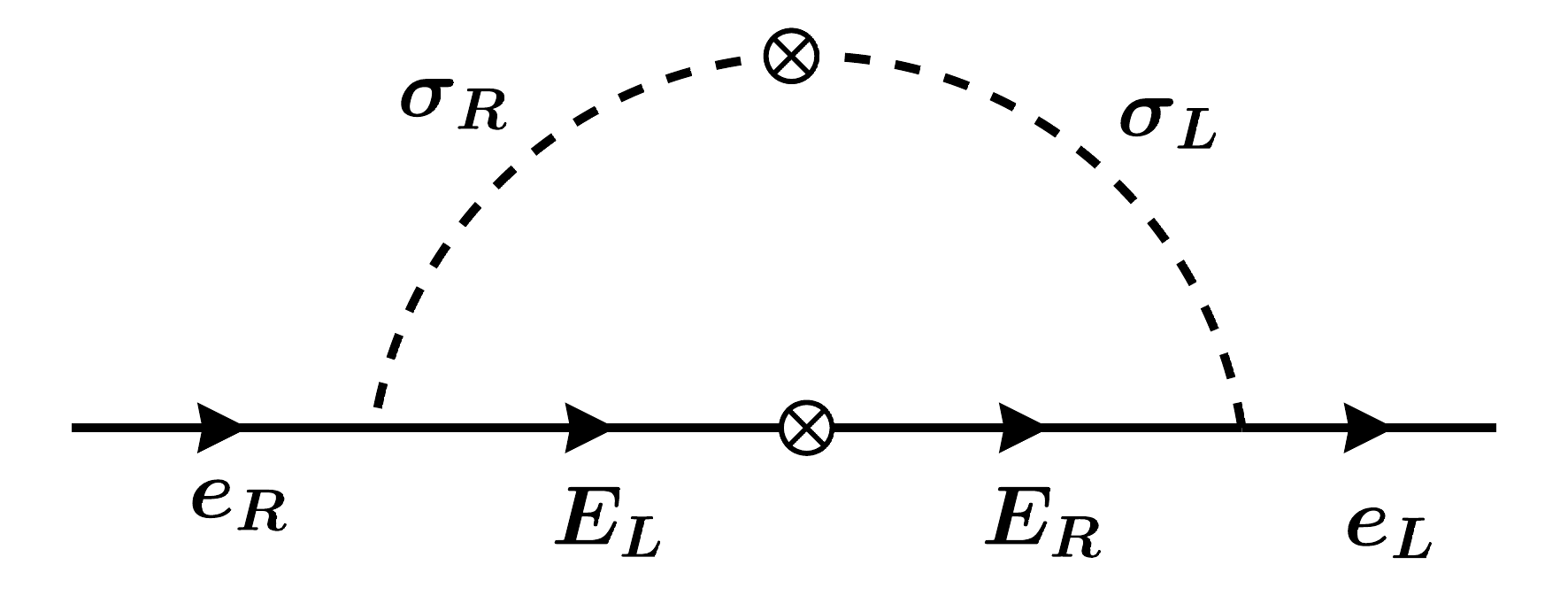}
        \caption{}
    \end{subfigure}
     \begin{subfigure}[b]{0.325\textwidth}
        \centering
        \includegraphics[width=\textwidth]{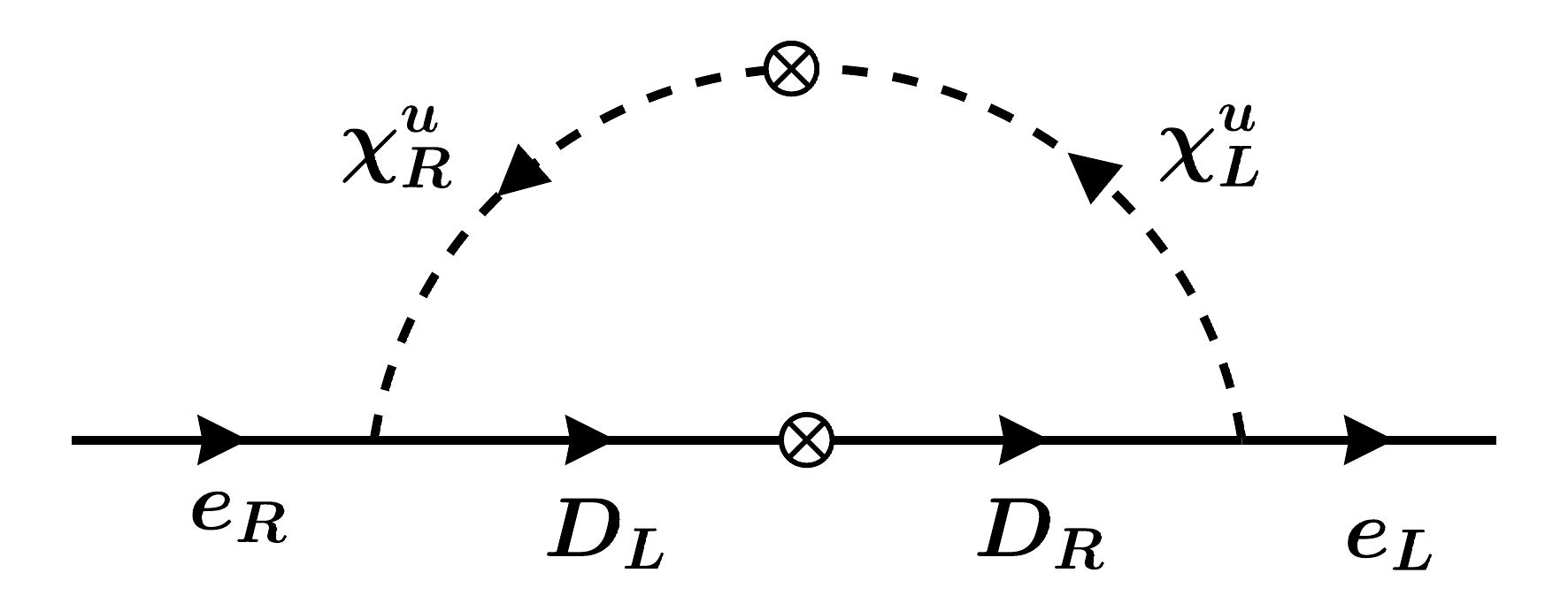}
        \caption{}
    \end{subfigure}
    \begin{subfigure}[b]{0.325\textwidth}
        \centering
        \includegraphics[width=\textwidth]{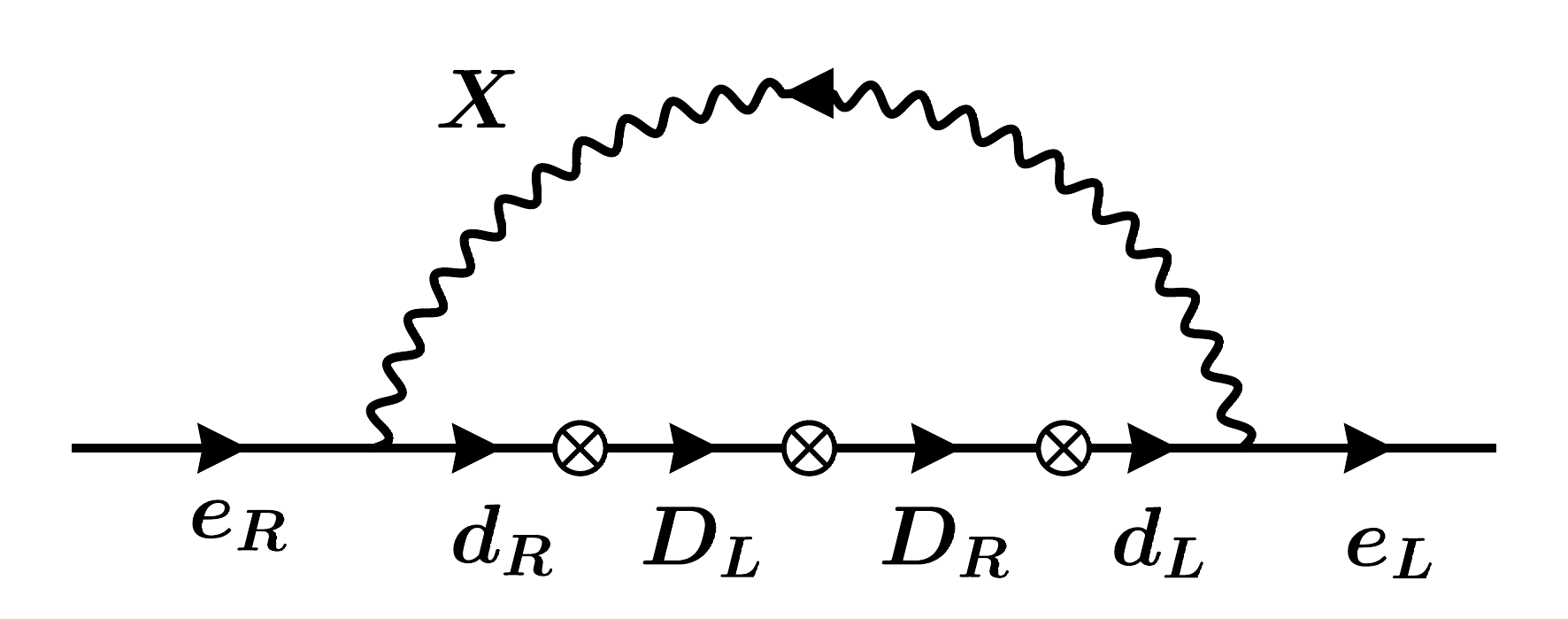}
        \caption{}
    \end{subfigure}

        \vspace{0.2cm} 

    \begin{subfigure}[b]{0.325\textwidth}
        \centering
        \includegraphics[width=\textwidth]{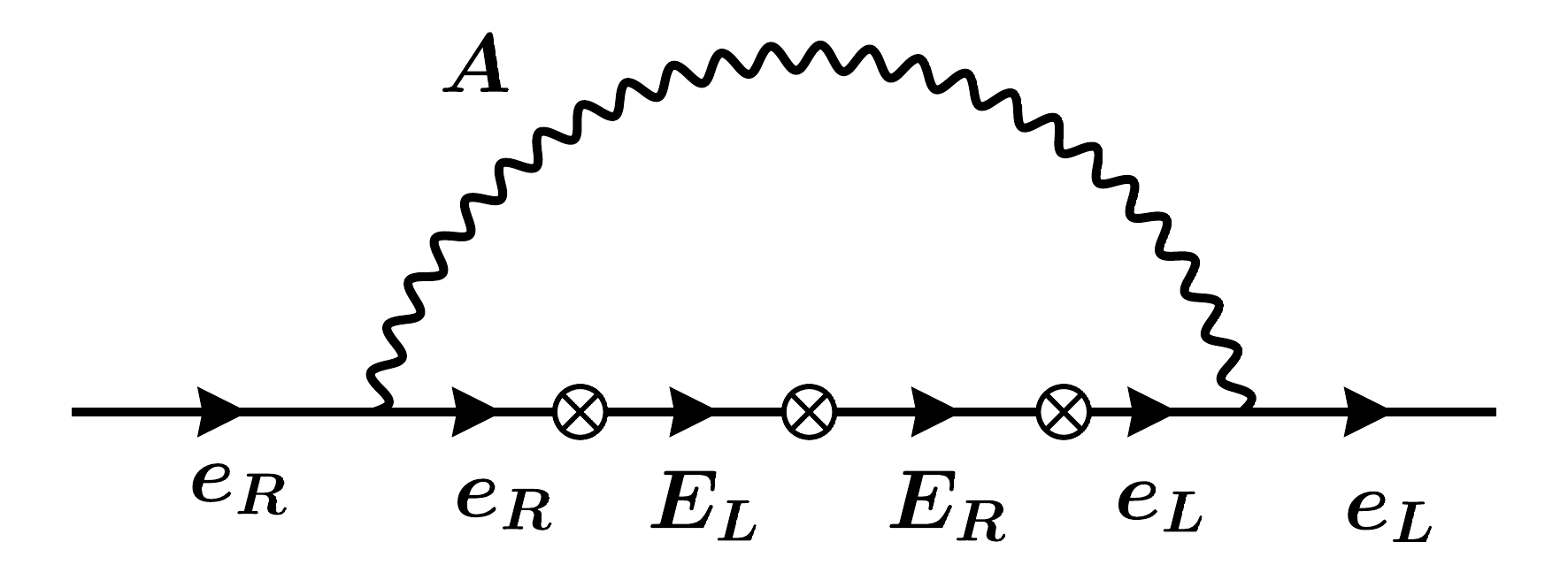}
        \caption{}
    \end{subfigure}
     \begin{subfigure}[b]{0.325\textwidth}
        \centering
        \includegraphics[width=\textwidth]{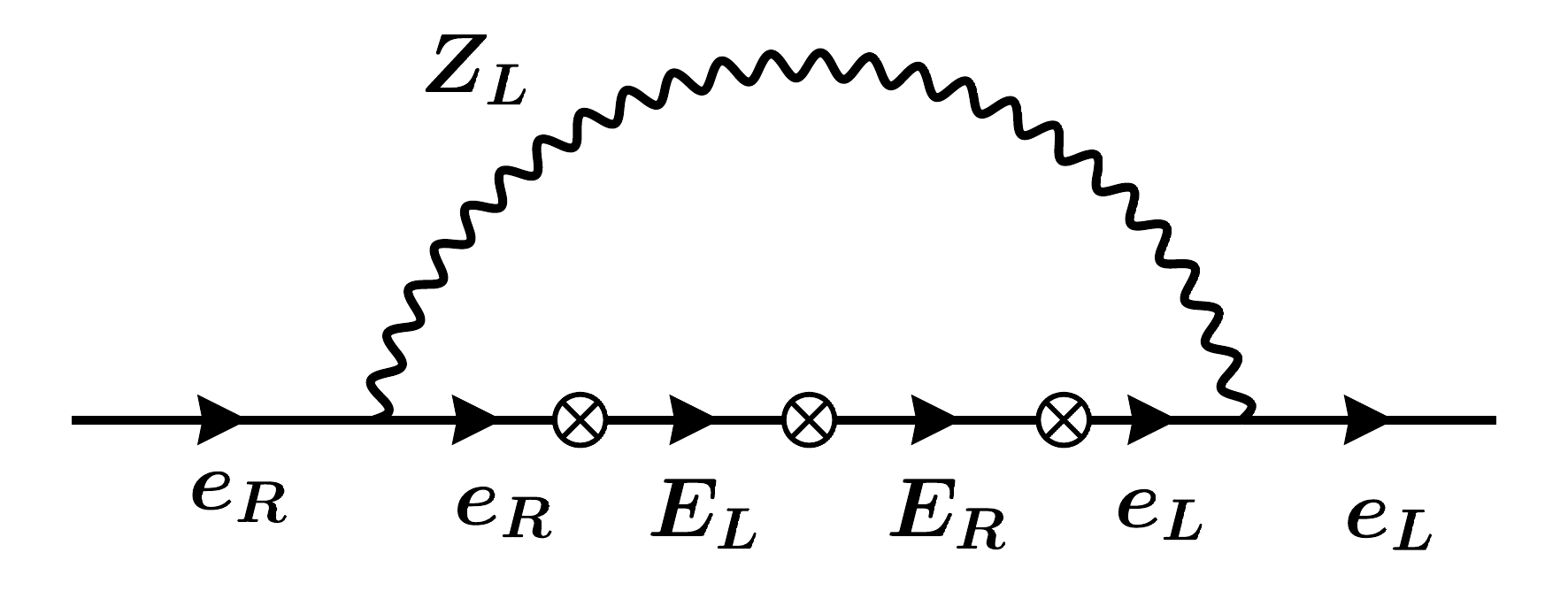}
        \caption{}
    \end{subfigure}
    \begin{subfigure}[b]{0.325\textwidth}
        \centering
        \includegraphics[width=\textwidth]{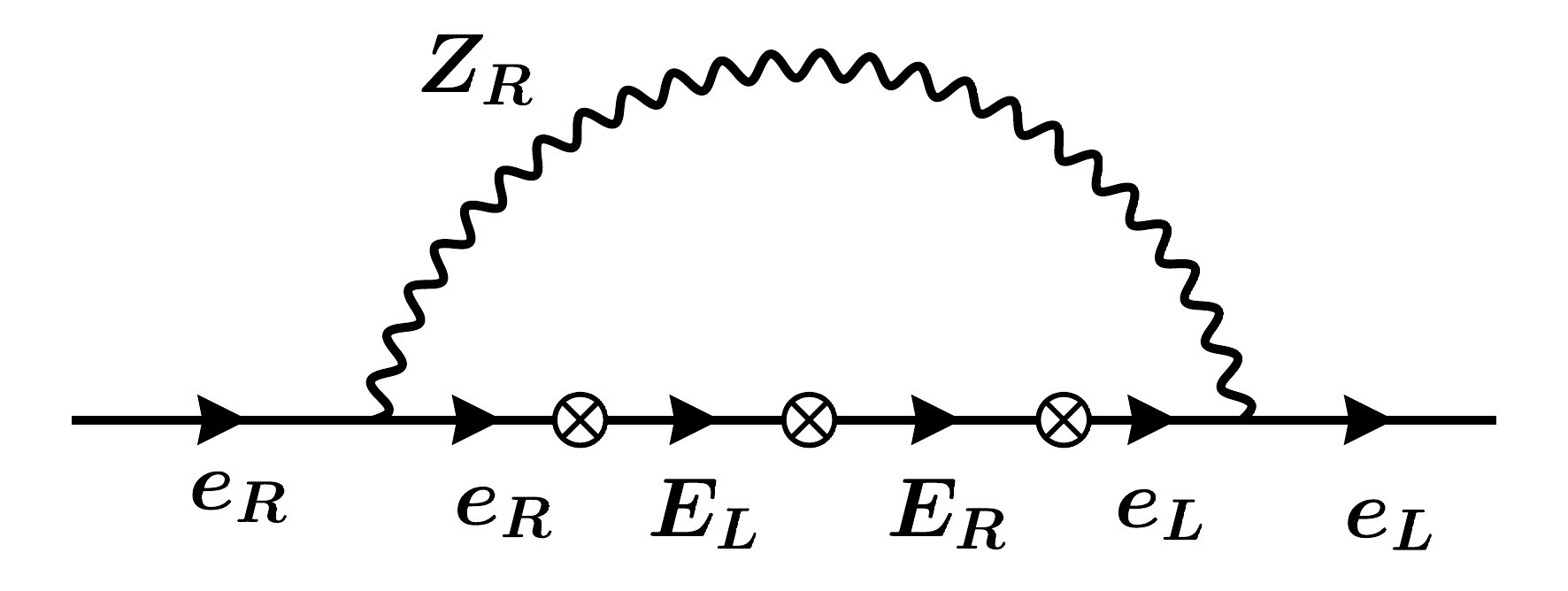}
        \caption{}
    \end{subfigure}
    \caption{One–loop radiative corrections to the down–quark and charged–lepton masses induced by the $Y_{10}$ coupling.  The diagrams involve the exchange of $\chi^u_{L,R}$ leptoquark scalars, leptoquark and neutral gauge bosons together with the charged lepton, the color sextet, and the down–type quarks.} 
    \label{fig:Y10_induced_mass}
\end{figure}

The Feynman diagrams that correct $M_d$ and $M_e$ in this scenario are shown in Fig.~\ref{fig:Y10_induced_mass}. In evaluating these diagrams, we assume, as in the previous benchmark point, that only terms proportional to $(M_{10})_3$ are significant.  The justification to this approximation is that $(M_{10})_1$ and $(M_{10})_2$ can be much larger than $(M_{10})_3$, in which case the terms proportional to $(M_{10})_{1,2}$ will be suppressed by a factor $\kappa_R/(M_{10})_{1,2}$, while those proportional to $(M_{10})_3$ will have a milder suppression factor of $\kappa_R/(M_{10})_3$. 

Rather than evaluating all the diagrams of Fig.~\ref{fig:Y10_induced_mass}, we shall focus on Fig.~\ref{fig:Y10_induced_mass} (b) and (h), which are mediated by $\chi_{L,R}^u$ leptoquark scalars. This assumption will be justified a posteriori. In fact, we shall see that the relevant quartic coupling for Fig.~\ref{fig:Y10_induced_mass} (b) and (h) is $\lambda_4 \sim 3$, which is larger than $g_4^2/2 \sim 1/4$ relevant to the $X_\mu$ gauge boson exchange diagrams by an order of magnitude. Similarly, we ignore the $\sigma_L-\sigma_R$ exchange diagrams by assuming that the relevant quartic coupling there, $(\lambda_3 + \lambda_4)$, is smaller than $\lambda_4$.

As in the previous benchmark point, we keep only the contributions proportional to $(M_{10})_3$ in evaluating these leptoquark scalar-mediated diagrams. The one-loop corrections to $M_d$ and $M_e$ are given by
\begin{eqnarray}
\Delta M_d &\simeq& (3) \frac{Y_{10} M_{10} Y_{10}^\dagger}{16 \pi^2}
(2 \lambda_4 \kappa_L \kappa_R) \,F(M_{10_3}^2, M_{\chi_1^u}^2, M^2_{\chi_2^u})
\label{eq:correct-3}\nonumber \\
\Delta M_e &\simeq& (5) \frac{Y_{10} M_{10} Y_{10}^\dagger}{16 \pi^2}
(2 \lambda_4 \kappa_L \kappa_R)\, 
\,F(M_{10_3}^2, M_{\chi_1^u}^2, M^2_{\chi_2^u})~
\label{eq:correct-4}
\end{eqnarray}
with the function $F$ given in Eq. (\ref{eqn:F(x,y,z)}). Note that $\chi_2^u$ is the Goldstone boson absorbed by the leptoquark gauge boson $X_\mu$ and the mass of $\chi_1^u$ is given in Eq. (\ref{eqn:chiu_goldstone}). We note here that this equation has been utilized in arriving at Eqs. (\ref{eq:correct-3})-(\ref{eq:correct-4}), where we have ignored the contributions proportional to $g_4^2$.

The full mass matrices $M_d$ and $M_e$ can now be written down as
\begin{eqnarray}
M_d = \left(\begin{matrix} m_{11} & m_{12} & m_{13} \cr m_{12}^* & m_{22} & m_{23} \cr m_{13}^* & m_{23}^* & m_{33}   \end{matrix}  \right) + 6 m_0 \left(\begin{matrix} |\delta_1|^2 & \delta_1 \delta_2^* & \delta_1 \delta_3^* \cr \delta_1^* \delta_2 & |\delta_2|^2 & \delta_2 \delta_3^* \cr \delta_1^* \delta_3 & \delta_2^* \delta_3 & |\delta_3|^2    \end{matrix}   \right) ,\nonumber\\
M_e = 2 \left(\begin{matrix} m_{11} & m_{12} & m_{13} \cr m_{12}^* & m_{22} & m_{23} \cr m_{13}^* & m_{23}^* & m_{33}   \end{matrix}  \right) + 10 m_0 \left(\begin{matrix} |\delta_1|^2 & \delta_1 \delta_2^* & \delta_1 \delta_3^* \cr \delta_1^* \delta_2 & |\delta_2|^2 & \delta_2 \delta_3^* \cr \delta_1^* \delta_3  & \delta_2^* \delta_3 & |\delta_3|^2    \end{matrix}   \right) .
\label{eq:paramet}
\end{eqnarray}
Here, the first matrices on the right-hand sides are the hermitian tree-level contributions, while the second ones are the loop corrections.  We have defined $\delta_i$ and $m_0$ as
\begin{eqnarray}
\delta_i &=& (Y_{10})_{i3}, \nonumber\\
m_0 &=& \left(\frac{1}{2}\right) \frac{(M_{10})_3} {{16 \pi^2}}
(2 \lambda_4 \kappa_L \kappa_R) \,F(M_{10_3}^2, M_{\chi_1^u}^2, M^2_{\chi_2^u}).
\label{eq:deltai-2}
\end{eqnarray}

As in the previous benchmark, we choose the off-diagonal $m_{ij}$ terms to make the charged lepton mass matrix diagonal. The resulting mass matrices are identical to the ones given Eq. (\ref{eq:simp}), but with the new identification of $\delta_i$ and $m_0$ as in Eq. (\ref{eq:deltai-2}).  Consequently Eq. (\ref{eq:radiative}) and Eq. (\ref{eq:deltavalues}) will follow.  We choose the Yukawa couplings $(Y_{10})_{3i}$ as in Eq. (\ref{eq:Yuk_suggest}), namely,
\begin{equation}
(Y_{10})_{33} = 1.434, ~~(Y_{10})_{23} = 0.481,~~ (Y_{10})_{13} = 0.068~.
\label{eq:Yuk_suggest-2}
\end{equation}
Unlike in the previous benchmark, here we should also fit the diagonal tau lepton mass, which is given by
\begin{equation}
m_\tau \simeq 2 \frac{|(Y_{10})_{33}|^2 \kappa_L \kappa_R}{(M_{10})_3} + 10 m_0 |\delta_3|^2~.
\end{equation}
A good fit to all the fermion masses is obtained by choosing the following parameters:
\begin{equation}
\lambda_4 = -2.19,~~~(M_{10})_3 = 146 \kappa_R,
\end{equation}
which yields $m_0 = -0.152$ GeV and $m_\tau = 1.76$ GeV.  Here we have used the mass relation  $M_{\chi_1^u} \simeq \sqrt{-\lambda_4} \kappa_R$.

To conclude this section, we have shown by two explicit constructions that the model can reproduce realistic quark and lepton masses once radiative corrections are included.

\section{Radiative neutrino mass generation}
\label{sec:sec4}
\begin{figure}[t!]
    \centering
    \begin{subfigure}[b]{0.32\textwidth}
        \centering
        \includegraphics[width=\textwidth]{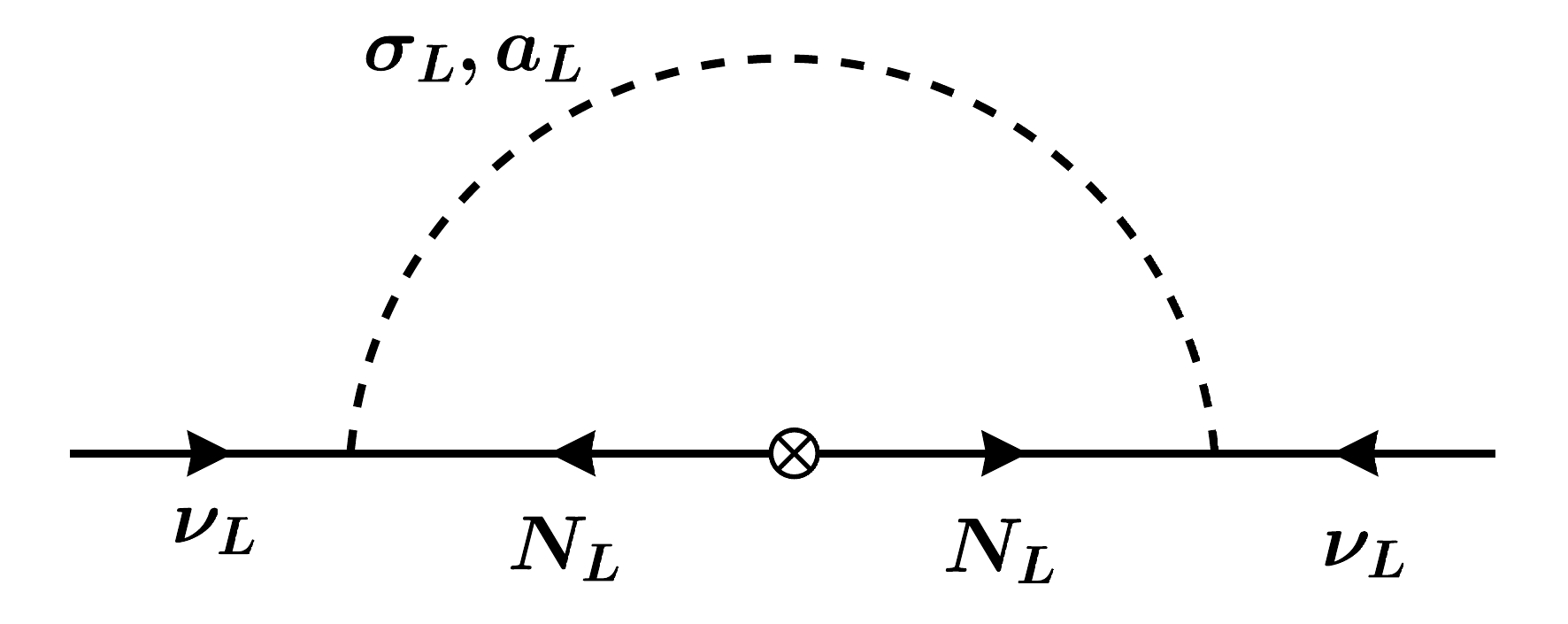}
        \caption{}
    \end{subfigure}
    \hfill
    \begin{subfigure}[b]{0.32\textwidth}
        \centering
        \includegraphics[width=\textwidth]{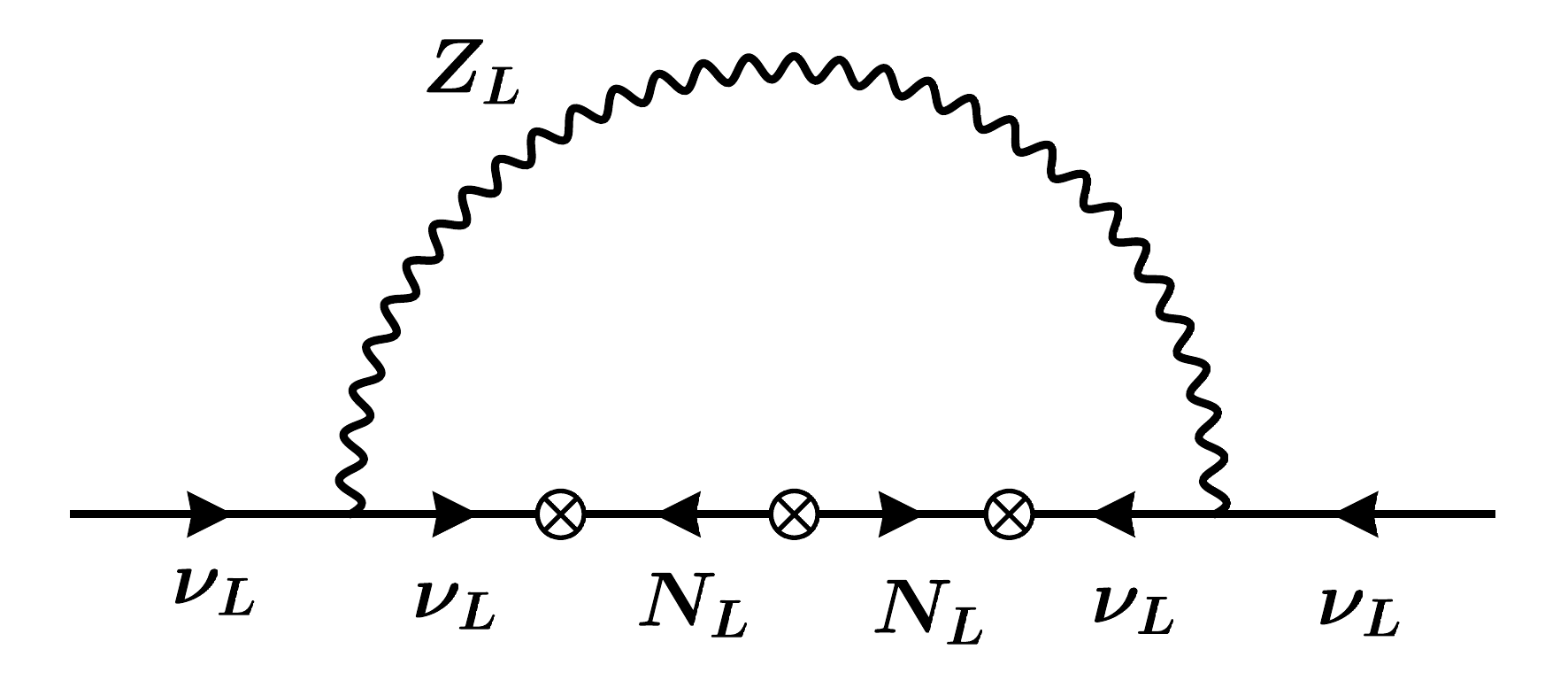}
        \caption{}
    \end{subfigure}
    \begin{subfigure}[b]{0.32\textwidth}
        \centering
        \includegraphics[width=\textwidth]{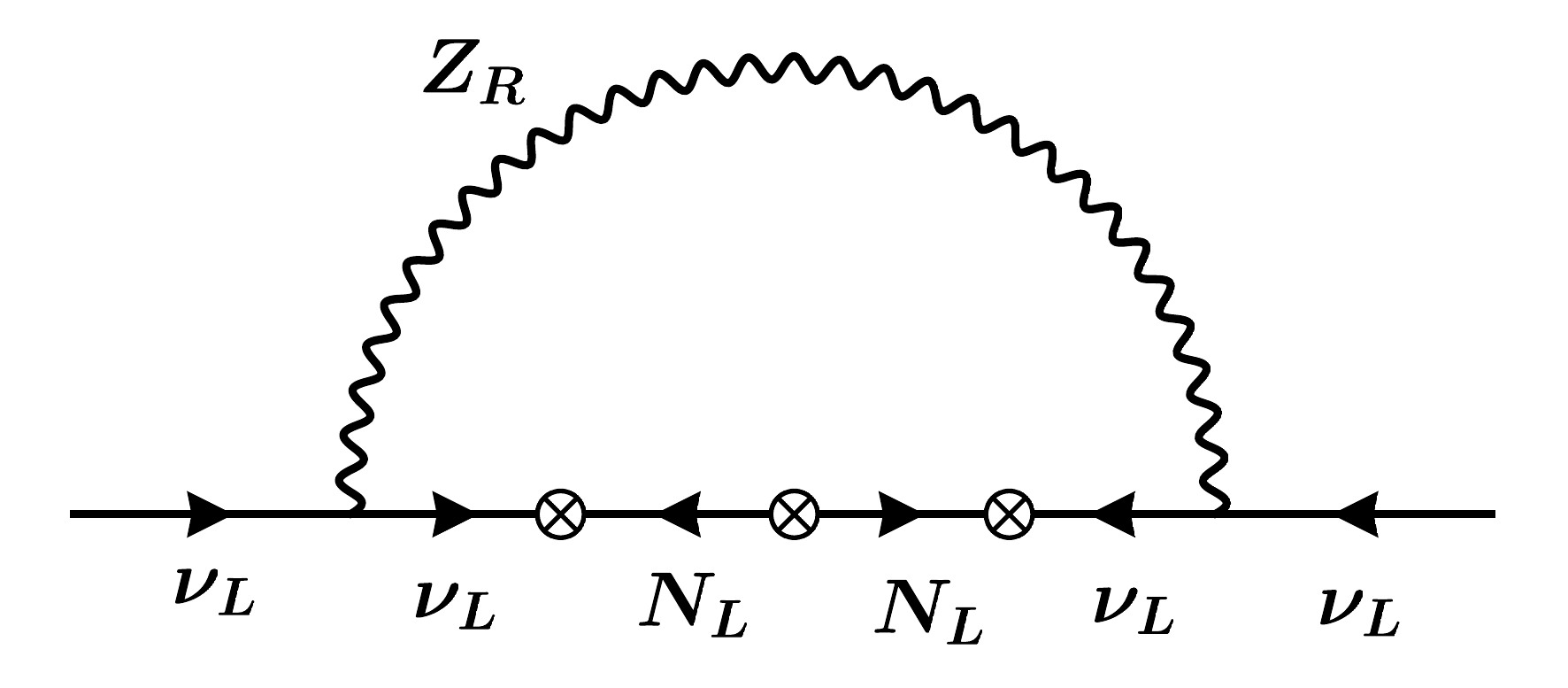}
        \caption{}
    \end{subfigure}

    \vspace{0.4cm} 

    \begin{subfigure}[b]{0.4\textwidth}
        \centering
        \includegraphics[width=\textwidth]{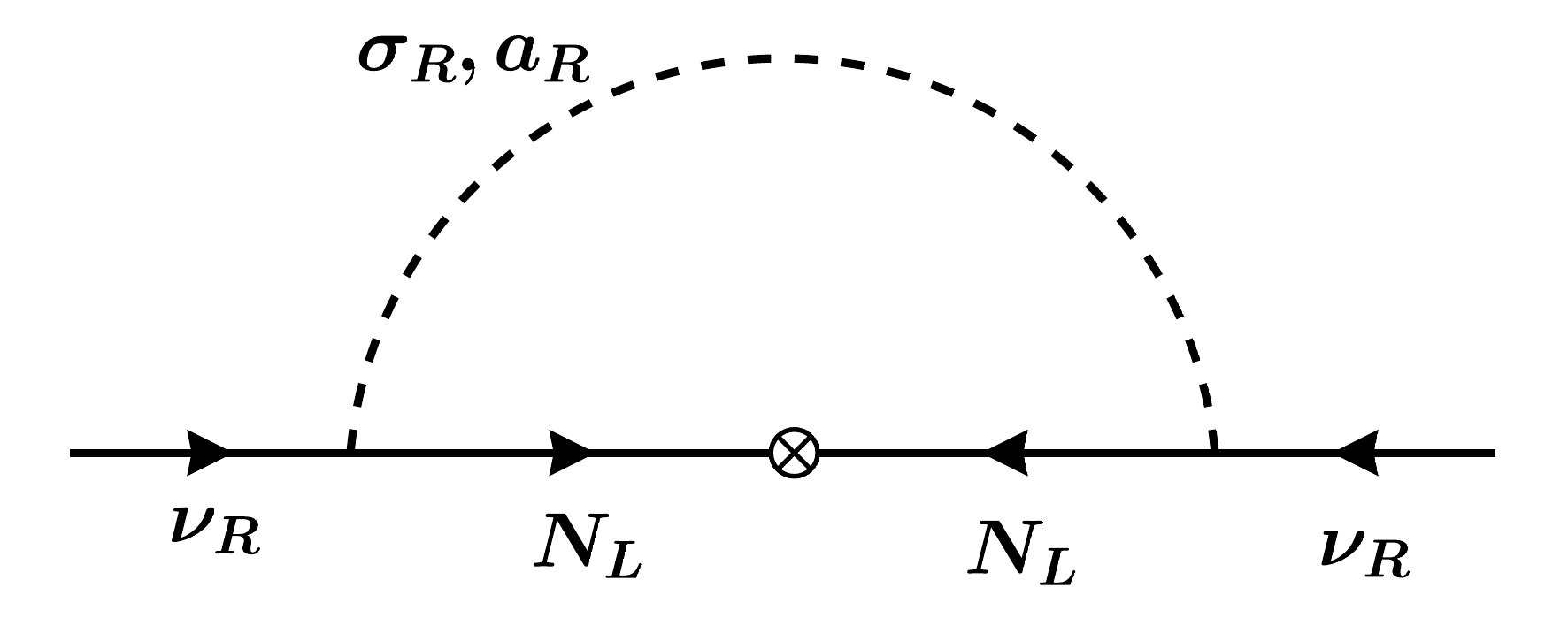}
        \caption{}
    \end{subfigure}
    \hfill
    \begin{subfigure}[b]{0.4\textwidth}
        \centering
        \includegraphics[width=\textwidth]{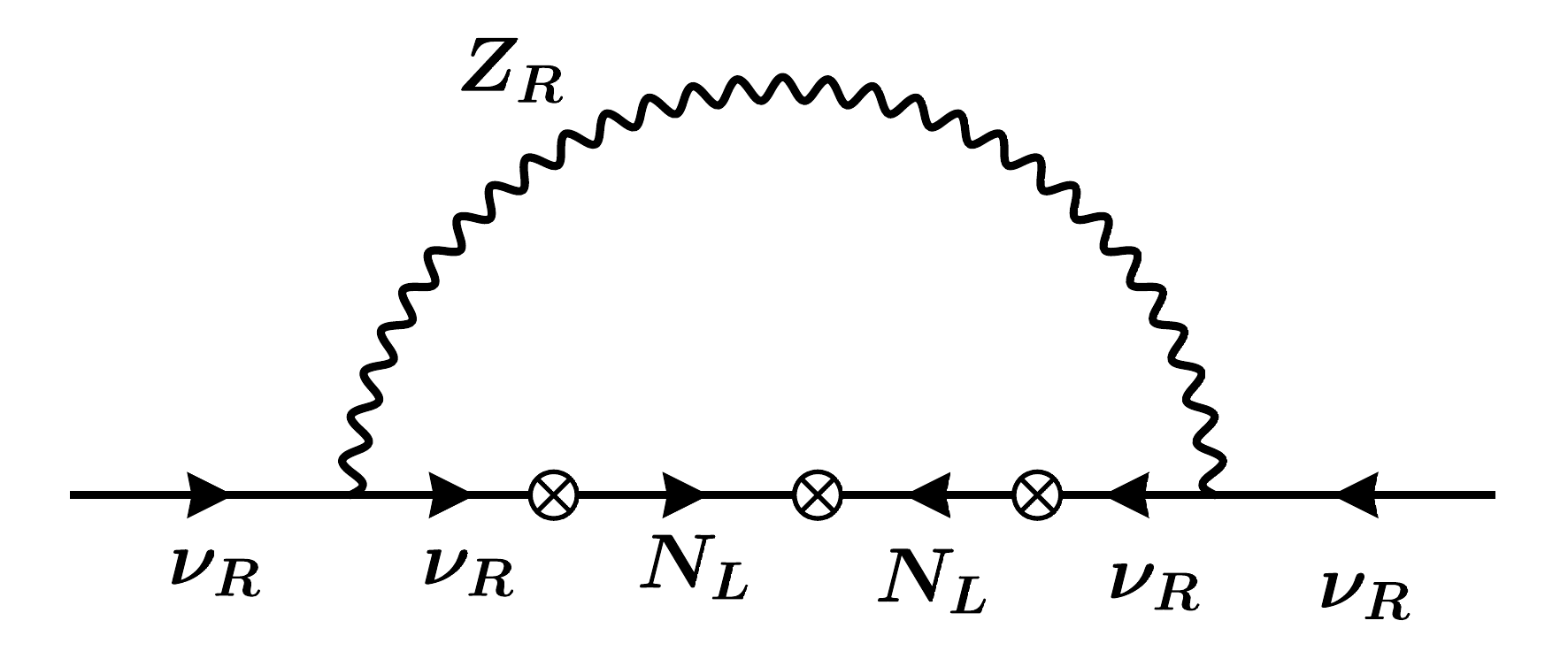}
        \caption{}
    \end{subfigure}
    
    \vspace{.4cm}
        \begin{subfigure}[b]{0.4\textwidth}
        \centering
        \includegraphics[width=\textwidth]{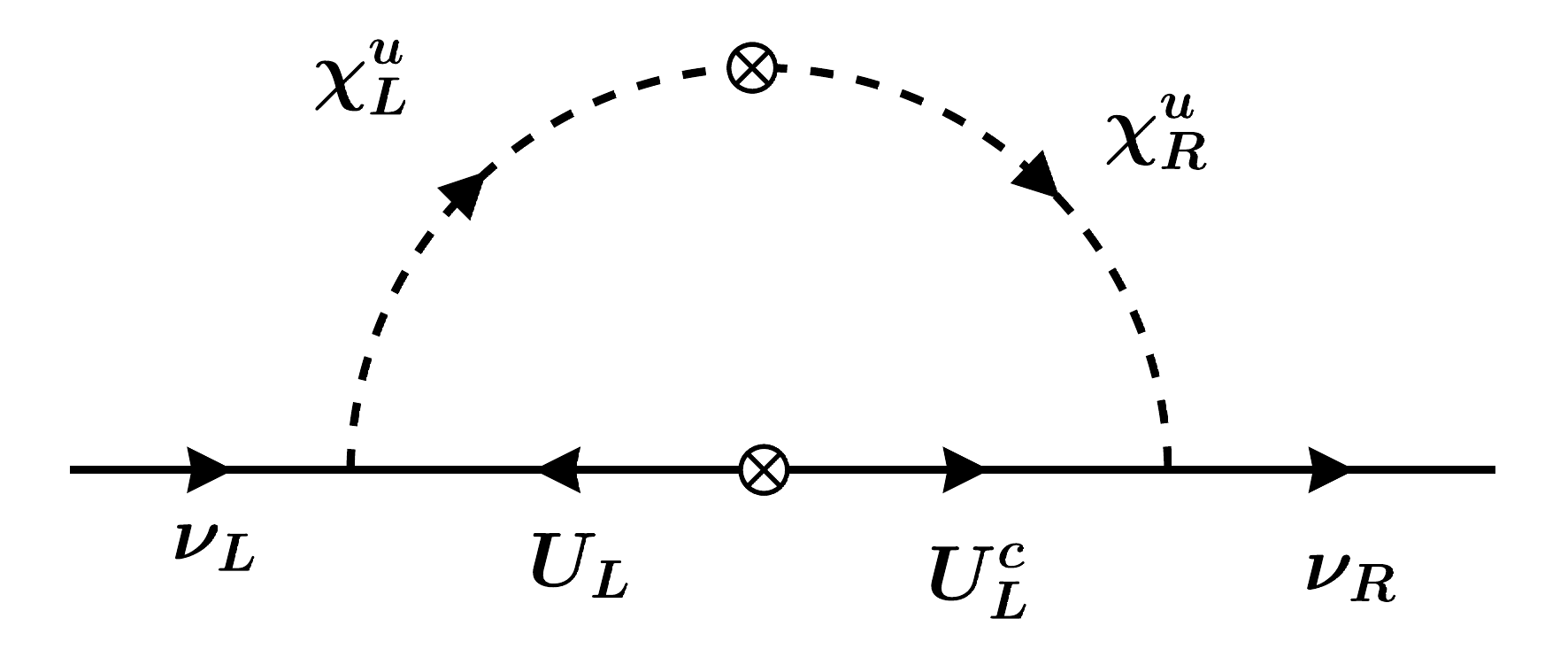}
        \caption{}
        \end{subfigure}
        \hfill
    \begin{subfigure}[b]{0.4\textwidth}
        \centering
        \includegraphics[width=\textwidth]{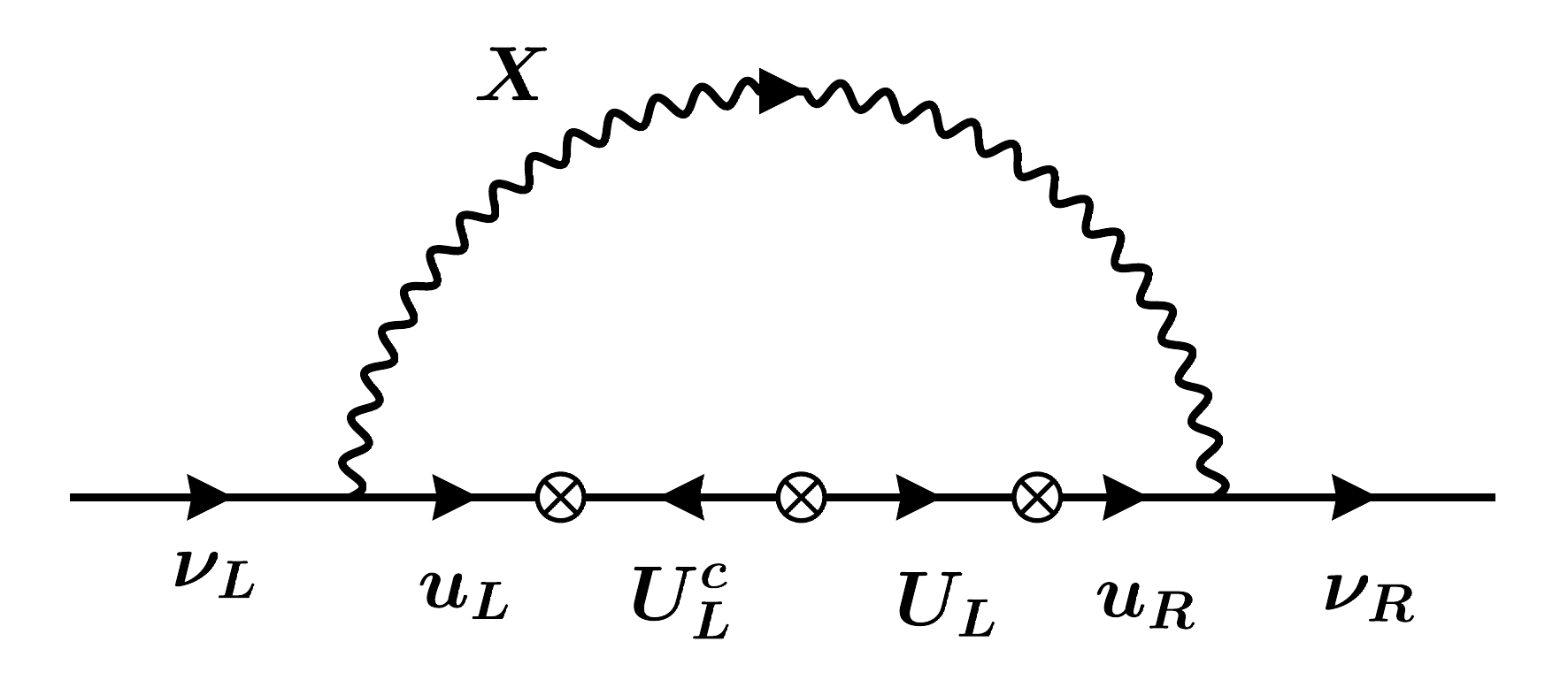}
        \caption{}
    \end{subfigure}

    \vspace{.4cm}
        \begin{subfigure}[b]{0.4\textwidth}
        \centering
        \includegraphics[width=\textwidth]{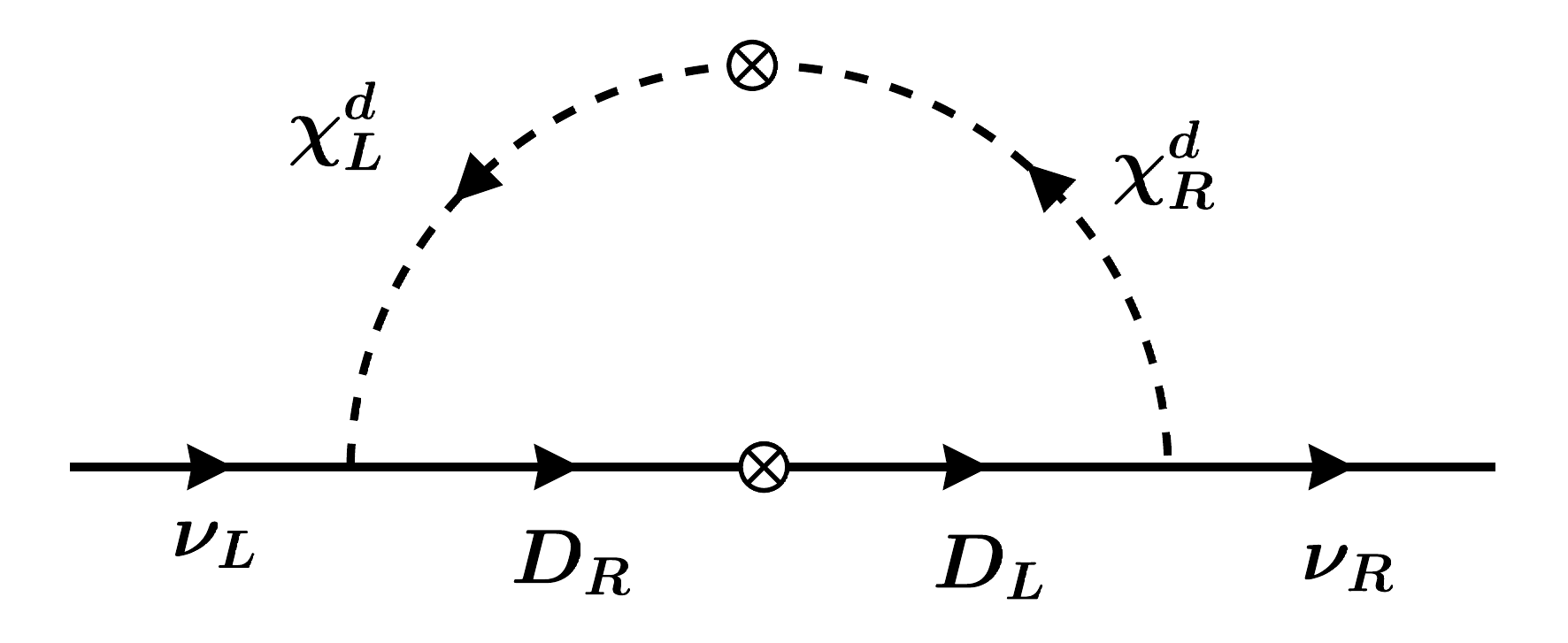}
        \caption{}
        \end{subfigure}
        \hfill
    \begin{subfigure}[b]{0.4\textwidth}
        \centering
        \includegraphics[width=\textwidth]{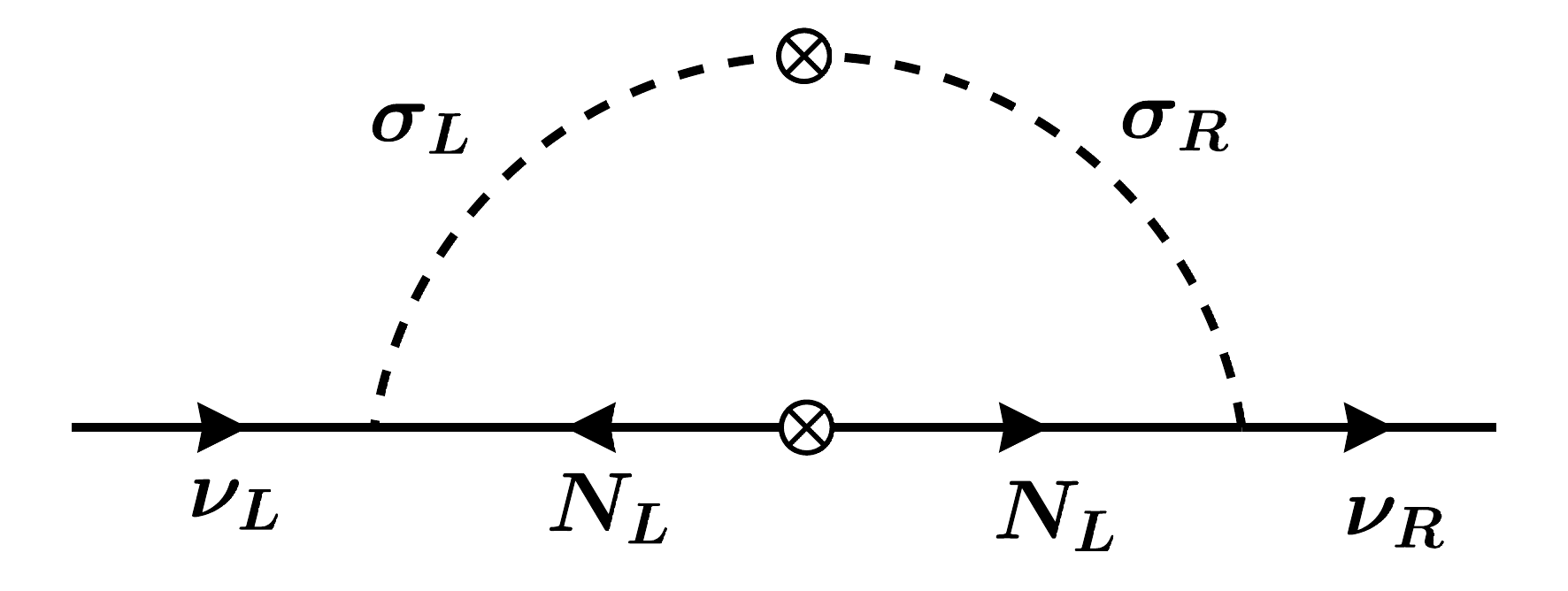}
        \caption{}
    \end{subfigure}
    \caption{Radiative neutrino mass generation through one-loop correction} 
    \label{fig:neutrino_mass_generation}
\end{figure}
In the neutral fermion sector the Majorana mass matrix spanning $(\nu_L, \nu^c_L, N_L)$ resulting from the Yukawa interactions of Eq. (\ref{eq:Yukexpand}) is given by
\begin{eqnarray}
{\cal M}_\nu = \left( \begin{matrix} 0 & 0 & -\frac{\sqrt{3}}{2} Y_{15}^*\kappa_L \cr 
0 & 0 & -\frac{\sqrt{3}}{2}Y_{15} \kappa_R \cr  -\frac{\sqrt{3}}{2} Y_{15}^{\dagger} \kappa_L & -\frac{\sqrt{3}}{2}Y_{15}^T \kappa_R & M_{15}
\end{matrix}  \right)~.
\label{eq:nu-tree}
\end{eqnarray}
It is clear from Eq. (\ref{eq:nu}) that three of the neutrinos remain massless at tree-level.  Quantum corrections, arising through one-loop diagrams,  will, however, generate nonzero entries in the zero blocks of ${\cal M}_\nu$, leading to finite and calculable light neutrino masses. We now turn to the discussion of the induced light neutrino masses.

Loop-induced corrections will populate the zero-blocks of Eq.~\eqref{eq:nu} so that the full matrix becomes
\begin{eqnarray}
{\cal M}_\nu = \left( \begin{matrix} m_L & m_\nu^D & -\frac{\sqrt{3}}{2} Y_{15}^* \kappa_L \cr 
(m_\nu^D)^T & m_R & -\frac{\sqrt{3}}{2}Y_{15} \kappa_R \cr  -\frac{\sqrt{3}}{2} Y_{15}^{\dagger} \kappa_L & -\frac{\sqrt{3}}{2}Y_{15}^T\kappa_R & M_{15}
\end{matrix}  \right)~.
\label{eq:nu}
\end{eqnarray}
with $m_{L,R} = m^T_{L,R}$. 
The light $3 \times 3$ neutrino mass matrix now becomes
\begin{align}
M_{\nu}^{\ell}
&\approx m_L-
\begin{pmatrix}
m_{\nu}^D  & -\dfrac{\sqrt{3}}{2}Y_{15}^* \kappa_L
\end{pmatrix}
\begin{pmatrix}
m_R & -\dfrac{\sqrt{3}}{2} Y_{15} \kappa_R\\
-\dfrac{\sqrt{3}}{2} Y_{15}^{T}\kappa_R & M_{15}
\end{pmatrix}^{-1}
\begin{pmatrix}
(m_{\nu}^D)^T\\
-\dfrac{\sqrt{3}}{2} Y_{15}^{\dagger}\kappa_L
\end{pmatrix}.
\label{eq:Mnu_light_approx-0}\\[2pt]
&\approx
-\dfrac{\kappa_L}{\kappa_R}\left(
m_{\nu}^D\,(Y_{15}^T)^{-1}Y_{15}^{\dagger}
+ Y_{15}^*\,Y_{15}^{-1}(m_{\nu}^D)^T\right)+\dfrac{\kappa_L^2}{\kappa_R^2}\,
\left(
Y_{15}^*\,Y_{15}^{-1}\,m_R\,(Y_{15}^T)^{-1}Y_{15}^{\dagger}
\right)
+ m_L.
\label{eq:Mnu_light_approx-1}
\end{align}
Eq.~\eqref{eq:Mnu_light_approx-1} is actually quite general. In going from
Eq.~\eqref{eq:Mnu_light_approx-0} to Eq.~\eqref{eq:Mnu_light_approx-1}, we have only assumed the hierarchy
\begin{equation}
M_{15}\ll \kappa_R\,m_R^{-1}\kappa_R.
\end{equation}

The loop-induced corrections $m_L$, $m_R$ and $m_\nu^D$ arise through the diagrams shown in Fig.~\ref{fig:neutrino_mass_generation}.  These diagrams can be evaluated to get
\begin{align}
m_L &\simeq
\frac{3}{4}Y_{15}^{\dagger}M_{15}\left[\frac{3}{32\pi^{2}}
\frac{M_{Z_L}^2}{M_{15_3}^{2}-M_{Z_L}^{2}}\ln\!\left(\frac{M_{15_3}^{2}}{M_{Z_L}^{2}}\right)
+ \frac{1}{32\pi^{2}}\frac{M_{\sigma_L}^{2}}{M_{15_3}^{2}-M_{\sigma_L}^{2}}\ln\!\left(\frac{M_{15_3}^{2}}{M_{\sigma_L}^{2}}\right)\right]Y_{15}^*,
\label{eq:mL}\\
m_R &\simeq \frac{3}{4}Y_{15}^{\dagger}M_{15}\left[\frac{3}{32\pi^{2}}
\frac{M_{Z_R}^2}{M_{15_3}^{2}-M_{Z_R}^{2}}\ln\!\left(\frac{M_{15_3}^{2}}{M_{Z_R}^{2}}\right)
+ \frac{1}{32\pi^{2}}\frac{M_{\sigma_R}^{2}}{M_{15_3}^{2}-M_{\sigma_R}^{2}}\ln\!\left(\frac{M_{15_3}^{2}}{M_{\sigma_R}^{2}}\right)\right]Y_{15}^*\,,
\label{eq:mR}\\
m_\nu^D &= m_{\nu}^D(\mathrm{f+g})\,+ m_{\nu}^D(\mathrm{h})\,+\,m_{\nu}^D(\mathrm{i}).
\label{eq:mD}
\end{align}
where we have defined,
\begin{align}
    m_{\nu}^D(\mathrm{f+g})&\simeq 3Y_{15}^{\dagger}M_{15}\left[\frac{3g_4^2}{32\pi^{2}}
\frac{\kappa_L\kappa_R}{M_{15_3}^{2}-M_{X}^{2}}\ln\!\left(\frac{M_{15_3}^{2}}{M_{X}^{2}}\right)
+\frac{\lambda_4 \kappa_L \kappa_R}{16 \pi^2} F(M_{15_3}^2, M_{\chi_1^u}^2,M_{X}^2)\right]Y_{15}\,,\\
 m_{\nu}^D(\mathrm{h})&\simeq \frac{3(2 \lambda_5 \kappa_L \kappa_R)}{16 \pi^2}\,Y_{10}^\dagger M_{10} Y_{10}\, F(M_{10_3}^2, M_{\chi_1^d}^2,M_{\chi_2^d}^2)\,,
 \label{eqn:m_nu_h}\\
  m_{\nu}^D(\mathrm{i})&\simeq \frac{2(\lambda_3+\lambda_4)\kappa_L \kappa_R}{16 \pi^2}\;
Y_{15}^\dagger M_{15} Y_{15}\,
F(M_{15_3}^2, M_{\sigma_L}^2, M_{\sigma_R^d}^2).
\label{eqn:m_nu_i}
\end{align}
The function \(F(x,y,z)\) has been defined in Eq.~\eqref{eqn:F(x,y,z)}. We have performed the loop calculations in ’t Hooft-Feynman gauge, following Refs.~\cite{AristizabalSierra:2011mn, Dev:2012sg} closely for self-energy diagrams. In the process, we have used the couplings \(g_{Z_L\nu\nu}=\sqrt{g_Y^2+g_{2L}^2}/2\) and \(g_{Z_R\nu_R\nu_R}=g_{2R}^2/(2\sqrt{g_{2R}^2-g_Y^2})\) which follow from the kinetic terms in Eq.~\eqref{eq:kinetic-terms} together with the definitions of \(Z_{L,R}\) in Eqs.~\eqref{eq:ZL basis}--\eqref{eq:ZR basis}. The approximate masses of \(Z_{L,R}\) can be obtained from Eq.~\eqref{eq:zprime}. The leptoquark gauge-boson coupling is specified in Eq.~\eqref{eq:X}, with its mass defined in Eq.~\eqref{eqn:charged_gauge_boson_masses}. Scalar couplings are obtained from the expanded Yukawa Lagrangian in Eq.~\eqref{eq:Yukexpand}, while the relevant scalar mass matrices are provided in Sec.~\ref{sec:sec3}.

To get an order of magnitude estimate for the neutrino mass, we consider two benchmark scenarios. One with \(M_{10}\sim \mu_P\) and another with \(M_{15}\sim \mu_P\). In each case, one multiplet needs to be taken relatively light while the other lies around the PS scale and generates neutrino mass, as it is favored by the parity solution to the strong CP problem, which will be discussed in Sec.~\ref{sec:Parity solution to the strong CP problem}.
\vspace{0.2 cm}

\noindent\textbf{First Benchmark:} In this scenario, we consider a case with heavy sextet and light octet fermions. With light octets, say \(M_{15}\sim 10^5 ~\rm{GeV}\), only diagram~\ref{fig:neutrino_mass_generation}(h) will generate neutrino mass of the right order; all the other corrections involving neutral leptons will be suppressed by a factor $M_{15}/\kappa_R$. Using Eq.~\eqref{eqn:m_nu_h}, we arrive at this benchmark:
\[
{\rm Input:}\qquad
\lambda_5 = 3,\qquad
Y_{10}=1,
\]
\[
M_{10_3}=1.5\times 10^{15}~{\rm GeV},\qquad
M_{\chi^d_1}=5\times 10^{12}~{\rm GeV},\qquad
M_{\chi^d_2}=7\times 10^{12}~{\rm GeV},
\]
\[
{\rm Output:}\qquad
m_{\nu}\approx 0.05\,\mathrm{eV}.
\]
The value \(\lambda_5=3\) in this benchmark should be understood as the value of the quartic coupling at the PS breaking scale. Although it is not a very small coupling, it is still perturbative: the relevant expansion parameter is
\(\lambda_5/(4\pi)\simeq 0.24\). It is also below the usual tree-level perturbative unitarity size for scalar quartic couplings, which is of order \(8\pi\). A rough one-loop estimate of the quartic self-renormalization,
\(16\pi^2 d\lambda_5/d\ln\mu\sim c_{\rm eff}\lambda_5^2\), with \(c_{\rm eff}\sim 10\) from scalar multiplicities in the \(\lambda_5\) interaction shows that the coupling does not become nonperturbative immediately above the PS scale.
Also note that this benchmark prefers \(\lambda_5\sim \mathcal{O}(1)\) for generating the neutrino mass. The effectiveness of seesaw in the down-quark sector allowed us to take \(Y_{10}\) to be of order one (see Eq.~\eqref{eq:uni2}).
\vspace{.2 cm}

\noindent \textbf{Second Benchmark:} In this scenario, we consider a case with heavy octets and light sextets. Now with heavy octets, all diagrams involving heavy neutral leptons in Fig.~\ref{fig:neutrino_mass_generation} will generate neutrino mass of the right order. We focus on one of the diagrams, say Fig~\ref{fig:neutrino_mass_generation}(i), and using Eq.~\eqref{eqn:m_nu_i}, we obtain a benchmark given by:
\[
{\rm Input:}\qquad
(\lambda_{3}+ \lambda_{4})= 1.5,\qquad
Y_{15}=1,
\]
\[
M_{15_3}=5\times 10^{13}~{\rm GeV},\qquad
M_{\sigma_L}=125~{\rm GeV},\qquad
M_{\sigma_R}=7\times 10^{13}~{\rm GeV},
\]
\[
{\rm Output:}\qquad
m_{\nu}\approx 0.05\,\mathrm{eV}.
\]

We note that the benchmarks discussed above are presented to show that the observed neutrino mass scale can be produced properly. We did not attempt a detailed fit to the oscillation data here since the model has enough freedom to accommodate them. In particular, the loop-induced matrix \(M_\nu^\ell\) in Eq.~\eqref{eq:Mnu_light_approx-1} depends on the flavor structures \(Y_{15}\) and \(Y_{10}\) through the independent contributions shown in Eqs.~\eqref{eq:mL}--\eqref{eq:mD}. Since these Yukawa matrices involve sufficient independent parameters (which include phases), one can, in principle, adjust both the overall neutrino mass scale and the flavor structure to produce the measured mass-squared differences and mixing angles. We therefore expect realistic fits to neutrino oscillation data within this setup.

\section{Parity solution to the strong CP problem}
\label{sec:Parity solution to the strong CP problem}
The QCD Lagrangian admits a CP-violating topological term,
\begin{equation}
\mathcal{L}_\theta^{\rm QCD}
= \frac{g_s^2}{32\pi^2}\,\theta_{\rm QCD}\, G^a_{\mu\nu}\,\widetilde G^{a\,\mu\nu},
\label{eq:Ltheta}
\end{equation}
where $G^a_{\mu\nu}$ the gluon field strength and 
$\widetilde G^{a\,\mu\nu}
=\tfrac12\epsilon^{\mu\nu\alpha\beta}G^a_{\alpha\beta}$.
$\theta_{\rm QCD}$ is unphysical by itself. In the presence of colored fermions, the physical strong CP phase receives contributions from the phases of all colored mass determinants after chiral rotations, and in our universal seesaw framework, it takes the form
\begin{equation}
  \bar\theta \;=\;
  \theta_{\rm QCD}
  + \mathrm {Arg Det}(\mathcal{M}_u \mathcal{M}_d)
  + 5\,\mathrm{Arg Det} \mathcal{M}_{\cal S}
  + 3\,\mathrm {Arg Det} \mathcal{M}_{\cal O},
    \label{eq:theta-bar-model}
\end{equation}
where \(\mathcal{M}_u\) and \(\mathcal{M}_d\) are the \(6\times 6\) up- and down-type quark mass matrices of
Eq.~\eqref{eq:matrices}, while \(M_{\cal S}\) and \(M_{\cal O}\) denote the bare mass matrices of the color–sextet and color–octet fermions defined in Eq.~\eqref{eqn:sextet-octet-mass}.

Since parity is an exact symmetry in our setup, it will forbid the parity-violating term in Eq.~\eqref{eq:Ltheta} in the fundamental Lagrangian, allowing us to set \(\theta_{\rm QCD}=0\). And from Eq.~\eqref{eq:matrices} it follows that parity forces \({\rm Det} \mathcal{M}_u\) and \({\rm Det} \mathcal {M}_d\) to be real and  since \(\cal M_{\cal S}\), \(\cal M_{\cal O}\) are hermitian, all arguments in Eq.~\eqref{eq:theta-bar-model} vanish and hence \(\bar\theta = 0\) at tree level. We now proceed to compute the loop-induced contributions to $\overline{\theta}$.

\subsection{Contributions to \(\bar{\theta}\) at one loop}

In this section, we compute the one-loop contributions to $\bar\theta$ in our model. We work in the weak (gauge) basis and follow the framework of Ref.~\cite{Babu:1989rb}. As shown in Ref.~\cite{Babu:1989rb} for universal left--right symmetric model, the one-loop corrections to $\bar\theta$ from neutral-scalars, \(\sigma_{L,R}\) and neutral-gauge bosons exchange $(W^0_{\mu L},\,W^0_{\mu R},\,G^{15}_\mu)$ vanish due to the seesaw structure together with parity. The same conclusion holds for our setup, so we do not
present those calculations here. Charged gauge bosons $W^\pm_{\mu L}$ and $W^\pm_{\mu R}$ also do not contribute at this order since they do not mix at the tree level. We instead focus on the new contributions coming from the quark--lepton symmetric structure, mediated by the vector leptoquark $X_\mu$ and the scalar leptoquarks $\chi^{u,d}_{L,R}$. We also take into account loop contributions arising from corrections to the masses of color octet and sextet fermions.

In the basis defined in Eqs.~\eqref{eq:matrices}--\eqref{eqn:sextet-octet-mass}, we can write the one-loop correction for each mass matrix ${\cal M}_i$ ($i=u,d,\cal S, \cal O$) as
\begin{equation}
{\cal M}_i^{\rm 1L}\equiv {\cal M}_i+\delta{\cal M}_i
= {\cal M}_i\,(1+\Sigma_i)\,,\qquad \Sigma_i={\cal M}_i^{-1}\delta{\cal M}_i\,.
\end{equation}
Since the tree-level determinants are real, the one-loop strong-CP phase follows from Eq.~\eqref{eq:theta-bar-model}:
\begin{equation}
\bar\theta^{(1)}=\Im\Tr\!\left[\Sigma_u^{(1)}+\Sigma_d^{(1)}+5\,\Sigma_{\cal S}^{(1)}+3\,\Sigma_{\cal O}^{(1)}\right].
\label{eq:btheta_1loop_all}
\end{equation}
\subsubsection{Corrections to up- and down-quark mass matrices}
\begin{figure}[t!]
    \centering
    \begin{subfigure}[b]{0.325\textwidth}
        \centering
        \includegraphics[width=\textwidth]{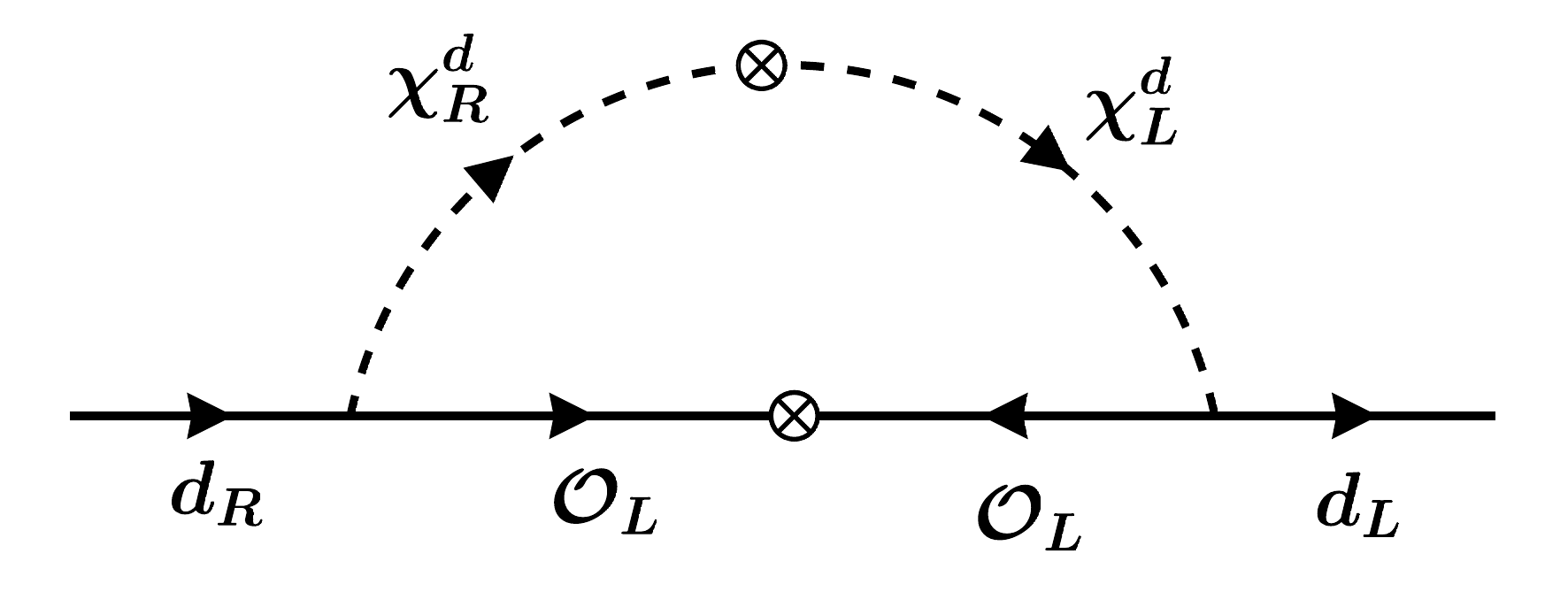}
        \caption{}
    \end{subfigure}
    \begin{subfigure}[b]{0.325\textwidth}
        \centering
        \includegraphics[width=\textwidth]{Figures/CPd_5.pdf}
        \caption{}
    \end{subfigure}
    \begin{subfigure}[b]{0.325\textwidth}
        \centering
        \includegraphics[width=\textwidth]{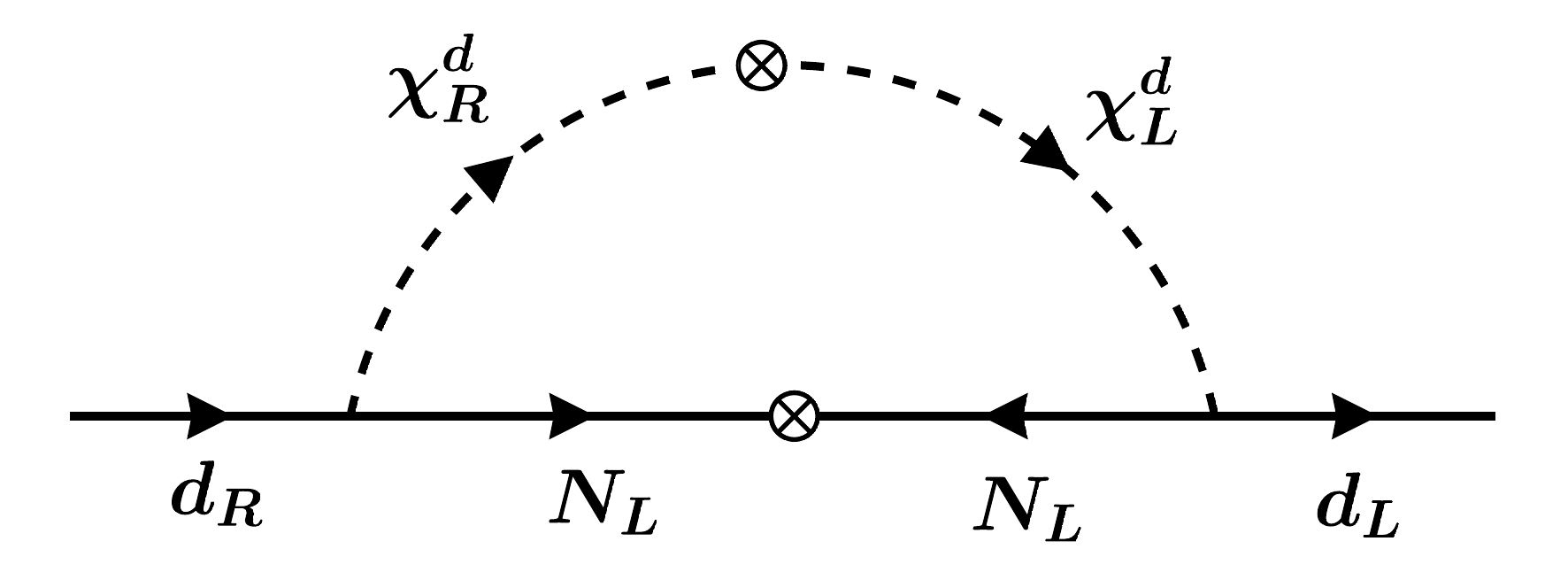}
        \caption{}
    \end{subfigure}

    \vspace{0.4cm} 

    \begin{subfigure}[b]{0.325\textwidth}
        \centering
        \includegraphics[width=\textwidth]{Figures/CPd_1.pdf}
        \caption{}
    \end{subfigure}
     \begin{subfigure}[b]{0.325\textwidth}
        \centering
        \includegraphics[width=\textwidth]{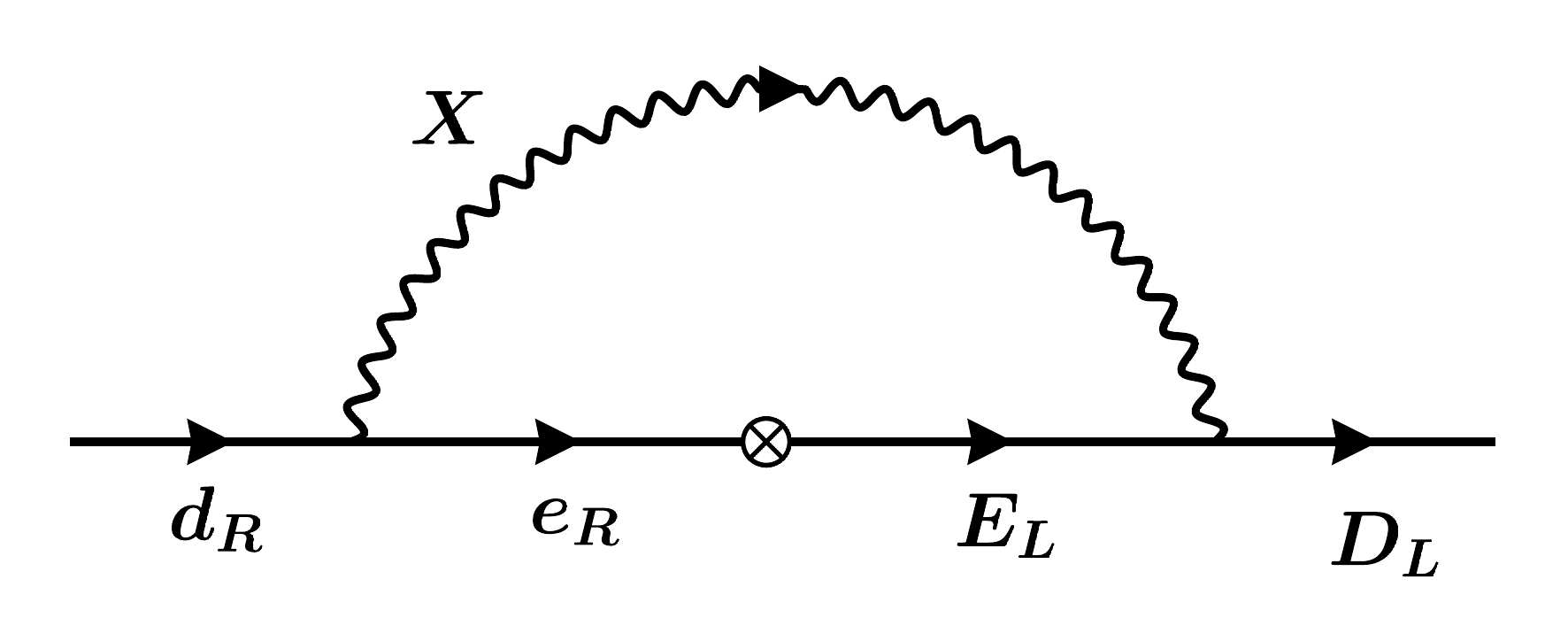}
        \caption{}
    \end{subfigure}
    \begin{subfigure}[b]{0.325\textwidth}
        \centering
        \includegraphics[width=\textwidth]{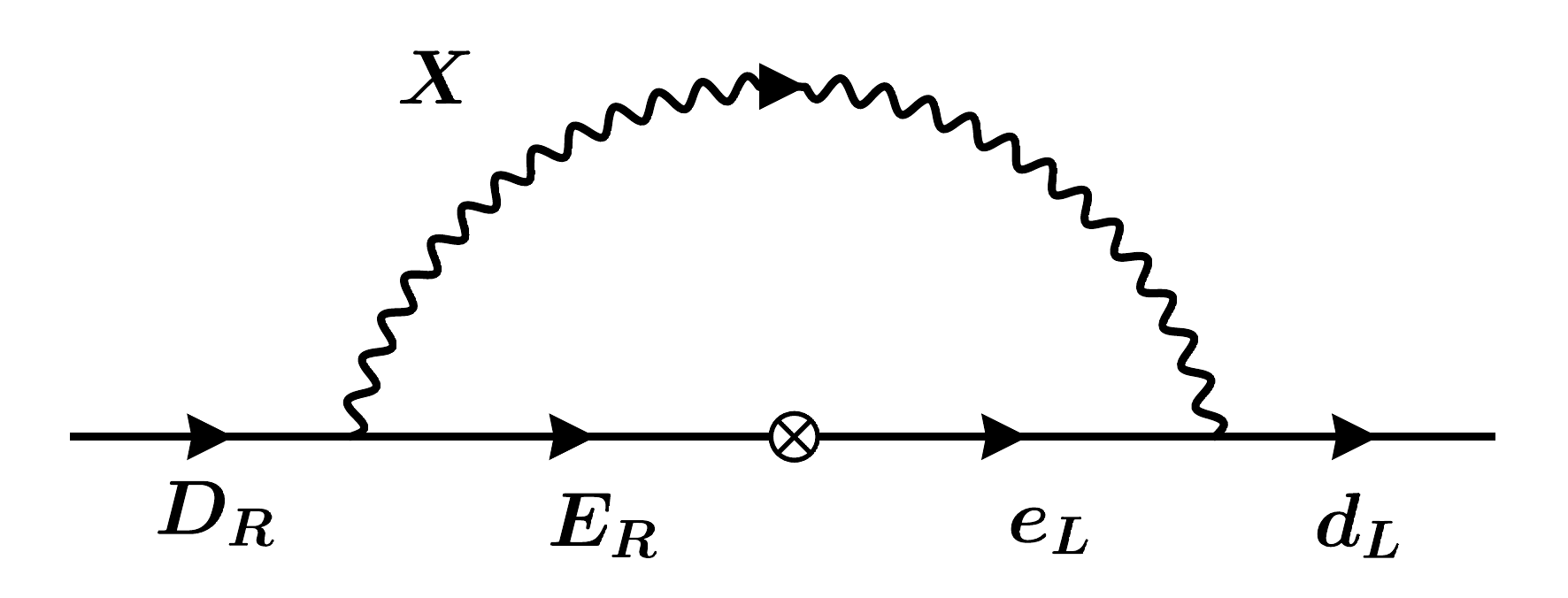}
        \caption{}
    \end{subfigure}
    \caption{One-loop radiative diagrams contributions to the down-quark mass matrix through leptoquark gauge bosons and leptoquark scalars.} 
    \label{fig:strong_cp_down_mass_correction}
\end{figure}

\begin{figure}[t!]
    \centering
    \begin{subfigure}[b]{0.325\textwidth}
        \centering
        \includegraphics[width=\textwidth]{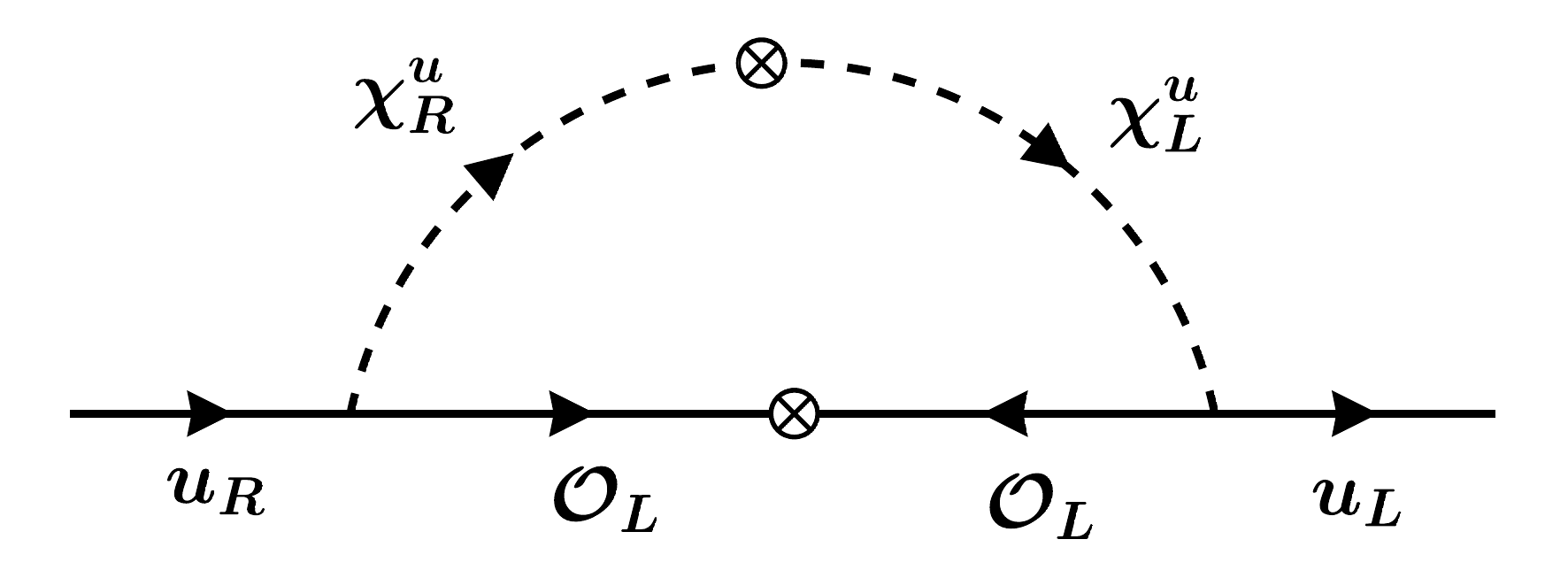}
        \caption{}
    \end{subfigure}
    \begin{subfigure}[b]{0.325\textwidth}
        \centering
        \includegraphics[width=\textwidth]{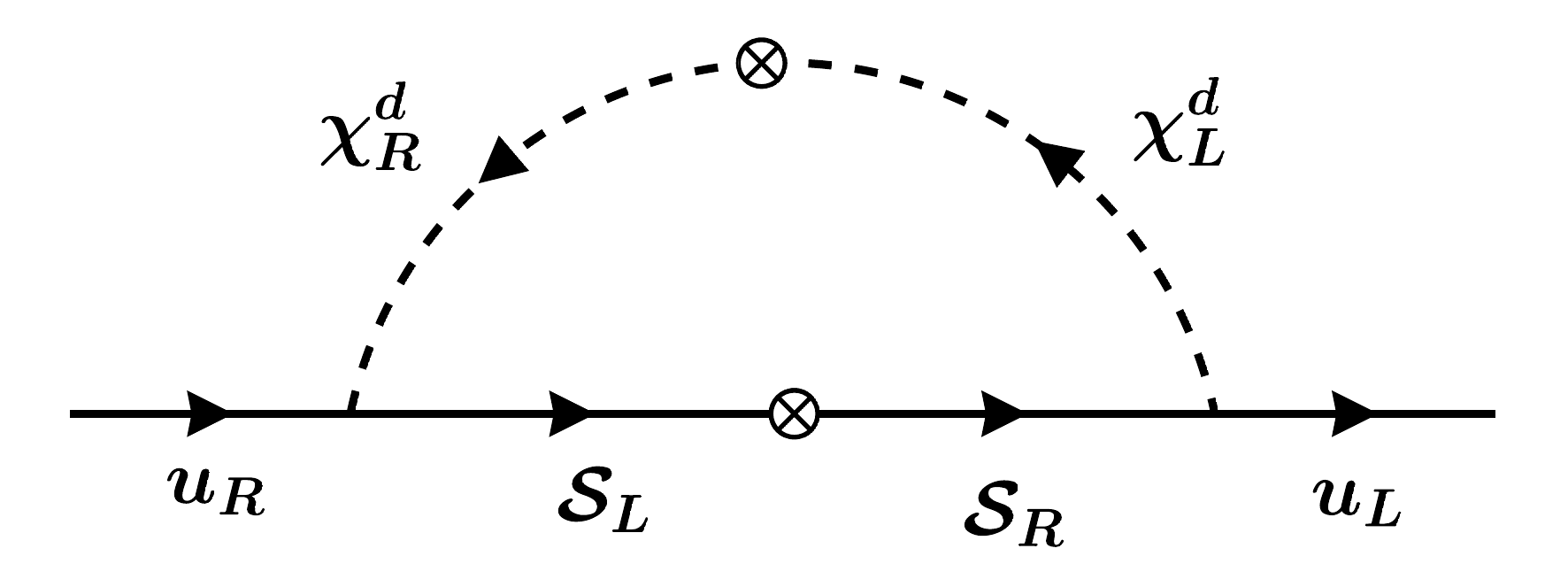}
        \caption{}
    \end{subfigure}
    \begin{subfigure}[b]{0.325\textwidth}
        \centering
        \includegraphics[width=\textwidth]{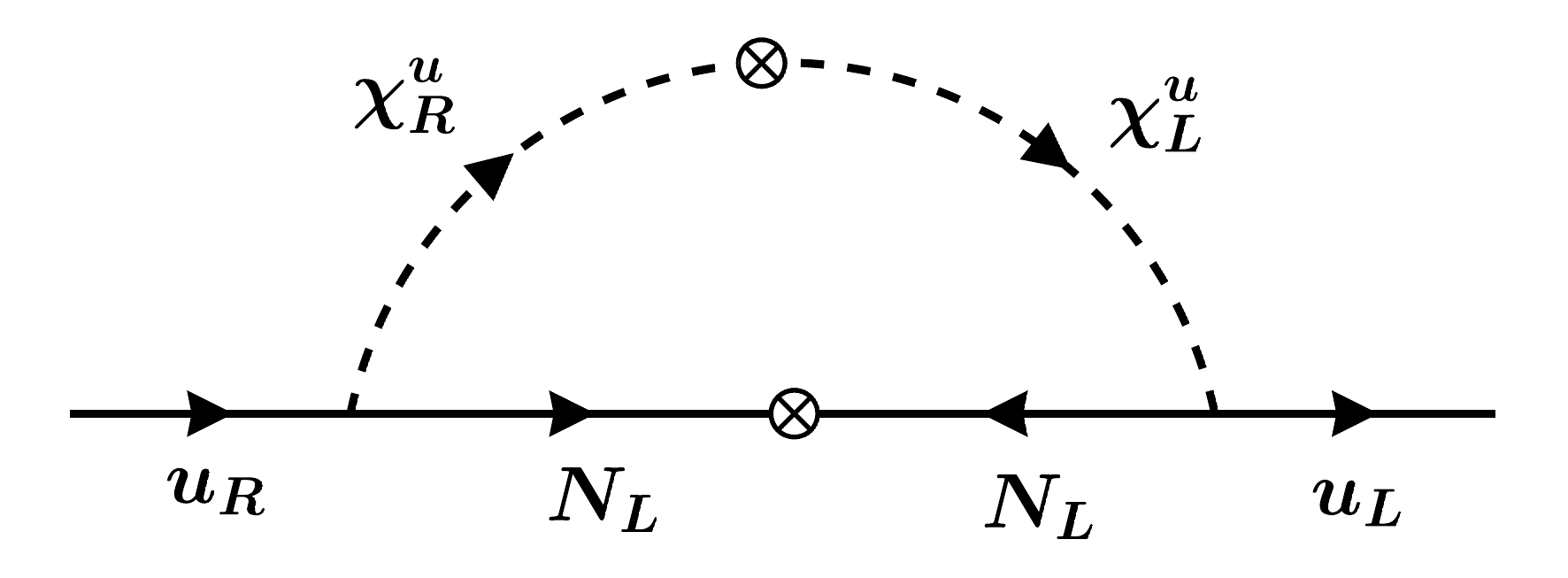}
        \caption{}
    \end{subfigure}

    \vspace{0.4cm} 

    \begin{subfigure}[b]{0.325\textwidth}
        \centering
        \includegraphics[width=\textwidth]{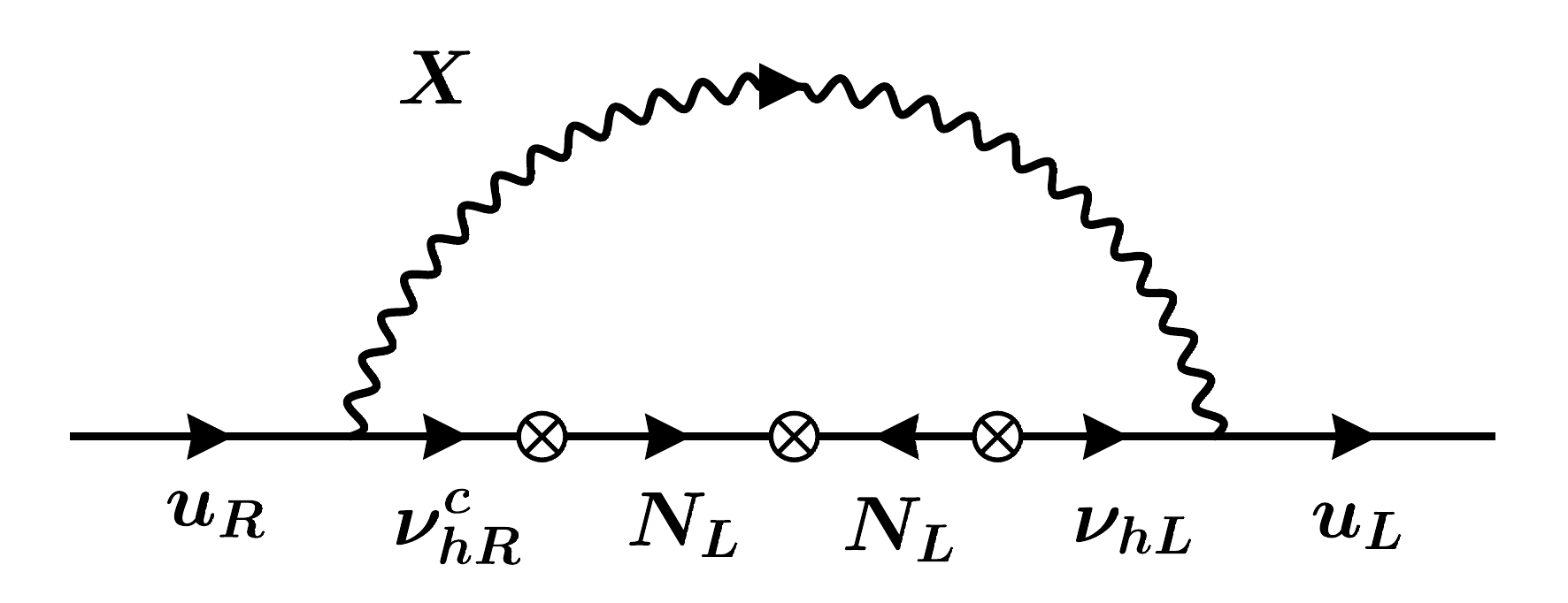}
        \caption{}
    \end{subfigure}
        \begin{subfigure}[b]{0.325\textwidth}
        \centering
        \includegraphics[width=\textwidth]{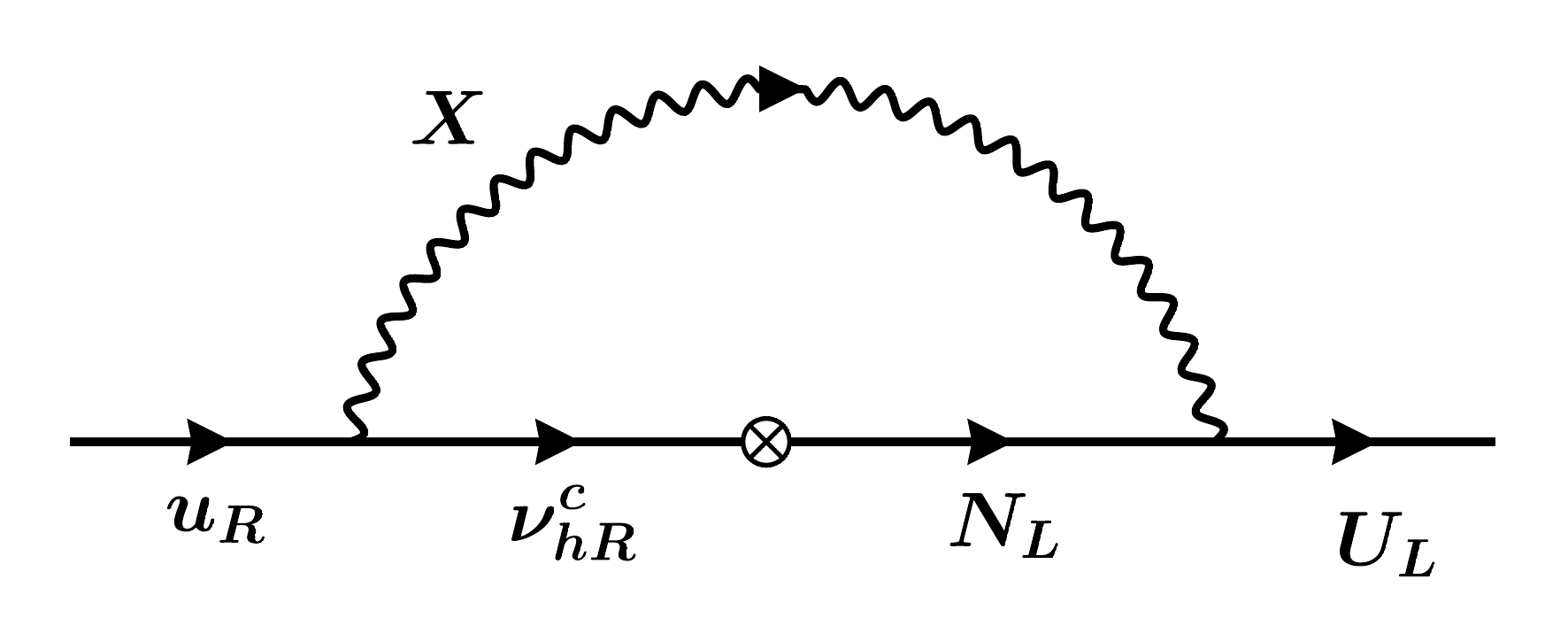}
        \caption{}
    \end{subfigure}
    \begin{subfigure}[b]{0.325\textwidth}
        \centering
        \includegraphics[width=\textwidth]{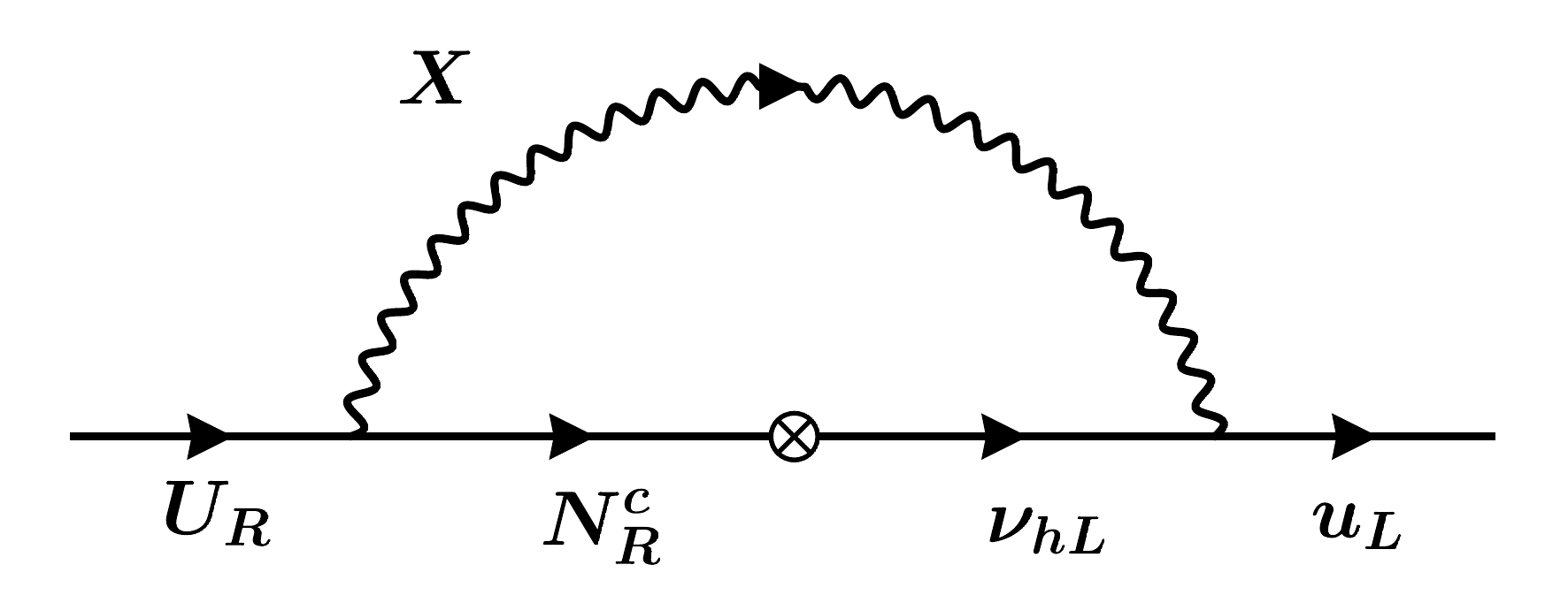}
        \caption{}
    \end{subfigure}
    \caption{One-loop radiative diagrams contributing to the up-quark mass matrix through leptoquark gauge bosons and leptoquark scalars.} 
    \label{fig:strong_cp_up_mass_correction}
\end{figure}

\begin{figure}[t!]
  \centering
  \includegraphics[width=0.45\textwidth]{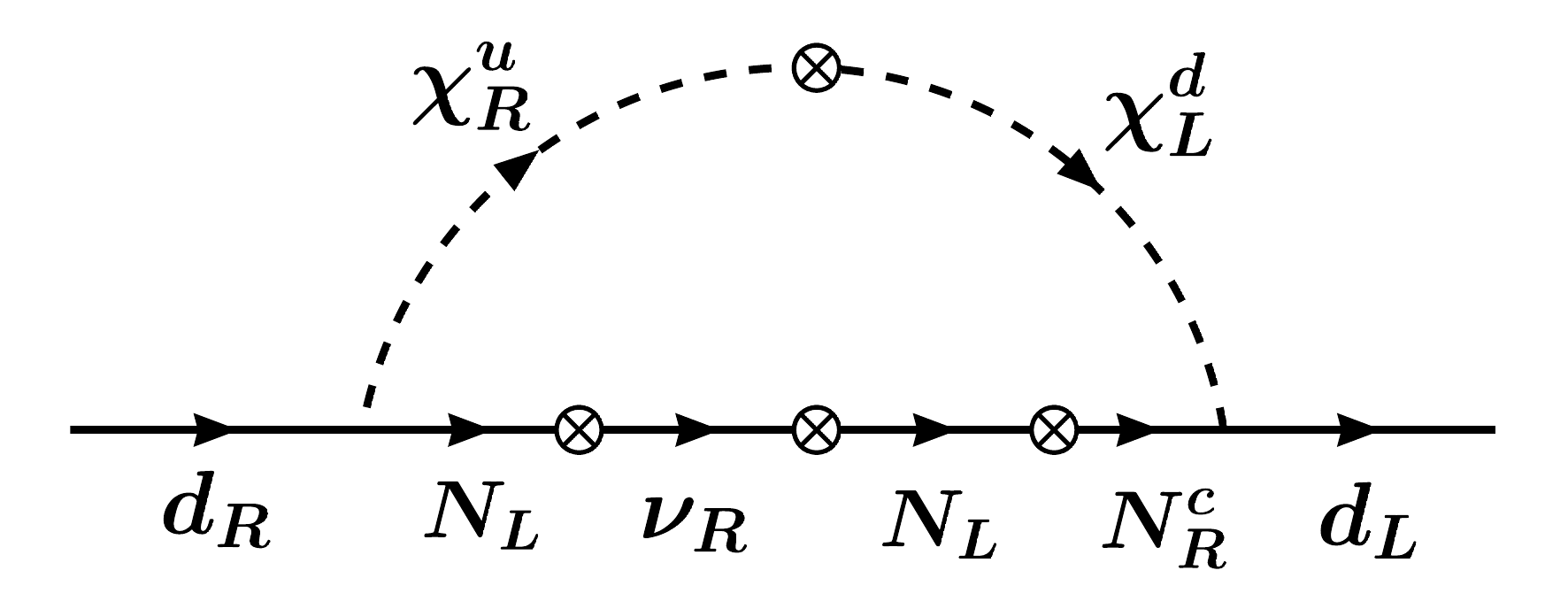}
  \includegraphics[width=0.45\textwidth]{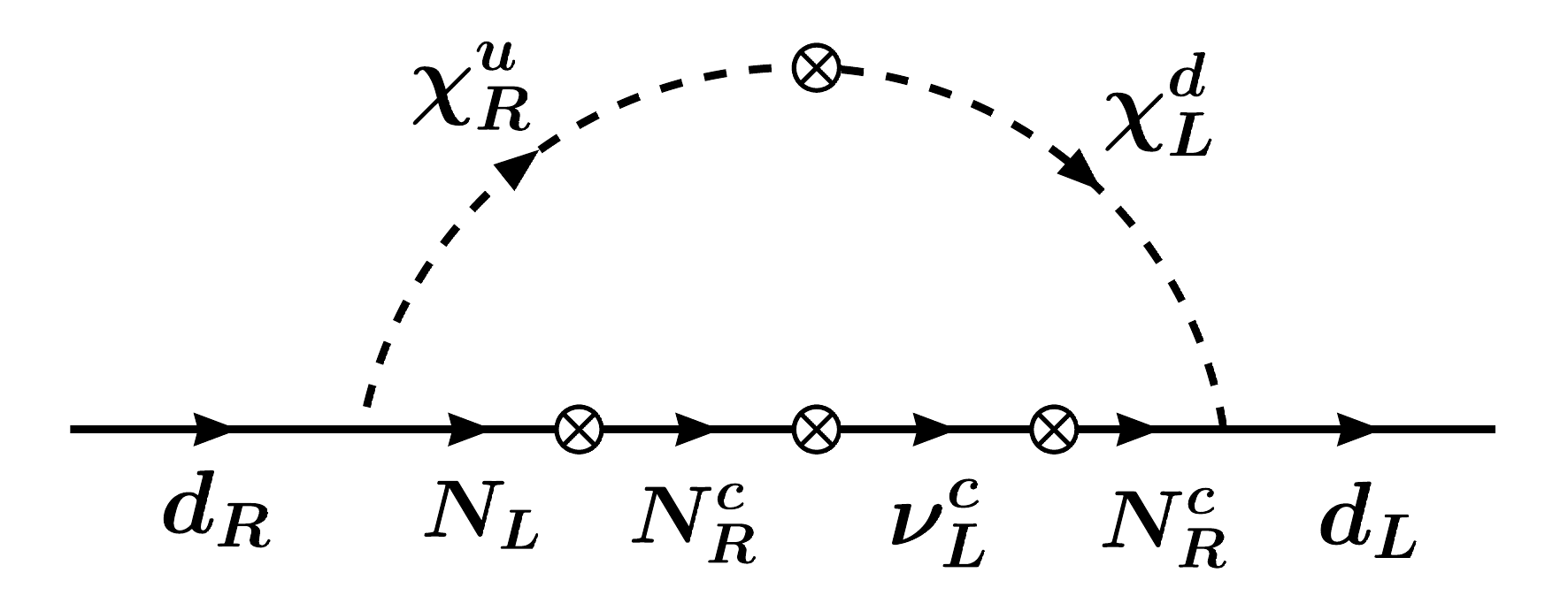}
  \caption{Leading one-loop contributions to $\bar\theta$ from the mass-insertion expansion of Fig.~\ref{fig:strong_cp_down_mass_correction}(c). The two diagrams show the two possible insertions of the vector-like neutral-lepton mass on the internal fermion line.}
  \label{Fig:theta_benchmark}
\end{figure}

\begin{figure}[h!]
\begin{center}
\includegraphics[width=0.35\textwidth]{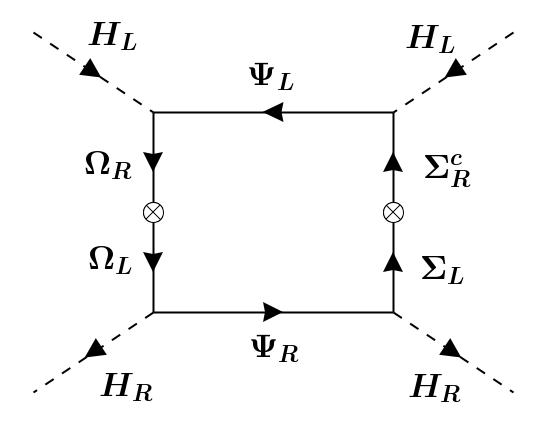}
\end{center}
\caption{Box diagram that radiatively induces the effective \(\lambda_5\) quartic coupling \(\bigl(H_L^T\tilde{H}_L^{*}\bigr)\bigl(H_R^{\dagger}\tilde{H}_R\bigr)\) after integrating out the heavy states.}
\label{Fig:induced_lambda5}
\end{figure}

First, we evaluate the corrections to the quark mass matrices using the $2\times2$ light--heavy block structure of ${\cal M}_{u,d}$. For $q=u,d$ we have $\Sigma_q={\cal M}_q^{-1}\delta{\cal M}_q$, so that
\begin{equation}
\bar\theta^{(1)}_{(u+d)}
=\Im\Tr\!\left[{\cal M}_u^{-1}\delta{\cal M}_u+{\cal M}_d^{-1}\delta{\cal M}_d\right].
\label{eq:btheta_1loop_quarks}
\end{equation}
Let us decompose the corrections to the mass matrix in the light/heavy basis as follows
\begin{equation}
\delta{\cal M}_q=
\begin{pmatrix}
\delta{\cal M}^{q}_{LL} & \delta{\cal M}^{q}_{LH}\\
\delta{\cal M}^{q}_{HL} & \delta{\cal M}^{q}_{HH}
\end{pmatrix},
\end{equation}
and using the inverse of ${\cal M}_q$\footnote {For the inverse of a 2$\cross$2 block matrix (assuming $(2,2)$ block is non-singular), we can implement the following recipe~\cite{lu2002inverses}:
\begin{align*}
    \left( \begin{matrix}
        A & B \\
        C & D 
    \end{matrix}
    \right)^{-1}=\begin{pmatrix}
        S^{-1} & -S^{-1}BD^{-1} \\
        -D^{-1}CS^{-1} &\quad  D^{-1}+D^{-1}CS^{-1}BD^{-1}\\
        \end{pmatrix}
\end{align*}
where the Schur complement $S=(A-BD^{-1}C)$ of $D$ is invertible.}, one finds for the down sector,
\begin{equation}
\bar\theta^{(1)}_d
=\mathrm{Im\,Tr}\left[
-\frac{1}{\kappa_L\kappa_R}\,\delta{\cal M}^{d}_{LL}\,(Y_{10}^\dagger)^{-1}\,M_{10}\,Y_{10}^{-1}
+\frac{1}{\kappa_L}\,\delta{\cal M}^{d}_{LH}\,Y_{10}^{-1}
+\frac{1}{\kappa_R}\,\delta{\cal M}^{d}_{HL}\,(Y_{10}^\dagger)^{-1}
\right],
\label{eq:btheta_master_d}
\end{equation}
where $M_{10}$ denotes the heavy down-quark mass block. Note that $\delta{\cal M}^{d}_{HH}$ drops out from  Eq.~\eqref{eq:btheta_master_d}, so corrections to the heavy--heavy block do not contribute to $\bar\theta$ at one loop. The corresponding equation for the up-sector can be obtained from Eq.~\eqref{eq:btheta_master_d} by the replacement $Y_{10}\to Y_{15}$ and $M_{10}\to M_{15}^{\dagger}$. Since we are treating the mass matrices as interaction vertices, the crosses in Figs.~\ref{fig:strong_cp_down_mass_correction}--\ref{fig:strong_cp_sextet/octet_mass_correction}
represent the geometric sum over all possible mass insertions on the internal fermion line. In addition to the universal left-right symmetric scenario, quark-lepton symmetry also allows leptons, color sextets, and octets to run in the self-energy diagrams to correct the quark masses, which we now turn to.
\vspace{0.2 cm}

\noindent \textbf{Charged-fermion as propagator:} Let us start with the contribution mediated through the charged-lepton line (see Figs.~\ref{fig:strong_cp_down_mass_correction}(d)--(f)). Following the definitions of the basis from Eq.~\eqref{eqn:basis_def}, the re-summed tree-level propagator connecting right- to left-handed fields with momentum $k$ can be written as
\begin{equation}
-{\cal L}^{e,
\,\rm tree}_{\rm eff}=
\left(\overline e_R\,\overline E_R\right)\,
{\cal M}_e^\dagger k^2\left(k^2-{\cal M}_e{\cal M}_e^\dagger\right)^{-1}
\binom{e_L}{E_L}\,,
\label{eq:resummed_lepton_prop}
\end{equation}
where \(\mathcal{M}_e\) has been defined in Eq.~\eqref{eq:matrices}. We write the $2\times2$ block decomposition as
\begin{equation}
\left({\cal M}_e{\cal M}_e^\dagger-k^2\right)^{-1}=
\begin{pmatrix}
P_e(k^2) & Q_e(k^2)\\
Q_e^\dagger(k^2) & R_e(k^2)
\end{pmatrix},
\quad \mathrm{with}\quad P_e(k^2)=P_e^\dagger(k^2),\; R_e(k^2)=R_e^\dagger(k^2) .
\label{eq:PQR_def}
\end{equation}
Inserting \(\mathcal{M}_e\) into Eq.~\eqref{eq:PQR_def} and multiplying out $\left({\cal M}_e{\cal M}_e^\dagger-k^2\right)\left({\cal M}_e{\cal M}_e^\dagger -k^2\right)^{-1}=\mathbb{1}$, we obtain the following relations,
\begin{align}
\Bigl(2\kappa_R^{2}\,Y_{10}^\dagger Y_{10}+M_{10}M_{10}^\dagger-k^{2}\Bigr)\,Q_e^\dagger(k^2)
&= -\sqrt{2}\,\kappa_L\,M_{10}\,Y_{10}^\dagger\,P_e(k^2),\\
2\kappa_L\,Y_{10}Y_{10}^\dagger\,P_e(k^2)+\sqrt{2}\,Y_{10}M_{10}^\dagger\,Q_e^\dagger(k^2)
&=\frac{1}{\kappa_L}\Bigl(\mathbb{1}+k^{2}P_e(k^2)\Bigr),\\
Q_e(k^2)= -\sqrt{2}\kappa_L\,H_e(k^2)\,Y_{10}\,& M_{10}^\dagger \,R_e(k^2),
\label{eq:PQR_identities}
\end{align}
with
\begin{align}
H_e(k^2)&=\Bigl(2\kappa_L^{2}\,Y_{10}Y_{10}^\dagger-k^{2}\Bigr)^{-1}
=H_e^\dagger(k^2),\\
R_e(k^2)&=\left[\left(2\kappa_R^{2}\,Y_{10}^\dagger Y_{10}+M_{10}M_{10}^\dagger-k^{2}\right)-2\kappa_L^{2}M_{10}Y_{10}^\dagger H_eY_{10}M_{10}^\dagger \right]^{-1}=R^{\dagger}_e(k^2).
\label{eq:H-R-def}
\end{align}
Here, though \(M_{10}\) is Hermitian due to parity, we have kept its hermitian conjugate as $M_{10}^\dagger$ for generality. Using these identities, the interaction corresponding to the cross on the electron line can be read off by expanding Eq.~\eqref{eq:resummed_lepton_prop}: 
\begin{align}
-&{\cal L}^{e,
\,\rm tree}_{\rm eff}
=\overline{E}_R\left[\frac{k^{4}}{\sqrt{2}\kappa_L}\,Y_{10}^{-1}\,Q_e(k^2)\right]E_L
+\overline{e}_R\left[\sqrt{2}k^{2}\,Y_{10}\,\kappa_R\,R_e(k^2)\right]E_L \notag\\
&
+\overline{E}_R\left[\frac{k^{2}}{\sqrt{2}\kappa_L}\,Y_{10}^{-1}\Bigl(\mathbb{1}+k^{2}P_e(k^2)\Bigr)\right]e_L
+\overline{e}_R\left[\sqrt{2}k^{2}\,Y_{10}\,\kappa_R\,Q_e^\dagger(k^2)\right]e_L
+\rm{h.c.}
\label{eq:E_prop_correction}
\end{align}
which we use to read the associated fermion line correction in Figs.~\ref{fig:strong_cp_down_mass_correction}(d)--\ref{fig:strong_cp_down_mass_correction}(f). Since ${\cal M}_{u,d}$ has the same seesaw structure as ${\cal M}_e$, we can use the same
re-summation and block-decomposition formalism to read off the corrections from up- and down-type quark lines (see Fig~\ref{fig:strong_cp_sextet/octet_mass_correction} for example).
\vspace{0.2 cm}

\noindent \textbf{Neutral lepton as propagator:}
For the neutrino line correction (see Figs.~\ref{fig:strong_cp_down_mass_correction}(c) and
\ref{fig:strong_cp_up_mass_correction}(c)--\ref{fig:strong_cp_up_mass_correction}(f)), we proceed similarly. Starting from the Majorana mass matrix in Eq.~\eqref{eq:nu-tree}, we may equivalently use the corresponding  $4$-component mass term $(\overline{\nu_L}\ \overline{\nu_L^c}\ \overline{N_L})\,\mathcal M_\nu'\,(\nu_R^c\ \nu_R\ N_R^c)^T$ and perform the following small rotation in the $(\nu_L,\nu_L^c)$ plane
that removes the $\nu_L$--$N$ mixing at ${\cal O}(\kappa_L/\kappa_R)$,
\begin{equation}
U \simeq \begin{pmatrix} \mathbb{1} & -\rho \\ \rho^\dagger & \mathbb{1} \end{pmatrix},
\qquad
\rho=\frac{\kappa_L}{\kappa_R}Y_{15}(Y_{15}^*)^{-1},
\label{eq:rho_explicit}
\end{equation}
and the $N$-coupling with $\nu_{\ell L}=\nu_L-\rho\,\nu_L^c$ vanishes up to ${\cal O}(\kappa_L^2/\kappa_R^2)$.
The rotated states can be written as
\begin{equation}
\nu_{\ell L}=\nu_L-\rho\,\nu_L^c,\qquad
 \nu_{hL}=\nu_L^c+\rho^\dagger\nu_L.
\label{eq:nu_rot_states}
\end{equation}
Keeping terms up to ${\cal O}(\kappa_L^2/\kappa_R^2)$, the tree-level neutral-lepton mass term gets reduced to a
$2\times2$ block in the heavy-sector subspace:
\begin{equation}
{\cal L}^{\nu_h}_{\rm mass}
=
\left(\overline{\nu_{hL}}\;\overline{N_L}\right)\,
{\cal M}_{\nu_h}\,
\binom{\nu_{hR}^{\,c}}{N_R^{\,c}}
\;+\;\text{h.c.},
\label{eq:Lmass_nuh}
\end{equation}
where we have defined
\begin{equation}
{\cal M}_{\nu_h}=
\begin{pmatrix}
0 & -\dfrac{\sqrt3}{2}\,\kappa_R\,Y_{15}^{*}\\[4pt]
-\dfrac{\sqrt3}{2}\,\kappa_R\,Y_{15}^{\dagger} & M_{15}^{\dagger}
\end{pmatrix}
+{\cal O}(\kappa_L^2/\kappa_R^2),
\label{eq:M_nuh_def}
\end{equation}
and $\nu_\ell$ remains massless at tree level. Though ${\cal M}_{\nu_h}$ has a similar structure as other mass matrices in Eq.~\eqref{eq:matrices}, there is an important difference. The (1,2) block and the (2,1) block are transposes of each other, unlike in  Eq.~\eqref{eq:matrices}, where the Yukawa matrices are hermitian conjugates.
Consequently, for the neutrino contributions to $\overline{\theta}$, we should also define the analogous \(2\times 2\) block decomposition for the projection operator and write down the identities carefully. We define
\begin{equation}
\bigl({\cal M}_{\nu_h}{\cal M}_{\nu_h}^\dagger-k^2\bigr)^{-1}=
\begin{pmatrix}
P_\nu(k^2) & Q_\nu(k^2)\\
Q_\nu^\dagger(k^2) & R_\nu(k^2)
\end{pmatrix},
\quad \mathrm{with}\quad P_{\nu}(k^2)=P_{\nu}^\dagger(k^2),\; R_{\nu}(k^2)=R_{\nu}^\dagger(k^2), 
\label{eq:PQR_nu_def}
\end{equation}
and adopt the same recipe as before to get the following relations:
\begin{align}
\Bigl(\frac{3}{4}\kappa_R^{2}\,Y_{15}^{\dagger}Y_{15}+M_{15}^{\dagger}M_{15}k^{2}\Bigr)\,Q_\nu^\dagger(k^2)
&=\frac{\sqrt{3}}{2}\,\kappa_R\,M_{15}^{\dagger}\,Y_{15}^{T}\,P_\nu(k^2),\\
\frac{3}{4}\kappa_R\,Y_{15}^{*}Y_{15}^{T}\,P_\nu(k^2)-\frac{\sqrt{3}}{2}\,Y_{15}^{*}M_{15}\,Q_\nu^\dagger(k^2)
&=\frac{1}{\kappa_R}\Bigl(\mathbb{1}+k^{2}P_\nu(k^2)\Bigr),\\
Q_\nu(k^2)
=\frac{\sqrt{3}}{2}\,\kappa_R\,H_\nu(k^2)\,Y_{15}^{*}&M_{15}\,R_\nu(k^2),
\label{eq:PQR_nu_identities}
\end{align}
with 
\begin{align}
H_\nu(k^2)
&=\Bigl(\frac{3}{4}\kappa_R^{2}\,Y_{15}^{*}Y_{15}^{T}-k^{2}\Bigr)^{-1}
=H_\nu^\dagger(k^2),
\label{eq:H-R-nu-def-0}\\
R_\nu(k^2)
&=\Biggl[
\Bigl(\frac{3}{4}\kappa_R^{2}\,Y_{15}^{\dagger}Y_{15}+M_{15}^\dagger M_{15}-k^{2}\Bigr)
-\frac{3}{4}\kappa_R^{2}\,M_{15}^\dagger Y_{15}^{T}H_{\nu}
Y_{15}^{*}M_{15}
\Biggr]^{-1}=R^{\dagger}_{\nu}(k^2).
\label{eq:H-R-nu-def}
\end{align}
Now, we can read the interaction terms corresponding to the cross on the neutrino line from the effective Lagrangian given by
\begin{align}
-&{\cal L}^{\nu,
\,\rm tree}_{\rm eff}
=\overline{N^c_R}\left[-\frac{2k^{4}}{\sqrt{3}\kappa_R}\,(Y_{15}^*)^{-1}\,Q_{\nu}(k^2)\right]N_L
+\overline{\nu_h^c}_R\left[-\frac{\sqrt{3}k^{2}}{2}\,Y_{15}\,\kappa_R\,R_{\nu}(k^2)\right]N_L \notag\\
&
+\overline{N^c_R}\left[-\frac{2k^{2}}{\sqrt{3}\kappa_R}(Y_{15}^*)^{-1}\Bigl(\mathbb{1}+k^{2}P_{\nu}(k^2)\Bigr)\right]\nu_{hL}
+\overline{\nu^c_h}_R\left[-\frac{\sqrt{3}k^{2}}{2}\,Y_{15}\,\kappa_R\,Q_{\nu}^\dagger(k^2)\right]\nu_{hL}
+\rm{h.c.}
\label{eq:Nu_prop_correction}
\end{align}
We use this equation to read the associated neutral lepton line correction in Fig.~\ref{fig:strong_cp_down_mass_correction}(c) and Fig.~\ref{fig:strong_cp_up_mass_correction}(c)--(f).
\vspace{.2 cm}

\noindent \textbf{Octet and sextet as propagator:}
Since the color octets, ${\cal O}_{L\beta}^{\alpha}$, do not mix with other fields, the insertion on the internal octet line involves only the Hermitian bare mass matrix $M_{15}$. Similar comment holds for the sextets, \(\mathcal{S_{\alpha\beta}}\), with mass \(M_{10}\).
\vspace{.2 cm}

\noindent \textbf{Null contributions to \(\bar{\theta}\) at one-loop:}
Now we are ready to evaluate the one-loop contributions to $\bar\theta$. Among the diagrams in Fig.~\ref{fig:strong_cp_down_mass_correction} and Fig.~\ref{fig:strong_cp_up_mass_correction}, only Fig.~\ref{fig:strong_cp_down_mass_correction}(c) can generate a nonzero contribution; all others give vanishing contributions to $\overline{\theta}$, as we shall show below.

Before writing the explicit contributions, we clarify our convention for the ordering of flavor matrices. We use the mass-matrix basis of Eqs.~\eqref{eqn:basis_def}--\eqref{eqn:sextet-octet-mass}, so that the loop corrections \(\delta M_f\) are written in the same left--right convention. The diagrams first fix the ordering inside \(\delta M_d\). After this correction is obtained, its insertion into \(\bar\theta\) is fixed by Eq.~\eqref{eq:btheta_master_d}. Let's start with Fig.~\ref{fig:strong_cp_down_mass_correction}(a). The corresponding correction to the down-quark light--light block is
\begin{equation}
\delta{\cal M}^{d}_{LL}
=\int \frac{d^{4}k}{(2\pi)^{4}}
\frac{Y_{15}M_{15}\,Y_{15}^{\dagger}\,(2\lambda_5\,\kappa_L\kappa_R)}
{\big(k^{2}-M_{15}^{2}\big)\big((p-k)^{2}-M^{2}_{\chi_R^{d}}\big)\big((p-k)^{2}-M^{2}_{\chi_L^{d}}\big)}\,,
\end{equation}
where the relevant vertices follow from Eqs.~\eqref{eq:Yukexpand} and \eqref{eq:Higgspot}. From Eq.~\eqref{eq:btheta_master_d}, we see that this diagram contributes to \(\bar{\theta}_d\) as follows:
\begin{equation}
\Im\,\Tr\!\left[
-2\lambda_5\!\int \frac{d^{4}k}{(2\pi)^{4}}\,
\frac{Y_{15}M_{15}\,Y_{15}^{\dagger}}
{\big(k^{2}-M_{15}^{2}\big)\big((p-k)^{2}-M^{2}_{\chi_R^{d}}\big)\big((p-k)^{2}-M^{2}_{\chi_L^{d}}\big)}
\,(Y_{10}^\dagger)^{-1}\,M_{10}\,Y_{10}^{-1}
\right].
\end{equation}
We don't need to do the explicit loop integration here. We can already see that the quantity inside the trace is a product of two Hermitian matrices,
\begin{equation}
Y_{15}M_{15}\,Y_{15}^{\dagger}
\qquad\text{and}\qquad
(Y_{10}^\dagger)^{-1}\,M_{10}\,Y_{10}^{-1},
\end{equation}
and therefore its trace is real. Hence this contribution to $\bar\theta$ from Fig.~\ref{fig:strong_cp_down_mass_correction}(a) vanishes. We follow the same techniques to check the one-loop contributions from other diagrams. 

Fig.~\ref{fig:strong_cp_down_mass_correction}(b) has the similar flavor structure, 
\begin{equation}
\Im\,\Tr\!\left[\left(Y_{10}\,M_{10}\,Y_{10}^\dagger\right)\left((Y_{10}^\dagger)^{-1}\,M_{10}\,Y_{10}^{-1}\right)\right]=0.
\end{equation}
Fig.~\ref{fig:strong_cp_down_mass_correction}(d) involves the leptoquark gauge boson \(X_\mu\). We evaluate this contribution in the unitary gauge. In that case, we don't need to compute the would-be Goldstone boson contribution (but, in a renormalizable gauge, it would carry the same flavor structure as the corresponding gauge boson interaction). Now let's use the charged-lepton mass insertion from Eq.~\eqref{eq:E_prop_correction} and pick the corresponding light--light down-type block contribution to \(\bar{\theta}\) from Eq.~\eqref{eq:btheta_master_d}. The relevant flavor structure can be written as
\begin{equation}
\Im\Tr\left[Q_eY_{10}^{\dagger}(Y_{10}^\dagger)^{-1}\,M_{10}\,Y_{10}^{-1}\right]=-\sqrt{2}\kappa_L\Im\Tr\left[\Bigl(2\kappa_L^{2}\,Y_{10}Y_{10}^\dagger-k^{2}\Bigr)^{-1}\left(M_{10}^\dagger \,R_e(k^2)M_{10}\right)\right],
\end{equation}
which is also the trace of a product of two Hermitian matrices, and hence zero. We have used the relevant definitions from Eqs.~\eqref{eq:PQR_identities}-\eqref{eq:H-R-def} to reach this conclusion.

Fig.~\ref{fig:strong_cp_down_mass_correction}(e) contributes to the 
$\delta M^d_{HL}$ block in Eq.~\eqref{eq:btheta_master_d}. Using the insertion in 
Eq.~\eqref{eq:E_prop_correction}, the flavor structure entering $\bar\theta$ reduces to
\begin{equation}
\Im\,\Tr\!\left[R_e^{\dagger}Y_{10}^{\dagger}(Y_{10}^{\dagger})^{-1}\right]
=0,
\end{equation}
where the fact $R_e=R_e^{\dagger}$ has been used.
 
Fig.~\ref{fig:strong_cp_down_mass_correction}(f) contributes to the 
$\delta M^d_{LH}$ block leading to 
\begin{equation}
\Im\,\Tr\!\left[\bigl(\mathbb{1}+k^{2}P_e(k^2)\bigr)^{\dagger}(Y_{10}^{-1})^{\dagger}Y_{10}^{-1}\right]=0.
\end{equation}
With the fact that \(P_e=P_e^{\dagger}\), it's the imaginary trace of a product of two hermitian matrices, which vanishes.

Now let's move to the up-type quark sector, Fig.~\ref{fig:strong_cp_up_mass_correction}(a) and (b), representing the up-quark mass correction through the octet and sextet, respectively, which give null contributions to \(\bar{\theta}\), similar to the down sector. Explicitly, for Fig.~\ref{fig:strong_cp_up_mass_correction}(a), we get
\begin{equation}
\Im\,\Tr\!\left[\left(Y_{15}\,M_{15}\,Y_{15}^\dagger\right)\left((Y_{15}^\dagger)^{-1}\,M_{15}^{\dagger}\,Y_{15}^{-1}\right)\right]=0,
\end{equation}
where \((Y_{15}^\dagger)^{-1}\,M_{15}^{\dagger}\,Y_{15}^{-1}\) appears from the \(\delta\mathcal{M}_{LL}^u\) insertion (the up-sector analogue of Eq.~\eqref{eq:btheta_master_d}). Similarly, Fig.~\ref{fig:strong_cp_up_mass_correction}(b) also gives no contribution,
\begin{equation}
\Im\,\Tr\!\left[\left(Y_{10}^\dagger\,M_{10}^{\dagger}\,Y_{10}\right)\left((Y_{15}^\dagger)^{-1}\,M_{15}^{\da}\,Y_{15}^{-1}\right)\right]=0.
\end{equation}

The flavor structure of Fig.~\ref{fig:strong_cp_up_mass_correction}(c) takes the form
\begin{equation}
\Im\,\Tr\!\left[
Y_{15}(Y_{15}^*)^{-1}\,Q_{\nu}(k^2)\,Y_{15}^{\dagger}
\Bigl((Y_{15}^\dagger)^{-1}\,M_{15}^{\dagger}\,Y_{15}^{-1}\Bigr)
\right],
\end{equation}
which is actually zero. This can be understood after recasting this structure using Eqs.~\eqref{eq:PQR_nu_identities}--\eqref{eq:H-R-nu-def-0} as
\begin{equation}
\Im\,\Tr\!\left[
\Bigl(\tfrac{3}{4}\kappa_R^{2}\,Y_{15}^{T}Y_{15}^{*}-k^{2}\Bigr)^{-1}
\,M_{15}^{\dagger}\,R_{\nu}(k^2)\,M_{15}^{\dagger}
\right]=0,
\label{eqn:cp-fig5c}
\end{equation}
since $R_{\nu}(k^2)$ is Hermitian and
$\bigl(\tfrac{3}{4}\kappa_R^{2}\, Y_{15}^{T}Y_{15}^{*}-k^{2}\bigr)^{-1}$ is also Hermitian, implying that the trace is real.

For Figs.~\ref{fig:strong_cp_up_mass_correction}(d)--\ref{fig:strong_cp_up_mass_correction}(f), we read off the leptoquark gauge couplings in the rotated basis using Eqs.~\eqref{eq:X} and \eqref{eq:nu_rot_states}. The flavor structure of Fig.~\ref{fig:strong_cp_up_mass_correction}(d) is
\begin{equation}
\Im\,\Tr\!\left[
\rho\,Q_{\nu}(k^2)\,Y_{15}^{\dagger}
\Bigl((Y_{15}^\dagger)^{-1}\,M_{15}^{\dagger}\,Y_{15}^{-1}\Bigr)
\right].
\end{equation}
Using Eq.~\eqref{eq:rho_explicit} for the definition of \(\rho\) together with
Eqs.~\eqref{eq:PQR_nu_identities}--\eqref{eq:H-R-nu-def-0}, this becomes
\begin{equation}
\Im\,\Tr\!\left[
\frac{\kappa_L}{\kappa_R}\,
\Bigl(\tfrac{3}{4}\kappa_R^{2}\,Y_{15}^{T}Y_{15}^{*}-k^{2}\Bigr)^{-1}
\,M_{15}^{\dagger}\,R_{\nu}(k^2)\,M_{15}^{\dagger}
\right]=0,
\end{equation}
which follows by the same logic as Eq.~\eqref{eqn:cp-fig5c}.

For Fig.~\ref{fig:strong_cp_up_mass_correction}(e), the correction to the
$\delta M^{u}_{HL}$ block provides a vanishing contribution to $\bar\theta$, because its flavor structure reduces to
\begin{equation}
\Im\,\Tr\!\left[
R_{\nu}^{\dagger}(k^2)\,Y_{15}^{\dagger}(Y_{15}^{\dagger})^{-1}
\right]=0,
\end{equation}
since $R_{\nu}(k^2)$ is Hermitian.

The remaining Fig.~\ref{fig:strong_cp_up_mass_correction}(f) contributes to the $\delta M^{u}_{LH}$ block with
\begin{equation}
\Im\,\Tr\!\left[
\rho \Bigl(\mathbb{1}+k^{2}P^{\dagger}_{\nu}(k^2)\Bigr)(Y_{15}^T)^{-1}Y_{15}^{-1}
\right]
=
\Im\,\Tr\!\left[
(Y_{15}^*)^{-1}\Bigl(\mathbb{1}+k^{2}P^{\dagger}_{\nu}(k^2)\Bigr)(Y_{15}^T)^{-1}
\right]
=0,
\end{equation}
where we used Eq.~\eqref{eq:rho_explicit} for $\rho$ and $P_{\nu}(k^2)=P_{\nu}^\dagger(k^2)$. Thus, we have explicitly shown that, apart from the neutral-lepton diagram correcting the down-quark mass (Fig.~\ref{fig:strong_cp_down_mass_correction}(c)), all other new one-loop contributions to $\bar{\theta}$ correcting down- and up-quark mass vanish individually within our setup (at least up to \(\mathcal{O}(\kappa_L^2/\kappa_R^2)\) for up-quark).

\vspace{.2 cm}

\noindent \textbf{Estimation of \(\bar{\theta}\) arising from Fig.~\ref{fig:strong_cp_down_mass_correction}(c):}
In this part, we present approximate one-loop estimations for the induced \(\bar{\theta}\). Unlike the diagrams discussed above, whose individual contributions vanish, the neutral-lepton contribution correcting the down quark mass in Fig.~\ref{fig:strong_cp_down_mass_correction}(c) is CP-violating at one loop in general. We organize the discussion as follows: first, we show that the relevant trace is no longer taken over a completely Hermitian flavor structure. We then write an approximate expression for \(\bar{\theta}\) at leading order to show the parametric dependence, and determine what is needed for a suppressed \(\bar{\theta}\).

The general flavor structure of Fig.~\ref{fig:strong_cp_down_mass_correction}(c) reads as
\begin{equation}
\mathrm{Im\,Tr}\!\left[
Y_{15}\,\xi\,Y_{15}^{\dagger}
\left((Y_{10}^\dagger)^{-1}\,M_{10}\,Y_{10}^{-1}\right)
\right],
\label{eqn:d-N-strong-cp}
\end{equation}
where we defined \(\xi=(Y_{15}^*)^{-1}\,Q_{\nu}(k^2)\). If \(\xi\) were Hermitian, the quantity inside the trace in Eq.~\eqref{eqn:d-N-strong-cp} would be a product of two Hermitian matrices and therefore would not contribute to \(\bar{\theta}\). To check this, we can write \(\xi\) explicitly in terms of original matrices using Eqs.~\eqref{eq:PQR_nu_identities}--\eqref{eq:H-R-nu-def}:
\begin{align}
\xi \;=\; \frac{\sqrt{3}}{2}\,\kappa_R\Biggl(
&\,k^{4}\,(M_{15}^{\dagger})^{-1}- k^{2}M_{15}^{\dagger}
+\frac{9}{16}\,\kappa^{4}\,Y_{15}^{\dagger}Y_{15}\,M_{15}^{-1}\,Y_{15}^{T}Y_{15}^{*} \notag\\
&-\frac{3}{4}\,k^{2}\kappa_R^{2}\,\Bigl(
M_{15}^{-1}\,Y_{15}^{T}Y_{15}^{*}+Y_{15}^{\dagger}Y_{15}\,M_{15}^{-1}
\Bigr)
\Biggr)^{-1}.
\end{align}
We can see that \(\xi\)  is not Hermitian in general unless we impose some additional flavor structures. For example, if \(Y_{15}^\dagger Y_{15}\) is real for all elements, then \(\xi\) becomes Hermitian. The reason Eq.~\eqref{eqn:d-N-strong-cp} is not Hermitian in general is that the down-sector mass receives a correction from neutral leptons whose mass matrix is complex symmetric, and the resulting flavor structure involves two independent Yukawa couplings, \(Y_{15}\) and \(Y_{10}\). We will discuss two scenarios where this contribution can be suppressed with \(\mathcal{O}(1)\) CP-violating phases.
\vspace{0.2 cm}

\noindent \textbf{First Benchmark:} We now estimate the contribution of Fig.~\ref{fig:strong_cp_down_mass_correction}(c), which gives the leading nonzero correction to $\bar\theta$ in the down-quark sector. In this benchmark, we work in the limit
\begin{equation}
       M_{10}\ll \kappa_R Y_{15}\ll M_{15},
\label{eq:theta_limit}
\end{equation}
which is compatible with the charged fermion benchmark of Sec.~\ref{sec:fermion_mass_benchmark_Y15} together with the second benchmark for neutrino mass generation from Sec.~\ref{sec:sec4}. In the basis where $M_{15}$ is diagonal, the correction to the light down-quark mass matrix is
\begin{align}
\delta M^d_{\alpha\beta}
&=
(-2\lambda_5\kappa_L\kappa_R)
\left({1\over 2\sqrt{3}}\right)^4 \kappa_R^2
\sum_{i,j}
(Y_{15})_{\alpha j}
\Big[
M_{15,j}(Y_{15}^\dagger Y_{15})_{ji}
+
(Y_{15}^T Y_{15}^*)_{ji}M_{15,i}
\Big]
(Y_{15}^\dagger)_{i\beta}
I_{ij},
\label{eq:deltaMd_theta}
\end{align}
where
\begin{equation}
I_{ij}=
{1\over 16\pi^2}
F_4\!\left(M_{15,i}^2,M_{15,j}^2,
M_{\chi_R^d}^2,M_{\chi_L^d}^2\right),
\label{eq:Iij_theta}
\end{equation}
with the definition
\begin{align}
F_4(a,b,c,d)
&=
{a\ln a\over (a-b)(a-c)(a-d)}
+{b\ln b\over (b-a)(b-c)(b-d)}
\nonumber\\
&\quad
+{c\ln c\over (c-a)(c-b)(c-d)}
+{d\ln d\over (d-a)(d-b)(d-c)} .
\label{eq:F4_theta}
\end{align}
The two mass-insertion diagrams corresponding to Eq.~\eqref{eq:deltaMd_theta} are shown in Fig.~\ref{Fig:theta_benchmark}. The induced contribution to $\bar\theta$ then can be read from Eq.~\eqref{eq:btheta_master_d}
\begin{align}
\bar\theta
&=
{\lambda_5\over 72}\,\kappa_R^2\,
{\rm Im}
\sum_{\alpha,\beta}\sum_{i,j}
(Y_{15})_{\alpha j}
\Big[
M_{15,j}(Y_{15}^\dagger Y_{15})_{ji}
+
(Y_{15}^{T}Y_{15}^{*})_{ji}M_{15,i}
\Big]
(Y_{15}^\dagger)_{i\beta}
\nonumber\\
&\hspace{3.0cm}\times
I_{ij}
\left[
(Y_{10}^\dagger)^{-1}M_{10}Y_{10}^{-1}
\right]_{\beta\alpha}.
\label{eq:theta_from_deltaMd}
\end{align}
For the benchmark values used below, the heavy masses entering the loop are all of the same order, with $M_{15}$ somewhat heavier than the scalar masses. To show the main suppression factors, it is helpful to expand the loop
function in the limit
\(M_{15,i}^2,M_{15,j}^2\gg M_{\chi_R^d}^2,M_{\chi_L^d}^2\).  In this limit
\begin{equation}
I_{ij}\simeq {1\over 16\pi^2}\,
{\ell(i,j)\over M_{15,i}^2M_{15,j}^2},
\label{eq:Iij_approx}
\end{equation}
where
\begin{equation}
\ell(i,j)=
{M_{15,j}^2\ln(M_{15,i}^2/\mu^2)
-M_{15,i}^2\ln(M_{15,j}^2/\mu^2)
\over
M_{15,i}^2-M_{15,j}^2}
+
{M_{\chi_R^d}^2\ln(M_{\chi_R^d}^2/\mu^2)
-M_{\chi_L^d}^2\ln(M_{\chi_L^d}^2/\mu^2)
\over
M_{\chi_R^d}^2-M_{\chi_L^d}^2}.
\label{eq:ellij_def}
\end{equation}
The scale \(\mu\) cancels in the sum, so \(\ell(i,j)\) is a dimensionless
threshold factor. Substituting Eq.~\eqref{eq:Iij_approx} into
Eq.~\eqref{eq:theta_from_deltaMd}, and using an approximate basis where \(Y_{10}\) and \(M_{10}\) are diagonal, gives
\begin{align}
\bar\theta
&\simeq
{\lambda_5\over 72}\,{1\over 16\pi^2}\,
{\rm Im}\sum_{\alpha=d,s,b}\sum_{i,j}\ell(i,j)\,
(Y_{15})_{\alpha j}(Y_{15})^*_{\alpha i}
\nonumber\\
&\quad\times
\left[
\left({\kappa_R\over M_{15,i}}\right)^2
{M_{10,\alpha}\over M_{15,j}y_\alpha^2}
(Y_{15}^\dagger Y_{15})_{ji}
+
\left({\kappa_R\over M_{15,j}}\right)^2
{M_{10,\alpha}\over M_{15,i}y_\alpha^2}
(Y_{15}^T Y_{15}^*)_{ji}
\right].
\label{eq:theta_simplified_benchmark}
\end{align}
This expression is meant only to appreciate the loop suppression, the explicit \(M_{10}/M_{15}\) suppression, and the parametric dependence on \((\kappa_R/M_{15})^2\).  The numerical scan below is performed using the exact expression in Eq.~\eqref{eq:theta_from_deltaMd}.

For the numerical illustration, we take the same rank-one structure used in the charged-fermion benchmark of Sec.~\ref{sec:fermion_mass_benchmark_Y15},
\begin{equation}
Y_{15}^{(0)}
=
\begin{pmatrix}
0 & 0 & 0.068\\
0 & 0 & 0.481\\
0 & 0 & 1.434
\end{pmatrix}.
\label{eq:Y15_rank_one_benchmark}
\end{equation}
This texture is then perturbed in the first two columns as
\begin{equation}
(Y_{15})_{\alpha a}
=
(Y_{15}^{(0)})_{\alpha a}
+\epsilon R_{\alpha a},
\qquad a=1,2,
\qquad
(Y_{15})_{\alpha 3}
=
(Y_{15}^{(0)})_{\alpha 3},
\label{eq:Y15_perturbation}
\end{equation}
where the nonzero entries of $R$ are taken to be generic complex numbers of order one.  This introduces the generic CP phases while keeping the texture close to the fermion mass benchmark. For the numerical estimate we take
\[
{\rm Input:}\qquad
\kappa_R=5.0\times 10^{13}~{\rm GeV},\qquad
\lambda_5=0.125,
\]
\[
M_{\chi_R^d}=6.0\times 10^{13}~{\rm GeV},\qquad
M_{\chi_L^d}=7.0\times 10^{13}~{\rm GeV},
\]
\[
M_{15}={\rm diag}(5.0\times 10^{15},\,2.5\times 10^{15},\,2.0\times 10^{14})~{\rm GeV},
\qquad
M_{10}={\rm diag}(10,\,50,\,80)~{\rm TeV}.
\]
The entries of $Y_{10}$ are fixed by the running down-quark masses at $\mu_P$, listed in Table~\ref{tab:mf-muP}.  This is a simple representative choice, because in this benchmark the down-quark masses are mainly generated by $Y_{10}\kappa_L$, while the $Y_{15}$ dependent terms are providing perturbative corrections.

For $\epsilon=0.1$, scanning over the complex order-one entries of $R$ gives
\begin{equation}
|\bar\theta|\sim 10^{-12}-10^{-10}.
\label{eq:theta_benchmark_result}
\end{equation}
Thus the neutral lepton contribution can naturally lie close to the present
neutron EDM sensitivity. In this benchmark the result is especially sensitive
to the first-generation sextet mass, due to the enhancement by \((M_{10})_1/|y_d|^2\) in Eq.~\eqref{eq:theta_simplified_benchmark}. This implies that first-generation sextets much heavier than the \(\mathcal{O}(1\)--\(100)~{\rm TeV}\) range can push \(\bar\theta\) above the present neutron EDM limit, while the heavier generation entries are less constraining.

We have used \(\lambda_5=0.125\), as in the charged-fermion benchmark of
Sec.~\ref{sec:fermion_mass_benchmark_Y15}. But smaller values of \(\lambda_5\) are also consistent and would
weaken the corresponding limit on the sextet masses. It is therefore useful to
estimate the radiatively induced size of this coupling. The box diagram shown
in Fig.~\ref{Fig:induced_lambda5} gives,
\begin{equation}
\lambda_{5}\sim \frac{3}{16\pi^{2}}\,
\Tr\!\left(Y_{10}^{\dagger}Y_{10}Y_{15}^{\dagger}Y_{15}\right)\,
\ln\!\left(\frac{M_{15_3}^2}{M_{\rm Pl}^2}\right)
\;\sim\;4.4\times 10^{-5},
\end{equation}
where we have used \(Y_{10}\sim 10^{-2}\) (effective bottom Yukawa), \(Y_{15}\sim 1\) (top Yukawa),
\(M_{15_3}\sim 10^{14}\,\mathrm{GeV}\), and \(M_{\rm Pl}\sim 10^{19}\,\mathrm{GeV}\).
\vspace{0.2 cm}

\noindent \textbf{Second Benchmark:}
We now estimate the contribution to $\bar\theta$ in the benchmark where the charged fermion masses are generated through the $Y_{10}$-induced radiative correction discussed in Sec.~\ref{sec:Y10 induced corrections}.  In this benchmark, we work in the limit where
\begin{equation}
        M_{15}\ll \kappa_R ,
\end{equation}
so, the up-quark masses are generated directly from the $Y_{15}$ coupling, while the down-quark and charged-lepton masses arise from the induced $Y_{10}$ contribution.  This is also compatible with the first benchmark of neutrino mass generation discussed in Sec.~\ref{sec:sec4}.  An important consequence is that the one-loop contribution to $\bar\theta$ is parametrically suppressed by the insertion of $M_{15}$. If we work in the basis where $Y_{15}$ is real and diagonal, we can write
\begin{equation}
        Y_{15} =
        {\rm diag}(y_u,y_c,y_t),
        \qquad
        D_i = y_i \kappa_R,
\end{equation}
and the tree-level down-quark mass matrix is given by Eq.~\eqref{eq:uni2}. The leading contribution to $\bar\theta$ from the diagrams of Fig.~\ref{Fig:theta_benchmark} is then
\begin{align}
\bar\theta
&=
2\lambda_5
\left({1\over 2\sqrt{3}}\right)^4
{\rm Im}
\Bigg[
\sum_{\alpha,\beta,i,j}
\left[
(Y_{10}^\dagger)^{-1}M_{10}Y_{10}^{-1}
\right]_{\beta\alpha}
(Y_{15})_{\alpha i}
\nonumber\\
&\hspace{1.3cm}\times
\Bigg\{
(M_{15})_{ij}D_j^2
\left[
J_{ij}^{(2)}
+3D_j(D_i+D_j)J_{ij}^{(1)}
+D_iD_j^3J_{ij}^{(0)}
\right]
\nonumber\\
&\hspace{1.6cm}
+
(M_{15})_{ji}D_i^2
\left[
J_{ji}^{(2)}
+3D_i(D_j+D_i)J_{ji}^{(1)}
+D_jD_i^3J_{ji}^{(0)}
\right]
\Bigg\}
(Y_{15}^{\dagger})_{j\beta}
\Bigg] .
\label{eq:theta_benchmark2_exact}
\end{align}
Here the loop functions can be evaluated by using the following definitions
\begin{equation}
J_{ij}^{(n)}
=
{ 1\over 16\pi^2}\,
{1\over 2}
{\partial^2\over \partial(D_j^2)^2}
\mathcal{F}_n(M_{\chi_R}^2,M_{\chi_L}^2,D_i^2,D_j^2),
\end{equation}
with
\begin{equation}
\mathcal{F}_n(a,b,c,d)
=
\sum_{x=a,b,c,d}
{x^{n+1}\ln x\over \prod_{y\neq x}(x-y)} .
\end{equation}
For the numerical estimate, we fix \(Y_{15}\) by the running up-quark masses at the PS scale, listed in Table~\ref{tab:mf-muP}, through
\begin{equation}
Y_{15}={\rm diag}\left(
{m_u\over \kappa_L},
{m_c\over \kappa_L},
{m_t\over \kappa_L}
\right)
=
{\rm diag}\left(
3.48\times 10^{-6},
1.67\times 10^{-3},
0.489
\right),
\label{eq:Y15_second_benchmark}
\end{equation}
which gives
\begin{equation}
D_i
=
\left(
1.74\times 10^8,\,
8.33\times 10^{10},\,
2.44\times 10^{13}
\right)\ {\rm GeV}.
\label{eq:Dvalues_second_benchmark}
\end{equation}
For the numerical illustration, we have used, 
\[
{\rm Input:}\qquad
\lambda_5=3, \qquad \kappa_L= 174~{\rm GeV},\qquad\kappa_R=5.0\times 10^{13}~{\rm GeV},\qquad
\]
\[
M_{\chi_R^d}=6.0\times 10^{13}~{\rm GeV},\qquad
M_{\chi_L^d}=7.0\times 10^{13}~{\rm GeV},
\]
We choose
\begin{equation}
Y_{10}
=
{\rm diag}(0.05,\,0.15,\,1.434),
\label{eq:Y10_second_benchmark}
\end{equation}
and fix \(M_{10}\) from Eq.~\eqref{eq:uni2}, so that the tree-level down-quark masses agree with the running values in Table~\ref{tab:mf-muP}. As a check, we take a generic complex symmetric \(M_{15}\) with entries in the few hundred TeV range,
\begin{equation}
M_{15}
=
\begin{pmatrix}
1.0\times 10^3
&
2.0\times 10^5 e^{0.73 i}
&
6.0\times 10^5 e^{1.20 i}
\\
2.0\times 10^5 e^{0.73 i}
&
3.0\times 10^5 e^{0.31 i}
&
5.0\times 10^3 e^{1.85 i}
\\
6.0\times 10^5 e^{1.20 i}
&
5.0\times 10^3 e^{1.85 i}
&
1.0\times 10^5 e^{0.57 i}
\end{pmatrix}
{\rm GeV}.
\label{eq:M15_second_benchmark}
\end{equation}
For this point we get
\begin{equation}
\bar\theta=5.1\times 10^{-13}.
\label{eq:theta_second_result}
\end{equation}
We want to clarify that this is not any special alignment choice. Rather, it is a generic situation once we work in the basis where $Y_{15}$ is real and diagonal. As a confirmation, we allow matrix $M_{15}$ to be a generic complex matrix with entries in the range
\begin{equation}
        |(M_{15})_{ij}| \sim 1-10^3~{\rm TeV},
\end{equation}
and do a random scan over generic complex $M_{15}$ matrices which gives typical values
\begin{equation}
        \bar\theta \sim 10^{-14}-10^{-11},
\end{equation}
with the specific value depending on the phases and on the overall size of the entries of $M_{15}$. Thus this benchmark naturally gives a suppressed but potentially observable contribution to $\bar\theta$. Compared with the first benchmark, the contribution contains the small insertion $M_{15}/\kappa_R$ which is the main reason for this parametric suppression, while the fermion mass fit itself remains controlled by the $Y_{10}$-induced radiative correction.

We can summarize the results from the two benchmark scenarios as follows. Realistic charged-fermion masses can be realized once the radiative corrections are included, if at least one of the vector-like mass scales, \(M_{10}\) or \(M_{15}\), lies near the PS-breaking scale. However, the strong CP constraint then correlates this choice with the mass scale of the other multiplet. For example, if the fermion masses are dominantly generated through the \(M_{15}\) sector, the neutral-lepton correction to \(M_d\), shown in Fig.~\ref{fig:strong_cp_down_mass_correction}(c), prefers the \(M_{10}\) to remain comparatively light. On the other hand, if the \(M_{10}\) sector plays the dominant role, the same argument favors a relatively light \(M_{15}\) sector. Quantitatively, the non-dominant multiplet is typically preferred to lie below about \(
\lesssim \mathcal{O}(10^5)\ {\rm GeV},
\) up to order-one flavor factors and phases with general alignments.

We find it as an interesting complementarity between collider searches and the neutron EDM. If the color sextets or octets are not found in the multi-TeV region, the model is pushed towards a region where the induced contribution to \(\bar\theta\) can be close to the present neutron EDM sensitivity. On the other hand, non-observation of the neutron EDM would favor these colored states being relatively light, and therefore potentially accessible at future collider searches. Thus the radiative origin of fermion masses connects the strong CP constraint directly to the spectrum of colored vector-like multiplets.

\subsubsection{Corrections to the sextet and octet fermion mass matrices}
\begin{figure}[t!]
    \centering
    \begin{subfigure}[b]{0.325\textwidth}
        \centering
        \includegraphics[width=\textwidth]{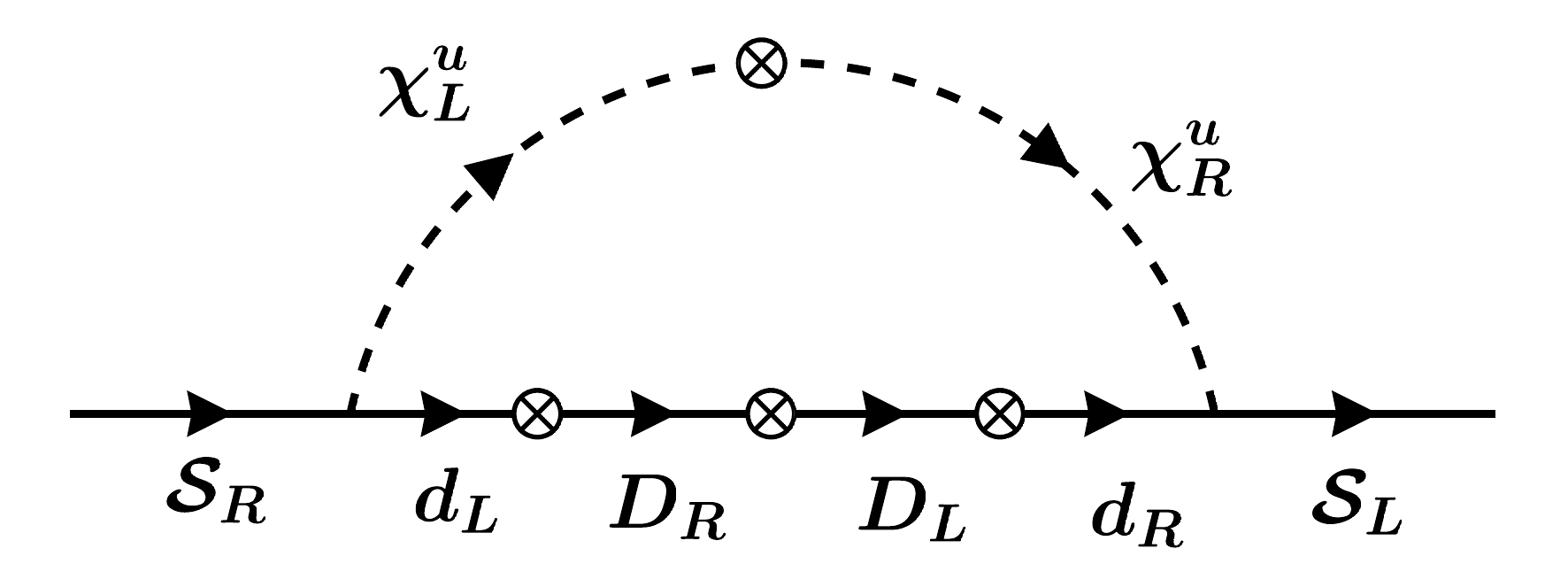}
        \caption{}
    \end{subfigure}
    \begin{subfigure}[b]{0.325\textwidth}
        \centering
        \includegraphics[width=\textwidth]{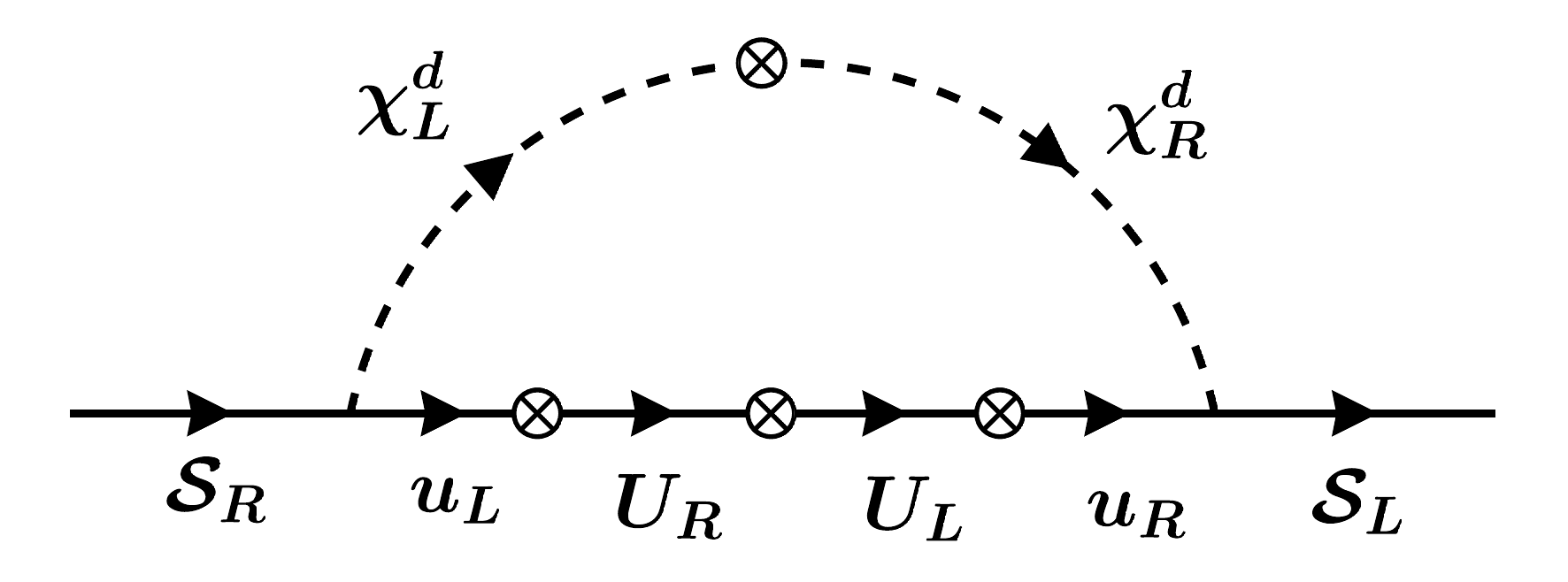}
        \caption{}
    \end{subfigure}
    \begin{subfigure}[b]{0.325\textwidth}
        \centering
        \includegraphics[width=\textwidth]{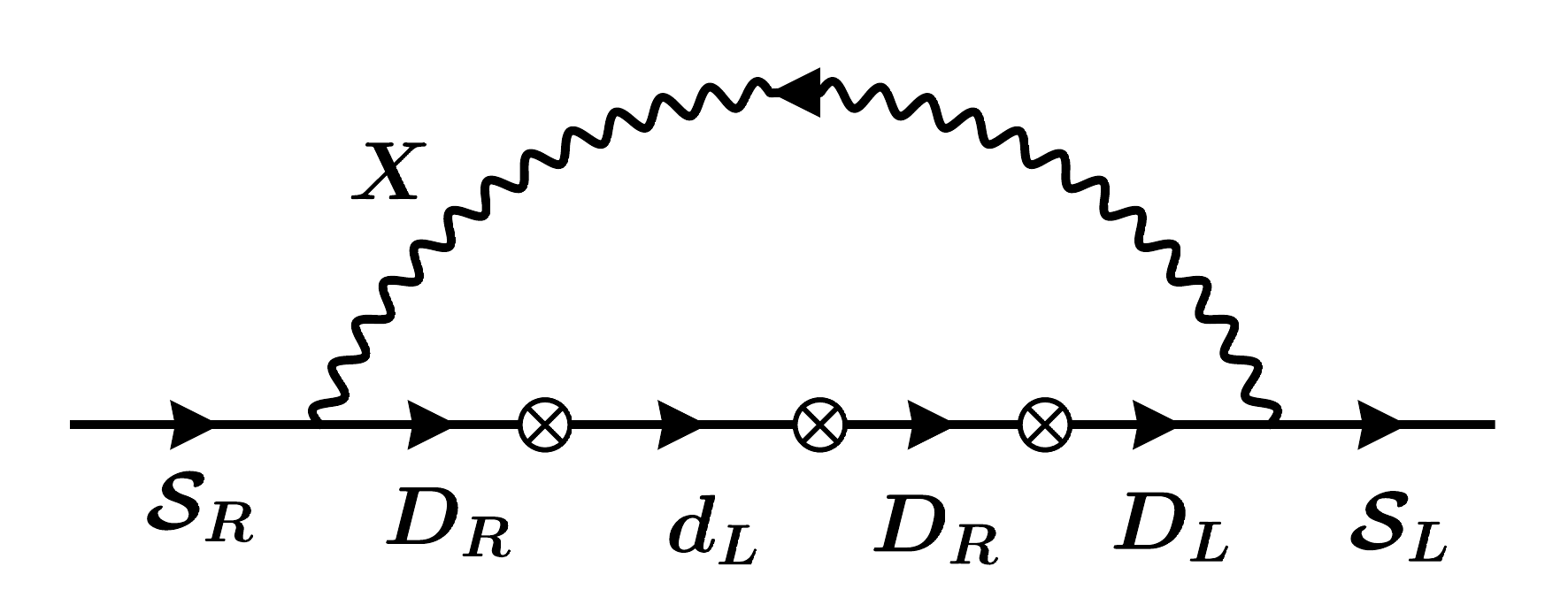}
        \caption{}
    \end{subfigure}

    \vspace{0.4cm} 

    \begin{subfigure}[b]{0.325\textwidth}
        \centering
        \includegraphics[width=\textwidth]{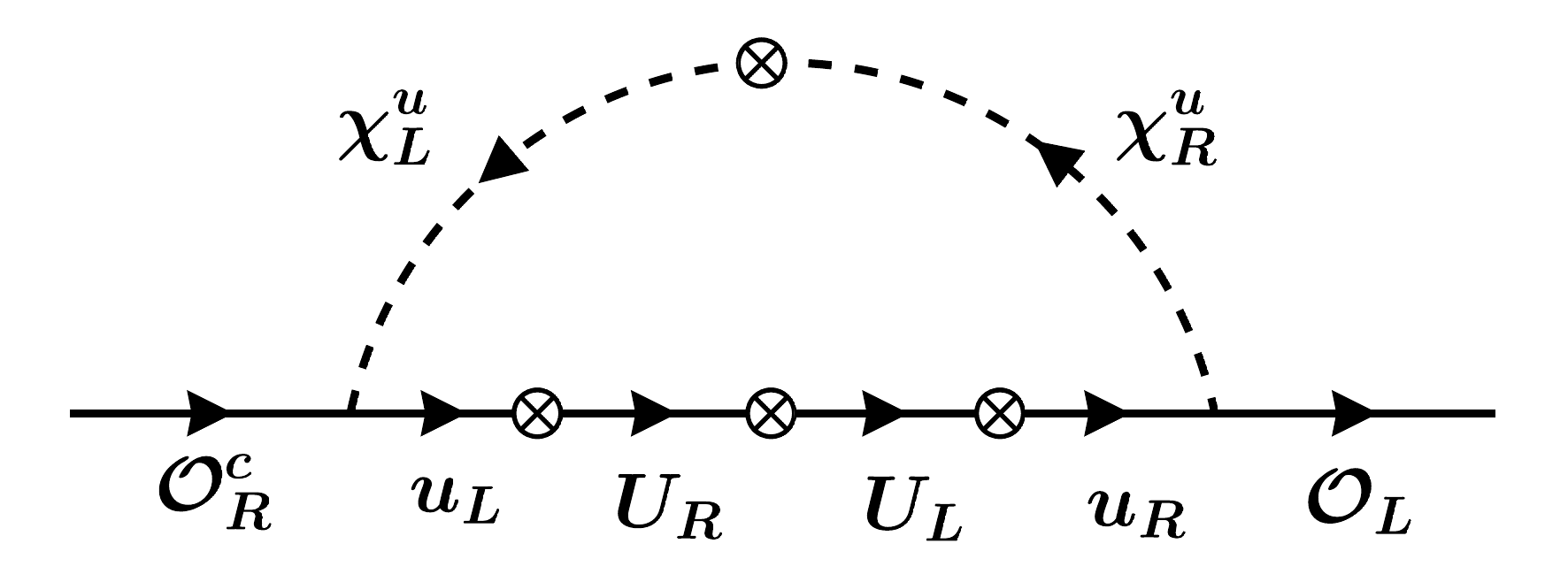}
        \caption{}
    \end{subfigure}
    \begin{subfigure}[b]{0.325\textwidth}
        \centering
        \includegraphics[width=\textwidth]{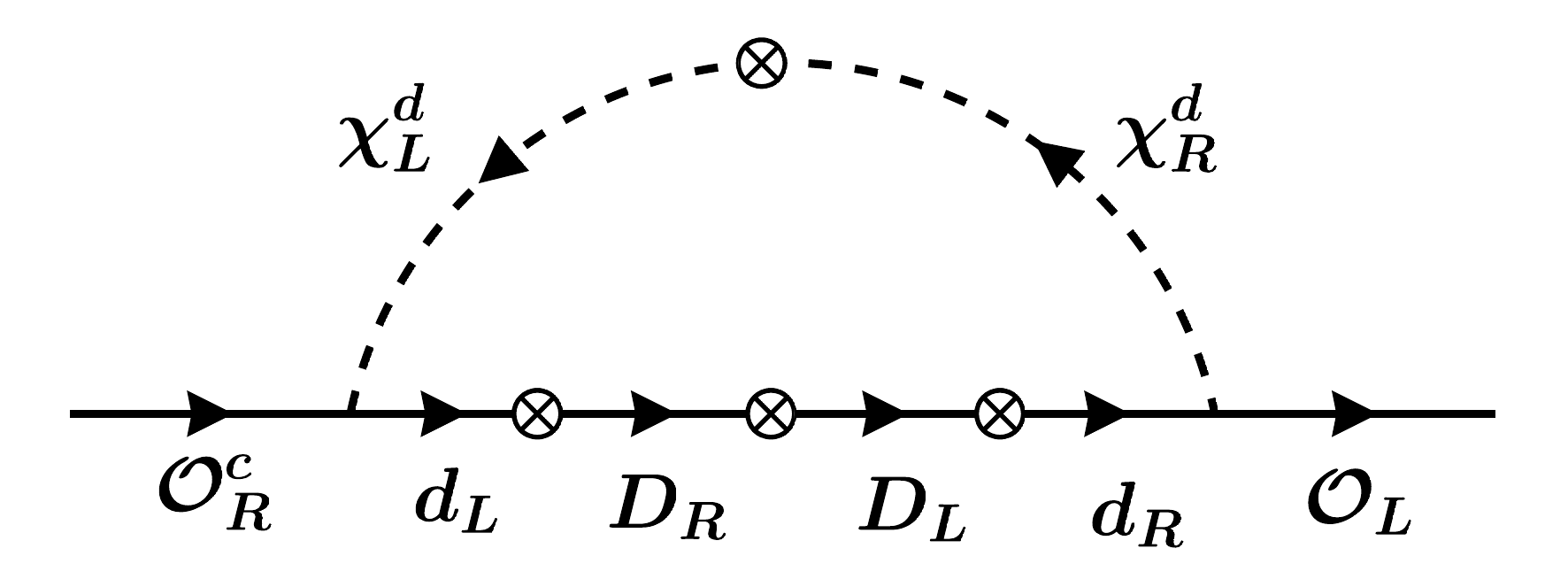}
        \caption{}
    \end{subfigure}
    \begin{subfigure}[b]{0.325\textwidth}
        \centering
        \includegraphics[width=\textwidth]{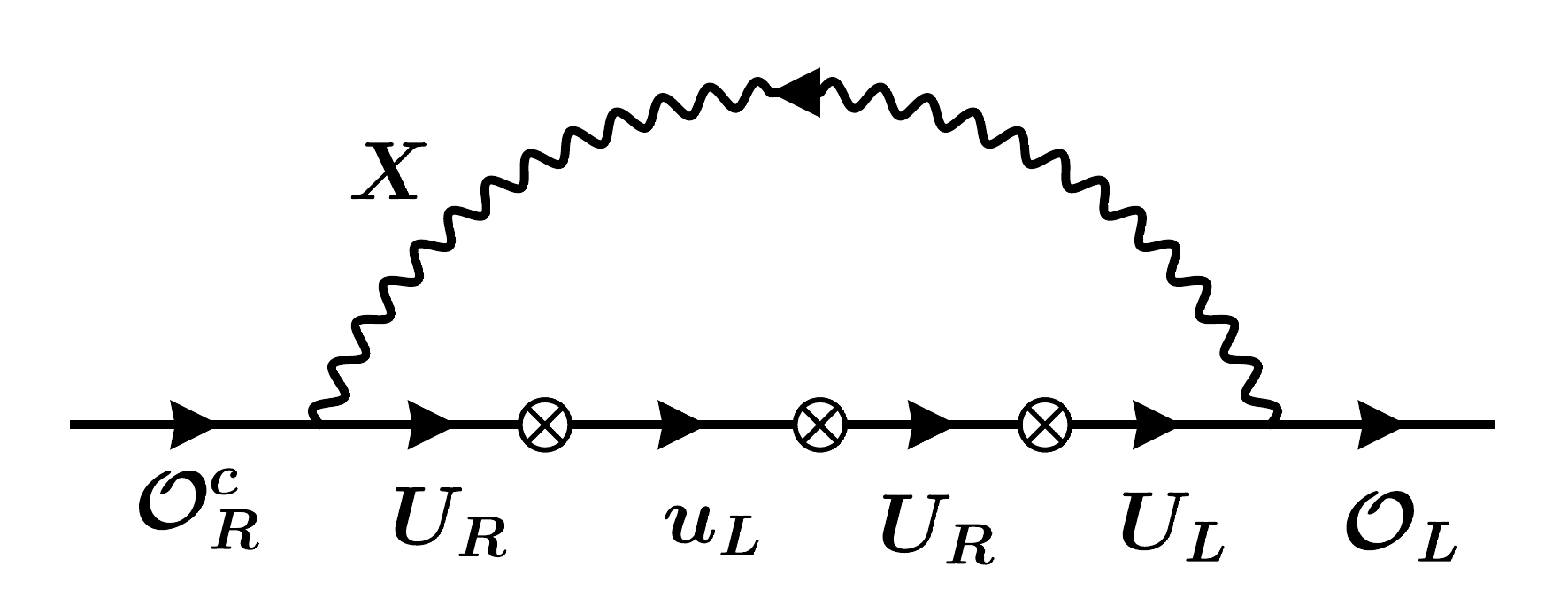}
        \caption{}
    \end{subfigure}
    \caption{One-loop radiative diagrams contributing to the sextet and octet mass matrix through leptoquark gauge bosons and leptoquark scalars.} 
    \label{fig:strong_cp_sextet/octet_mass_correction}
\end{figure}

Here, we evaluate the \(\bar{\theta}\) arising from the corrections to the sextet and octet mass matrices:
\begin{equation}
\bar\theta^{(1)}_{(\mathcal{S}+\mathcal{O})}
=\Im\Tr\!\left[{5\,\cal M}_{\mathcal{S}}^{-1}\delta{\cal M}_{\mathcal{S}}+3\,{\cal M}_{\mathcal{O}}^{-1}\delta{\cal M}_{\mathcal{O}}\right],
\label{eq:btheta_1loop_sextet-octet}
\end{equation}
where the bare masses \(\mathcal M_{S}\) and \(\mathcal {M_{O}}\) have been defined in Eq.~\eqref{eqn:sextet-octet-mass}. The relevant one-loop diagrams are shown in Fig.~\ref{fig:strong_cp_sextet/octet_mass_correction}. We will focus on the sextet-mass corrections shown in Fig.~\ref{fig:strong_cp_sextet/octet_mass_correction}(a)--(c). The corresponding octet-mass corrections have the same flavor structure, so they follow directly from the sextet discussion, and we will not repeat them.

Let's consider Fig.~\ref{fig:strong_cp_sextet/octet_mass_correction}(a) first. The associated flavor structure is
\begin{equation}
5\,k^2\,\kappa_R\Im\Tr\!\left[
M_{10}^{-1}\zeta_a\right].
\label{eqn:Fig-6(a)}
\end{equation}
where \(\zeta_a=\,\left(
Y_{10}^{\dagger}\,Y_{10}\,R_d^{\dagger}\,M_{10}\,Y_{10}^{\dagger}\,H_d\,Y_{10}
\right)\). We have read the insertion on the down-type fermion line in direct analogy with the electron-type correction in Eq.~\eqref{eq:E_prop_correction}. Using identities similar to the down-type quarks as in Eqs.~\eqref{eq:PQR_nu_identities}--\eqref{eq:H-R-nu-def},  \(\zeta_a\) can be expressed as 
\begin{align}
   \zeta_a=-\kappa_L&\left(k^4Y_{10}^{-1}(Y_{10}^{\dagger})^{-1}M_{10}^{-1}Y_{10}^{-1}(Y_{10}^{\dagger})^{-1}-k^2Y_{10}^{-1}(Y_{10}^{\dagger})^{-1}M_{10}^{\dagger}Y_{10}^{-1}(Y_{10}^{\dagger})^{-1}+\kappa_L^2\kappa_R^2M_{10}^{-1}\right.\notag\\
   &\left. -k^2\left(\kappa_L^2M_{10}^{-1}Y_{10}^{-1}(Y_{10}^{\dagger})^{-1}+\kappa_R^2Y_{10}^{-1}(Y_{10}^{\dagger})^{-1}M_{10}^{-1}\right)\right)^{-1}
   \label{eqn:zeta}
\end{align}
If \(\zeta_a\) were Hermitian, then \(\Tr(M_{10}^{-1}\zeta_a)\) would be real and Eq.~\eqref{eqn:Fig-6(a)} would not contribute to \(\bar{\theta}\). But we see the non-Hermiticity appearing in the last term of Eq.~\eqref{eqn:zeta}.  It is useful to estimate the size of \(\bar{\theta}\) using Eq.~\eqref{eq:btheta_1loop_sextet-octet} to get an idea of the numeric:
\begin{equation}
\bar{\theta}_{\mathcal{S}(a)}^1 \sim \frac{5\lambda_4}{8\pi^2}\,
Y_{10}^{\dagger}\left(\frac{\kappa_L^2}{M_{10_3}^{2}-M_{\chi_1^u}^{2}}\,
\ln\!\left(\frac{M_{10_3}^{2}}{M_{\chi_1^u}^{2}}\right)-\frac{\kappa_L^2}{M_{10_3}^{2}-M_{X}^{2}}\,
\ln\!\left(\frac{M_{10_3}^{2}}{M_{X}^{2}}\right)\right)\,Y_{10}\,.
\end{equation}
We see a huge suppression of \(\mathcal{O}(\kappa_L^2/\kappa_R^2)\) for \(\bar{\theta}\) which is well below the experimental limit. 

Now let's move to Fig.~\ref{fig:strong_cp_sextet/octet_mass_correction}(b). We take the same approach. We first check the Hermiticity of the flavor structure, 
\begin{equation}
5\,k^2\,\kappa_R\Im\Tr\!\left[
M_{10}^{-1} Y_{10}^{\dagger}\zeta_bY_{10}\right].
\label{eqn:Fig-6(b)}
\end{equation}
where we can explicitly write down \(\zeta_b\) as
\begin{align}
   \zeta_b=-\kappa_L\left(k^4(Y_{15}^{\dagger})^{-1}(M_{15}^{\dagger})^{-1}Y_{15}^{-1}-k^2(Y_{15}^{\dagger})^{-1}M_{15}Y_{15}^{-1}+\kappa_L^2\kappa_R^2Y_{15}(M_{15}^{\dagger})^{-1}Y_{15}^{\dagger}\right.\notag\\
   \left.-k^2\left(\kappa_L^2Y_{15}(M_{15}^{\dagger})^{-1}Y_{15}^{-1}+\kappa_R^2(Y_{15}^{\dagger})^{-1}(M_{15}^{\dagger})^{-1}Y_{15}^{\dagger}\right)\right)^{-1}~.
   \label{eqn:zeta-b}
\end{align}
And again, we observe a similar pattern of non-Hermiticity. We proceed to get a rough estimation for \(\bar{\theta}\), given as
\begin{equation}
\bar{\theta}_{\mathcal{S}(b)}^1 \sim \frac{5\lambda_5\kappa_L^2}{8\pi^2M_{10}}\,
Y_{10}^{\dagger}\left(\frac{M_{15}}{M_{15_3}^{2}-M_{\chi_1^d}^{2}}\,
\ln\!\left(\frac{M_{15_3}^{2}}{M_{\chi_1^d}^{2}}\right)-\frac{M_{15}}{M_{15_3}^{2}-M_{\chi_2^d}^{2}}\,
\ln\!\left(\frac{M_{15_3}^{2}}{M_{\chi_2^d}^{2}}\right)\right)\,Y_{10}.
\end{equation}
which is also highly suppressed. 

Now, we move to the gauge boson contribution in Fig.~\ref{fig:strong_cp_sextet/octet_mass_correction}(c). The relevant flavor structure is
\begin{equation}
\frac{5\,k^4}{\kappa_L}\Im\Tr\!\left[
M_{10}^{-1}\zeta_c\right].
\label{eqn:Fig-6(c)}
\end{equation}
where \(\zeta_c\) can be written explicitly as the following:
\begin{equation}
\begin{aligned}
\zeta_c
&=-\kappa_L\Bigl(
k^4 M_{10}^{-1}-k^2 M_{10}^{\dagger}
+\kappa_L^2\kappa_R^2\,Y_{10}^{\dagger}Y_{10}\,M_{10}^{-1}\,Y_{10}^{\dagger}Y_{10} \\
&\hspace{2.0cm}
-k^2\Bigl(\kappa_L^2\,Y_{10}^{\dagger}Y_{10}\,M_{10}^{-1}
+\kappa_R^2\,M_{10}^{-1}\,Y_{10}^{\dagger}Y_{10}\Bigr)
\Bigr)^{-1}.
\end{aligned}
\label{eqn:zeta-c}
\end{equation}
One comment about diagram~\ref{fig:strong_cp_sextet/octet_mass_correction}(c). The sextet can receive radiative corrections only through insertions of \(M_{10}\), or through insertions involving \(M_{10}\) together with \(\kappa_R\) before electroweak symmetry breaking (\(\kappa_L=0\)). However, from Eq.~\eqref{eqn:zeta-c} we see that setting \(\kappa_L=0\) does not induce any CP violation through this diagram, in this limit \(\dfrac{1}{\kappa_L}\Tr\!\left[\zeta_c\, M_{10}^{-1}\right]\) is real. So, CP violation can arise only after \(\langle\chi_L^\nu\rangle\neq 0\), i.e. when \(\kappa_L\) is turned on. This is why we have shown the leading representative diagram involving the insertions of \(\kappa_L\) in Fig.~\ref{fig:strong_cp_sextet/octet_mass_correction}(c). A rough estimate for \(\bar{\theta}\) from Fig.~\ref{fig:strong_cp_sextet/octet_mass_correction}(c) can be written as   
\begin{equation}
\bar{\theta}_{\mathcal{S}(c)}^{\,1} \sim \frac{15\,g_4^2}{32\pi^2}\,
Y_{10}^{\dagger}\,
\frac{\kappa_L^2}{M_{10_3}^{2}-M_{X}^{2}}\,
\ln\!\left(\frac{M_{10_3}^{2}}{M_X^{2}}\right)\,
Y_{10}\,,
\end{equation}
where the expected suppression \(\mathcal{O}(\kappa_L^2/\kappa_R^2)\) is manifest.

The same analysis holds for the octet-mass corrections. One can simply replace
\(Y_{10}\to Y_{15}\) and \(M_{10}\to M_{15}^{\dagger}\), and use the corresponding couplings from Eqs.~\eqref{eq:Yukexpand} and~\eqref{eq:X}. The takeaway message is the same: all diagrams contribute to
\(\bar{\theta}\) with an overall \(\mathcal{O}(\kappa_L^2/\kappa_R^2)\) suppression.

We conclude this section by noting that the new one-loop diagrams contributing to $\overline{\theta}$ in the PS model are either suppressed by a factor of $(\kappa_L/\kappa_R)^2 \sim 10^{-23}$, or by a factor $(M_{10}/\kappa_R)$ or $(M_{15}/\kappa_R)$. By choosing either $M_{10}$ or $M_{15}$ to be well below the PS breaking scale of $5 \times 10^{13}$ GeV, the model provides an excellent solution to the strong CP problem based on parity symmetry.

\section{Other aspects of the model}

In this section, we comment on some other aspects of the model.

\subsection{Exact parity symmetric version}
\label{sub:exact parity}
In the Higgs potential of Eq. (\ref{eq:Higgspot}), we allowed for the soft breaking of parity symmetry by the dimension-two scalar mass terms with $\mu_L^2 \neq \mu_R^2$.  A fully realistic model can arise if we set $\mu_L^2 = \mu_R^2$, in which case parity is only spontaneously broken~\cite{Hall:2018let}. This scenario requires the parity symmetry breaking scale to coincide with the scale $\mu_c$ where the SM Higgs quartic coupling becomes zero when extrapolated to high energies.  This occurs around $\mu_c \sim 10^{11}$ GeV. As emphasized in Ref.~\cite{Hall:2018let}, such a version of the model has the attractive feature that, despite having multiple scales and Higgs bosons, it does not introduce an additional independent Higgs-sector tuning beyond the usual electroweak tuning. More precisely, even in the absence of additional particle physics thresholds, the SM is expected to be embedded in a theory including gravity, with a high cutoff such as the Planck scale. Keeping the weak scale much smaller than this cutoff already requires the usual tuning of the Higgs mass parameter. In the exact parity-symmetric version of the present model, \(H_L\) and \(H_R\) share a single parity-symmetric quadratic mass parameter. The hierarchy \(\kappa_L\ll \kappa_R\) still requires a tuning, but this tuning is tied to the same Higgs sector tuning rather than being an additional independent one. Schematically, this gives
\begin{equation}
    \frac{\kappa_R^2}{\Lambda^2}\times
    \frac{\kappa_L^2}{\kappa_R^2}
    = \frac{\kappa_L^2}{\Lambda^2}.
\end{equation}
Thus the total tuning is of the same order as the usual SM electroweak tuning with
cutoff \(\Lambda\).

The same setup also explains why the parity-breaking scale is tied to \(\mu_c\). In the exact parity limit, the potential has an approximate enlarged symmetry near the parity scale, and after \(H_R\) gets the large VEV, the SM Higgs doublet appears as the pseudo-Goldstone direction. This gives the matching condition \(\lambda_{\rm SM}(\mu_P)\simeq 0\), up to threshold corrections.

How does this critical value $\mu_c$ compare with the parity restoration scale $\mu_P = 5.2 \times 10^{13}$ GeV inferred from the equality $g_{2L} = g_{2R}$?  The small difference in the two scales $\mu_c $ and $\mu_P$  can be explained by threshold corrections near the PS scale.  The SM Higgs doublet has couplings with the physical scalars of the full theory that have masses of order the PS scale.  These include $\chi^u$, $\chi^d_{1,2}$ and $\sigma_R$. Such cross-couplings can increase the value of $\mu_c$ from its SM value by a factor of $(10-100)$, provided that some of the masses of these heavy fields are slightly below the PS scale.  Examining the scalar mass spectrum specified in Sec.~\ref{sec:sec3}, we see no contradiction with such a choice. Thus, the model proposed here can be realized with exact parity symmetry.

\subsection{Baryon number violation}
\label{Baryon number violation}
\begin{figure}[h!]
\begin{center}
\includegraphics[width=0.45\textwidth]{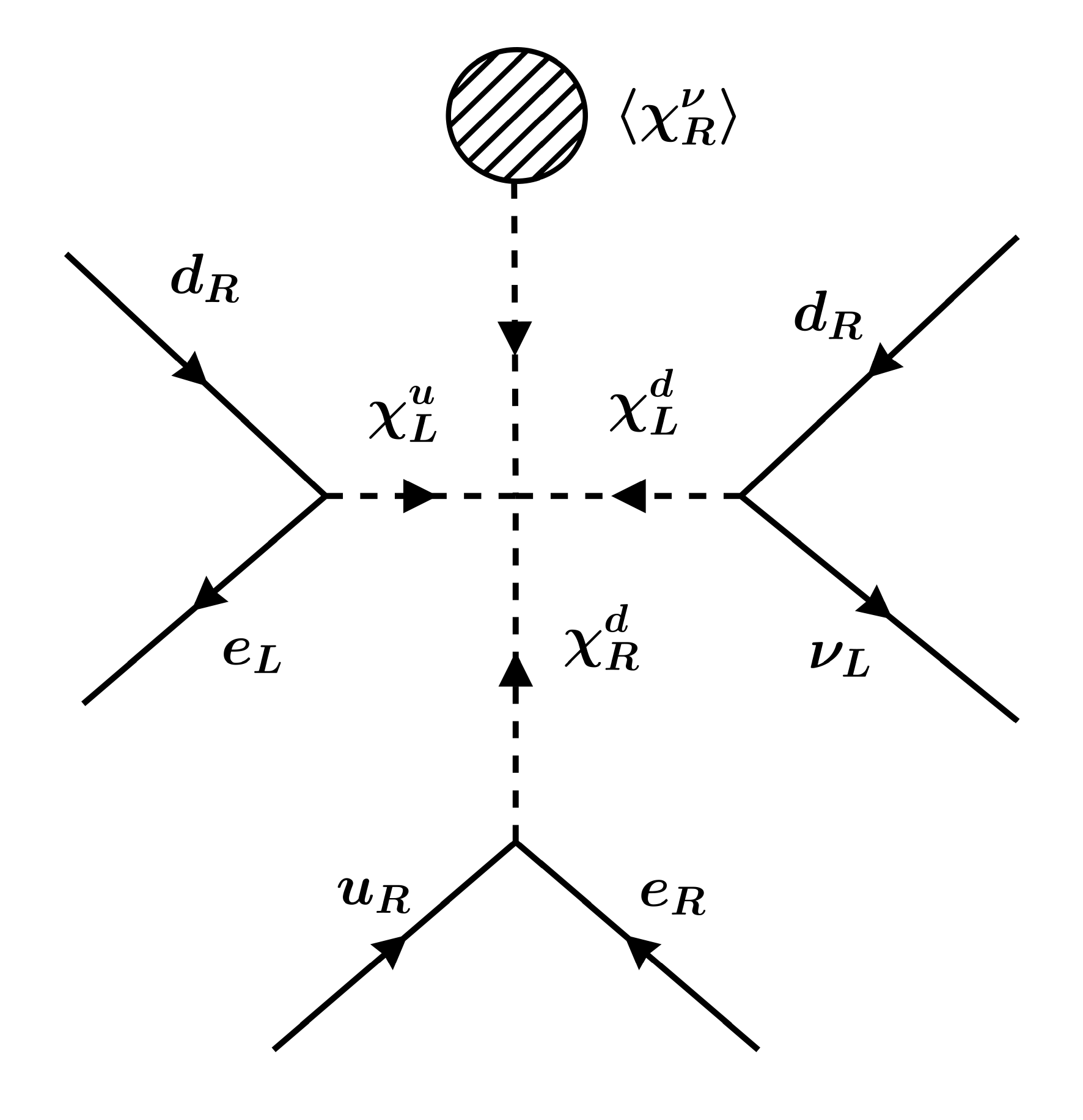}
\end{center}
\caption{Feynman diagram for baryon–number–violating nucleon decay induced by the $\lambda_6$ term in the Higgs potential and mediated by $\chi^{u,d}_{L,R}$ leptoquarks.}
\label{fig:fig4}
\end{figure}
There is no gauge boson-induced baryon number violation in Pati-Salam models.  This can be seen by assigning $B$ of $+2/3$ to the PS gauge boson $X_\mu(3,1,\frac{2}{3})$ with all color-triplet quarks and leptoquarks carrying baryon number of $+1/3$.  All color singlet fields carry zero $B$. While the gauge and Yukawa interactions are seen to conserve $B$, the scalar sector exhibits $B$ violation through the $\lambda_6$ term in the Higgs potential of Eq. (\ref{eq:Higgspot}). This term, when expanded, reads as 
\begin{equation}
V(\lambda_6) = 4 \lambda_6 \left(\chi^{u\alpha} _L \chi^{d\beta} _L \chi^{d\gamma}_R \chi^\nu_R \epsilon_{\alpha\beta\gamma} + \chi^{d\alpha}_L \chi^{u\beta}_R \chi^{d\gamma}_R \chi_L^\nu\epsilon_{\alpha\beta\gamma} \right) + {\rm h.c.}
\label{eq:lambda6}
\end{equation}
Here we have set $\chi_{L,R}^e = 0$, which are the Goldstone bosons.  We see that Eq. (\ref{eq:lambda6}) breaks baryon number by one unit.  This, in turn, leads to nucleon decay into leptons with or without additional meson final states, which we discuss now.

When the couplings of Eq.~ (\ref{eq:lambda6}) are inserted into the Yukawa couplings of Eq. (\ref{eq:Yukexpand}), along with the VEVs of $\chi_L^\nu$ and $\chi_R^\nu$, and the mixing between ordinary fermions with exotic fermions are included, nucleon decay will arise through diagrams such as the one shown in Fig.~\ref{fig:fig4}. This diagram arises from the first term of Eq. (\ref{eq:lambda6}). There are other diagrams arising from the second term of Eq. (\ref{eq:lambda6}); however, we find them to be suppressed by an additional factor of $(\kappa_L/\kappa_R)^3$.  This suppression factor arises since one of the light-heavy quark mixing angles involves left-handed quarks,  one needs to use the VEV of $\chi_L^\nu$, and since the physical non-Goldstone component of $\chi_R^u$ is proportional to $\kappa_L/\kappa_R$. Hence, we focus on the diagram shown in Fig.~\ref{fig:fig4}. The dominant nucleon decay has an amplitude given by
\begin{equation}
A[(\overline{\nu}_L d_R)(\overline{e}_L d_R) (\overline{e^c}_L u_R)] = \frac{4 \lambda_6\, \kappa_R}{M_{\chi_L}^4 M^2_{\chi^d_{_R}}} \left(\frac{Y_{10}^2\, \kappa_R}{M_{10}} \right)^2 \left(\frac{Y_{15}^2 \,\kappa_R}{M_{15}}  \right)~.
\label{eq:pdk}
\end{equation}
Here, the couplings $Y_{10}$ and $Y_{15}$ refer to the first-generation Yukawa couplings, and the masses $M_{10}$ and $M_{15}$ stand for the masses of the first-generation vector-like fermions. The operator of Eq. (\ref{eq:pdk}) will lead to the decays
\begin{eqnarray}
n &\rightarrow& e^+ e^- \nu \nonumber \\
p &\rightarrow& \pi^+\pi^0 e^+ e^- \nu~.
\end{eqnarray}
These decays would violate $(B-L)$ by two units. Such decays have been suggested in the context of Pati-Salam models with somewhat different Higgs sectors in Ref.~\cite{Pati:1983jk,Pati:1983zp,ODonnell:1993kdg}. The operators of Eq. (\ref{eq:pdk}) are of dimension-nine, and consequently the lifetime of the nucleon arising from them are rather long.  We estimate the lifetime to be of order $10^{126}$ yrs, if the PS breaking scale is identified as the parity restoration scale of $\mu_P \sim 5.2 \times 10^{13}$ GeV.  In this estimate, we assumed that the Yukawa couplings are all of order unity, the scalar masses are at the PS scale, and the vector-like fermion masses $M_{10}$ and $M_{15}$ are $10^3 \kappa_R$, a choice dictated by the fermion masses of the first family. We note, however, that if the PS scale is taken to be of order $10^4$ GeV, the lifetime of the nucleon will be of order $10^{33}$ yrs. 

We note in passing that, unlike in the Pati-Salam model with conventional Higgs fields, there is no neutron-antineutron oscillation~\cite{Mohapatra:1980qe} induced in the present model.  That process is replaced by leptonic decay of the nucleon, owing to differences in selection rules between the conventional Higgs sector and the ones employed here.

\subsection{Quality of the parity solution to the strong CP problem}

The model presented here can be extrapolated all the way to the Planck scale, as it remains perturbative throughout.  To see this, we note that for $\mu > \mu_P \simeq 5.2 \times 10^{13}$ GeV, the one-loop beta function coefficients for the $SU(2)_{L,R}$ and $SU(4)_c$ are given by
\begin{equation}
b_{2 L} = b_{2R} = -\frac{8}{3},~~~b_4 = + 10,
\end{equation}
where $b_i$ are defined through $dg_i/d\,{\rm ln}\mu = b_i g_i^3/(16 \pi^2)$. The quoted values for $b_i$ include contributions from the chiral fermion families as well as three families of vector-like fermions and the Higgs scalars, which are assumed to have a common mass at $\mu_P$. Thus, the two $SU(2)$ gauge groups remain asymptotically free, while $SU(4)_c$ is not.  However, using $\alpha_4(\mu_P) \simeq 1/39.1$, we see that its value at the Planck scale $M_{\rm P}  = 1.22 \times 10^{19}$ GeV is $\alpha_4(M_{\rm P}) \simeq 1/19.7$, which is well within the perturbative regime. This allows us to explore higher-dimensional operators that may be induced by quantum gravity. In particular, such operators need not respect parity symmetry, which may disturb the parity solution to the strong CP problem.

An interesting aspect of the model is that higher-dimensional operators induced by gravity, which only have the Higgs scalars $H_L$ and $H_R$, do not upset the strong CP solution based on parity.  This is because one can choose the VEVs $\langle \chi_{L,R}^\nu\rangle$ of these fields to be real by gauge rotations. The lowest-dimensional operators that can potentially affect the strong CP parameter are of dimension-five involving $H_{L,R}$ and the chiral fermion fields $\psi_{L,R}$, given by
\begin{equation}
{\cal L}_{\rm gravity}^{(1)} = \frac{A}{M_{\rm P}} {\rm Tr}(\overline{\Psi}_LH_L) {\rm Tr}(H_R^\dagger \Psi_R) +  \frac{B}{M_{\rm P}} {\rm Tr}(\overline{\Psi}_LH_L H_R^\dagger \Psi_R)~.
\label{eq:MP}
\end{equation}
If parity is a global symmetry, quantum gravitational effects are expected to violate it, in which case the $3 \times 3$ matrices $A$ and $B$ in Eq. (\ref{eq:MP}) will be non-Hermitian.  These terms would then contribute to the strong CP parameter $\overline{\theta}$. 
In this case their coefficients should obey the conditions $|A_{ij}|, |B_{ij}| \leq 10^{-6}$ when $\kappa_R \sim 5 \times 10^{13}$ GeV is used, to be consistent with the constraint $|\overline{\theta}| < 10^{-10}$.

If the PS breaking scale obeys $\kappa_R \leq 10^7$ GeV, the operators of Eq. (\ref{eq:MP}) would be harmless, even with order one coefficients~\cite{Berezhiani:1992pq}. This situation can be easily realized in the left-right symmetric version of the universal seesaw model, which is based on the gauge group $SU(3)_c \times SU(2)_L \times SU(2)_R \times U(1)$~\cite{Babu:1989rb}. Since this gauge group has three independent gauge couplings, once parity symmetry is imposed, an arbitrary scale for parity restoration can be realized in this context.  In contrast, in the PS model developed here based on $SU(2)_L \times SU(2)_R \times SU(4)_c$, there are only two gauge couplings when parity symmetry is imposed. This sets the parity restoration scale $\mu_P \simeq 5.2 \times 10^{13}$ GeV, which makes the operators of Eq. (\ref{eq:MP}) non-negligible.  However, in comparison to the quality of the axion solution to the strong CP problem, where one must choose the coefficient of the $d=5$ operator $|\Phi|^4 \Phi/M_{\rm P}$ (where $\Phi$ is a SM singlet scalar which acquires a VEV of order $f_a \sim 10^{11}$ GeV) to be less than $10^{-50}$, the tuning needed in the PS model is very modest.

It is also possible that parity can emerge as a discrete gauge symmetry upon compactification of higher-dimensional theories. For the realization of CP as a discrete gauge symmetry, see Refs.~\cite{Choi:1992xp,Dine:1992ya}. When parity is realized as a discrete gauge symmetry, quantum gravity would conserve it, implying that the matrices $A$ and $B$ of Eq. (\ref{eq:MP}) would be hermitian, giving zero contributions to $\overline{\theta}$. There are parity-conserving operators at dimension-six which can contribute to $\overline{\theta}$ with the leading operator given by~\cite{Hall:2018let}
\begin{equation}
{\cal L}_{\rm gravity}^{(2)} = \frac{c}{M_{\rm P}^2} G_{\mu \nu} G_{\alpha \beta} \epsilon^{\mu \nu \alpha \beta} \left({\rm Tr}(H_L^\dagger H_L) - {\rm Tr}(H_R^\dagger H_R) \right)~
\label{eq:MP2}
\end{equation}
where $G_{\mu\nu}$ stand for the gluon field strength.  Inserting the VEVs $\kappa_L$ and $\kappa_R$ in Eq. (\ref{eq:MP2}) will induce a nonzero $\overline{\theta}$.  The experimental limit $|\overline{\theta}| \leq 10^{-10}$ requires $v_R < 10^{13}$ GeV (with the coefficient $c \sim 1$ in Eq. (\ref{eq:MP2}))~\cite{Hall:2018let}, which is nicely consistent with the parity restoration scale inferred from low energy data.

\section{Conclusions}

In this paper, we have developed a universal seesaw Pati-Salam model that has many attractive features. Quarks and leptons are unified in common multiplets of the gauge symmetry $SU(2)_L \times SU(2)_R \times SU(4)_c$ with lepton number identified as the fourth color~\cite{Pati:1974yy}. Parity is realized as a spontaneously broken symmetry with the scale of parity restoration inferred to be $\mu_P \simeq 5.2 \times 10^{13}$ GeV from extrapolating low-energy gauge couplings of the SM to high energies. As such, new particles of the model are not directly accessible to experiments.  However, the model developed here has many noteworthy features, which we summarize here.

In the universal seesaw version of the PS model, all fermion masses arise through mixing of the chiral families with heavy vector-like families of fermions. The Higgs sector of the model is very simple, consisting of a single pair of fields $\{H_L(2,1,4)+H_R(1,2,4)\}$. We have shown that vector-like fermions transforming as $(1,1,15)$ and $\{(1,1,10)_L+(1,1,10)_R\}$ can generate all quark and lepton masses consistently.  Potentially unacceptable mass relations among down-type quarks and charged leptons are evaded by the inclusion of one-loop radiative corrections induced by the up-type or down-type quark sector.  Explicit benchmark points have been presented that accommodate realistic fermion masses, which is highly nontrivial.

The structure of the theory is such that the neutrinos are massless at tree-level. This is because only one combination of $\nu_L$ and $\nu^c_L$ acquires mass via mixing with the SM singlet fermion contained in the $(1,1,15)$ fermion multiplet. Small Majorana neutrino masses are induced via one-loop corrections, which are shown to be compatible with neutrino oscillation data. One interesting aspect is that the model realizes a minimal radiative mechanism for neutrino mass generation~\cite{Grimus:2002nk,
AristizabalSierra:2011mn,Dev:2012sg,Lopez-Pavon:2015cga}.

One of the main motivations for constructing the universal seesaw version of the Pati-Salam model is that it can solve the strong CP problem by parity symmetry. The QCD $\theta$ term is zero by parity, and the quark contribution to $\overline{\theta}$ is zero due to the parity-symmetric Yukawa couplings and the seesaw structure of the quark mass matrices. We have examined the induced one-loop contributions to $\overline{\theta}$ in the model and found that a vast majority of diagrams vanish. However, unlike in the left-right symmetric theories based on the gauge group $SU(3)_c \times SU(2)_L \times SU(2)_R \times U(1)$ where all one-loop contributions vanish to $\overline{\theta}$~\cite{Babu:1988mw}, here neutral lepton contributions yield non-vanishing $\overline{\theta}$ at one-loop.  These contributions are parametrically suppressed by the ratios $(M_{10}/\kappa_R)$ or $(M_{15}/\kappa_R)$, which can be chosen to be sufficiently small, consistent with phenomenology. The small and finite $\overline{\theta}$ induced at the loop level is not far below the constraint from the neutron electric dipole moment.  Observation of a nonzero neutron EDM will provide further credence to the proposed model. Finally, the gauge and Yukawa interactions of the model conserve baryon number, but the Higgs self-interactions do not.  Nucleon decay of the type $n \rightarrow e^+ e^- \nu$ and $p \rightarrow e^+ e^- \nu \pi^+ \pi^0$ are predicted to occur, but with lifetimes well outside of the sensitivity of current experiments.

\section*{Acknowledgments}
This work is supported in part by the U.S. Department of Energy under grant no. DE-SC0016013. We thank Jogesh Pati for discussions and encouragement.

\bibliographystyle{JHEP}
\bibliography{biblio}

\end{document}